\documentclass[onecolumn,superscriptaddress,nofootinbib,prd,preprintnumbers,amsmath,amssymb]{revtex4}

\usepackage[pdftex]{graphicx}
\usepackage{dcolumn}
\usepackage{bm}
\usepackage{graphics}
\usepackage{rotating}
\usepackage{amssymb} 
\usepackage{comment}
\usepackage{multirow}
\usepackage{footnote}
\usepackage{comment}

\newcommand{\nc}{\newcommand}  

\def\beq{\begin{equation}}
\def\eeq#1{\label{#1}\end{equation}}
\def\eeqn{\end{equation}}

\newenvironment{Eqnarray}%
   {\arraycolsep 0.14em\begin{eqnarray}}{\end{eqnarray}}
\def\beqa{\begin{Eqnarray}}
\def\eeqa#1{\label{#1}\end{Eqnarray}}
\def\eeqan{\end{Eqnarray}}

\nc{\ra}{\rightarrow}  
\nc{\slsh}{\slash\hspace*{-0.22cm}}
\def\Re{{\cal R \mskip-4mu \lower.1ex \hbox{\it e}\,}}
\def\Im{{\cal I \mskip-5mu \lower.1ex \hbox{\it m}\,}}

\nc{\vev}[1]{ \left\langle {#1} \right\rangle }
\nc{\bra}[1]{ \langle {#1} | }
\nc{\ket}[1]{ | {#1} \rangle }
\nc{\fb}{\,{\rm fb}^{-1}}
\nc{\ev}{{\rm eV}}
\nc{\kev}{{\rm keV}}
\nc{\Mev}{{\rm MeV}}
\nc{\gev}{{\rm GeV}}
\nc{\tev}{{\rm TeV}}
\nc{\mev}{{\rm MeV}}

\def\del{\partial}
\def\Dslash{\not{\hbox{\kern-4pt $D$}}}
\def\dslash{\not{\hbox{\kern-2pt $\del$}}}
\def\pslash{\not{\hbox{\kern-2pt $p$}}}
\def\ETmiss{ \not{\hbox{\kern-4pt $E$}}_T }

\def\msb{{\bar{\ssstyle M \kern -1pt S}}}

\newcommand{\GeV}{\rm{ GeV}}

\begin{document}

\title{Prospects for a Heavy Vector-Like Charge 2/3 Quark $T$ search at the LHC with $\sqrt{s}$=14 TeV and 33 TeV
 \\A Snowmass 2013 Whitepaper}

\affiliation{Brown University, Providence, RI}
\affiliation{The State University of New York at Buffalo, NY}
\affiliation{Purdue University Calumet, Hammond, IN}

\author{Saptaparna Bhattacharya}\affiliation{Brown University, Providence, RI}
\author{Jimin George}\affiliation{The State University of New York at Buffalo, NY}
\author{Ulrich Heintz}\affiliation{Brown University, Providence, RI}
\author{Ashish Kumar}\affiliation{The State University of New York at Buffalo, NY}
\author{Meenakshi Narain}\affiliation{Brown University, Providence, RI}
\author{John Stupak III}\affiliation{Purdue University Calumet, Hammond, IN}



\begin{abstract}
\vspace{50 pt}
\begin{center}
{\bf Abstract}
 We present the prospects for the discovery or exclusion of heavy vector-like charge 2/3 quarks, $T$,
 in proton-proton collisions at two center-of-mass energies, 14 and 33 TeV at the LHC. 
 In this note, the pair production of $T$ quark and its antiparticle, with decays to 
 $W$ boson and a $b$ quark ($Wb$), a top quark and the Higgs boson ($tH$), and a top quark 
 and $Z$ boson ($tZ$) are investigated. Higgs boson decays 
 to $b\bar{b}$ and $W^{+}W^{-}$ final states are selected for this study. 
\end{center}

\end{abstract}

\maketitle

\newpage

\section{Introduction}
\label{intro}
The standard model comprises of three generations of chiral quarks. Many theoretical extensions 
of physics beyond the standard model posit the existence of vector-like quarks. 
Such quarks occur in models like the Little Higgs, extra dimensions or the minimally 
supersymmetric standard model. These quarks have the same left and right quantum numbers. 
They could be an SU(2) singlet or a doublet with the same left and right couplings. 
This condition on the couplings makes the interactions of these quarks purely vector-like. 
We look at a minimal extension of the standard model by introducing a vector-like top-like 
(of charge 2/3) heavy quark that couples to the third generation \cite{topprime_original}. 
The model we consider has only two parameters: the mass of the new heavy quark and the 
mixing with the third generation, parametrized as an angle (set to $\arcsin$ 0.02 during 
signal generation).

The non-chiral nature of these fermions makes considerations of anomaly cancellations redundant. 
Such vector-like fermions have been studied as part of the Little Higgs model where the new 
heavy fermion is introduced to cancel the Higgs mass quadratic divergence which 
results from the interaction with the top quark. We consider decays of the this heavy vector-like 
quark T into three possible decay modes: T $\rightarrow$ bW, T $\rightarrow$ t H and T $\rightarrow$ Z t . 
In the asymptotic region, where the mass of T is large, the model decouples from the standard model and 
the decay modes are equally shared by the Goldstone modes following the principle of Goldstone equivalence.
The decay topology of the the T quark leads to the presence of multiple W and Z bosons in the final state 
that produce leptons and multiple b-jets. The Higgs (at 125 GeV) is forced to decay into 
either $b\bar{b}$ or $W^{+}W^{-}$ since these are the modes that have the highest branching 
fraction and lead to final states that are sensitive to the analysis strategy.

\section{Signal and Background Samples}

The model for production and decay of the $T$ quark signal samples is implemented using 
the MadGraph~\cite{Madgraph} event generator. The {\sc pythia}~\cite{Pythia6},
Monte Carlo generator was used to perform parton fragmentation and hadronization of quarks and gluons.
Forty eight samples (six decay modes and eight mass points) corresponding to values of the $T$ mass from 500 to 1900 \GeV are generated for the analysis at $\sqrt{s}=14$~TeV, while sixty samples between $T$ mass of 700 to 3500 GeV were generated for the  analysis at $\sqrt{s}=33$~TeV.
The production cross section for $T$ quark signal are calculated using HATHOR~\cite{HATHOR} 
and corresponds to  an  (approximate) next-to-leading order calculation. 
The dominant Standard Model (SM) background processes for this analysis are 
$t \overline{t}$, $V+jets$, Drell Yan, $W^{\pm}W^{\pm}$, $WWW$, $t \overline{t}$W and $t \overline{t}$Z.
These were simulated with {\sc Madgraph} event generator as described in the snowmass SM background 
generation whitepaper~\cite{SnowmassBKG} and {\sc pythia} is used 
to take care of the parton fragmentation and hadronization. 
The background cross sections were obtained from the snowmass 
twiki~\cite{SnowmassBKGTwiki} and snowmass background whitepaper~\cite{SnowmassBKG}. 
For both background and signal samples, the events were processed through the 
the Snowmass detector using the  snowmass Delphes fast simulation and object reconstruction 
software~\cite{Delphes,SnowmassPerf}. The large statistics background samples were generated using the 
Open Science Grid infrastructure for snowmass~\cite{Snowmass-OSG}.

\section{Analysis Strategy}
\label{strategy}

\subsection{Single lepton Channels}
\label{onelep}

For the ``single lepton channel'', the study is performed in events in which one of 
the $W$ bosons (originating either from the decay of the heavy quark, 
or from the subsequent decay of a top quark) decays leptonically, while the other bosons decay into 
quark-antiquark pairs. These events have a large number of jets originating from $b$-quarks and 
hadronic $W$, $Z$, or $H$ boson decays.
For large $T$ masses it is likely that the jets from one or more boson decay are not resolved
which gives rise to jets that have substructure and a large invariant mass. 
In particular, we use ``top-tagging'' and ``W-tagging''  variables. 
The jets from highly boosted top quarks and W/Z bosons are clustered using the Cambridge-Aachen 
algorithm~\cite{CAjets} with a distance parameter of $R$=0.8. We call jets that are consistent with
originating from boosted $W$ or $Z$ boson decays as $W-jets$ and those consistent with a top quark as top-jets.

Selected events are required to have exactly one charged lepton (electron or muon) with $p_T>$30 GeV, large missing transverse 
energy $MET>$20 GeV, and at least three jets with $|\eta|<$2.5 and $p_T>$200, 90, 50 GeV.
The dominant standard-model (SM) background is $t\bar{t}$ production that results in the same signature.
Other SM background contributions include electroweak processes: $W$+jets, $Z$+jets, single top quark, 
and diboson production, as well as multijet events.
These background processes are characterized by smaller lepton and jet transverse momenta and 
lower jet multiplicities than those in heavy quark decays.

The search for heavy quark is performed by classifying selected events based on the number of final-state jets.
\begin{itemize}
\item 3-jets: $p_T^{jet1}>$200 GeV, $p_T^{jet2}>$90 GeV, and $p_T^{jet3}>$50 GeV. At least one jet with being consistent with $W-jet$ is required. 
If there is $b$-jet in the event, the highest $p_T$ $b$-jet should have $p_T>$150 GeV.
\item 4-jets: at least 4 jets with $p_T^{jet1}>$200 GeV, $p_T^{jet2}>$90 GeV, $p_T^{jet3}>$50 GeV, and $p_T^{jet4}>$35 GeV. 
If there is $b$-jet in the event, the highest $p_T$ $b$-jet should have $p_T>$150 GeV.
\end{itemize}
These events are further subdivided into categories based on multiplicity of b-jets : (a) no b-jet, (b) 1 b-jet,
(c) 2 b-jets, and (d) at least 3 b-jets. The analysis thus considers a total of sixteen channels split into the two
leptonic final states which differ in their selection criteria. With the combination of electron and muon channels, we have eight different categories.
For each of these categories, the scalar sum ($H_T$) of the transverse momenta of the jets is used to test for 
the presence of the signal.

\begin{itemize}
\item $e3+\mu3$: $\geq$ 3 jets with $p_T^{jet1}>$200 GeV, $p_T^{jet2}>$90 GeV, and $p_T^{jet3}>$50 GeV. 
At least one jet with being consistent  with $W-jet$ is required. 
 These events are further subdivided into four categories: 
\begin{itemize}
\item $e3+\mu3,0b$: No b-jet in the event.
\item $e3+\mu3,1b$: One b-jet in the event.
\item $e3+\mu3,2b$: Two b-jets in the event.
\item $e3+\mu3,3b$: $\geq$ 3 b-jets in the event.
\end{itemize}
\item $e4+\mu4$: $\geq$ 4 jets with$p_T^{jet1}>$200 GeV, $p_T^{jet2}>$90 GeV,
$p_T^{jet3}>$50 GeV, and $p_T^{jet4}>$35 GeV, but no jet consistent
with being a $W$-jet.
These events are further subdivided into four categories: 
\begin{itemize}
\item $e4+\mu4,0b$: No b-jet in the event.
\item $e4+\mu4,1b$: One b-jet in the event.
\item $e4+\mu4,2b$: Two b-jets in the event.
\item $e4+\mu4,3b$: $\geq$ 3 b-jets in the event.
\end{itemize}

\end{itemize}

The above mentioned event selections are valid for the 14 TeV analysis.
The analysis based on $\sqrt(s)=$33 TeV employs similar strategy with
some of the object selections optimized for better signal versus background
discrimination. The 33 TeV analysis uses significantly tighter selections
on the jet kinematics and missing transverse energy. The selections which are
different from the 14 TeV analysis are the following.

\begin{itemize}
\item $p_T^{jet1}>$350 GeV, $p_T^{jet2}>$200 GeV
\item In the presence of  $b$-tagged jet(s) in the event, the highest $p_T$ $b$-jet should have
$p_T>$200 GeV
\item Missing transverse energy $>$75 GeV
\end{itemize}

\subsection{Multi-lepton Channels}
\label{multilep}

We look at final states that contain at least two leptons in four mutually exclusive event categories (defined later in the section). 

In order to  suppress backgrounds from SM processes, and enhance the signal discrimination power, 
we use a few kinematic variables, defined as follows:
\begin{itemize}
\item{$H_T$:} is defined as the scalar sum of the transverse momenta of all the selected AK5 jets. 
\item{$S_T$:} is the scalar sum of the $H_T$, leptons and missing transverse energy.
\item{minM$_{lb}$:} is the minimum invariant mass of a lepton and a b-jet and sensitive to the mass of the $T$ quark.
\end{itemize}

In addition, we use jet substructure variables to enhance the signal yield. 
In the following sections, we use the phrase jet-constituent to designate the number of constituents 
from any given jet. Therefore, an AK5 jet (a jet clustered using the anti-K$_{T}$ algorithm 
with a distance parameter, $R$ of 0.5) has one constituent and a ``W-jet'' has two constituents while a 
``top-jet'' has three constituents. 
Our selection criteria involves the requirement of a minimum number of jet constituents, 
instead of requiring a certain number of independent AK5 or $W$-tagged ot a top-tagged jet. 

The details of the selection criteria for each category are given below.

\begin{itemize}
\item Opposite signed (OS) leptons: Here we require specifically two leptons with an opposite sign. 
The major irreducible background is from $t\bar{t}$ and Drell-Yan processes. In this channel, we 
further sub-divide final states into two categories:

\item OS23: This category is constructed to be solely sensitive to the TT$\rightarrow$ bWbW mode. 
In this category, we require 2 or 3 jets constituents, veto on events within the Z-boson mass window and require at least 1 b-tagged jet. 
In addition to the above requirements, we make additional restrictive cuts on the following event variables: $S_T$, $H_T$, and minM$_{lb}$ which is sensitive to the mass of the T quark. We require minM$_{lb} > 180 GeV$, H$_{T} >$ 700 GeV and  S$_{T} >$ 900 GeV.

\item OS5+: This category is constructed to be sensitive to modes that have multiple jets in the final state 
(from T$\rightarrow$ tH decay modes). 
We do not require a Z-boson mass veto in this category so as to be sensitive to final states that arise from T $\rightarrow$ tZ decays. We require at least two b-tagged jets and a minimum of 5 jet constituents in this category to minimize the Drell-Yan background. We find that the variables that give us the best discriminating power between signal and background are H$_{T}$ and S$_{T}$. We require H$_{T} >$ 900 GeV and  S$_{T} >$ 1000 GeV.

\item Same signed (SS) leptons: Here we specifically require two leptons with the same sign. 
The dominant background in this channel is from fake leptons that arise due to misidentification of jets as leptons. 
CMS searches at 7TeV and 8TeV show that these instrumental backgrounds are non-negligible and in fact are 
the dominant backgrounds in this channel \cite{cms_search1}, \cite{cms_search2} . Other sources of same signed 
leptons are from rare standard model processes like $t\bar{t}Z$, $t\bar{t}W$, diboson and triboson processes. In this category, we require, at least one btagged jet, missing E$_{T}$$>$ 30 GeV and a minimum of 2 jet constituents.  We further make stringent requirements on H$_{T}$ and S$_{T}$. We require H$_{T} > 600$ GeV and S$_{T} > 800$ GeV.

\item Multi-leptons: Here we specifically require three or more leptons. The dominant backgrounds are standard model processes that lead to three-lepton final states, such as diboson decays and contributions from fake leptons. We require at least one b-tagged jet, missing E$_{T}$$> $30 GeV and a minimum of 2 jet constituents. We put harsh requirements on H$_{T}$ and S$_{T}$ in this category as well. We require H$_{T} > 800$ GeV and S$_{T} > 900$ GeV.
\end{itemize}

Since these studies are solely based on simulated events, we do not account for instrumental backgrounds at this stage. The backgrounds estimated for an analysis carried out at 8TeV~\cite{cms_search2} could, in principle, be scaled but the pitfall associated with such a scaling is that it is not a good indicator of the relative contribution of the background, given that different sets of cuts were applied at different center of mass energies.

The cuts in each category were optimized by computing Signal/$\sqrt{\rm Background}$ and calculating efficiencies for particular sets of cuts and comparing these efficiencies to the published results at 8TeV~ \cite{cms_search2}.

In the multilepton channel, the following changes were made to optimize the analysis for $\sqrt s$=33 TeV:

\begin{itemize}

\item OS23: In the OS23 category, the 4 constituent bin is used. We require minM$_{lb} >$ 220 GeV, H$_{T} >$ 1000 GeV and S$_{T} >$ 1200 GeV.

\item OS5+: Here H$_{T}$ and S$_{T}$ are set to 1800 GeV and 2000 GeV respectively.  

\item SS: We require H$_{T} >$ 900 GeV and S$_{T} >$ 1000 GeV.  

\item Multi-leptons: We require H$_{T} >$ 900 GeV and S$_{T} >$ 1000 GeV. 

\end{itemize}

\section{Event Yields at $\sqrt{s}$=14 TeV }
\label{yields300}

\begin{itemize}
\item{\bf Single-lepton Channels}
Table~\ref{tab:rates14tev300PU0} shows expected signal and background event yields determined assuming an
integrated luminosity of 300 fb$^{-1}$ of data at $\sqrt{s}$=14 TeV and no additional pileup events
for  the different event categories. 
The background contributions from different electroweak processes are combined into a single background category.
In tables~\ref{tab:rates14tev300PU50} and ~\ref{tab:rates14tev3000PU140},
 the expected signal and background yields for two different run scenarios are quoted: 
(a) 300 fb$^{-1}$  with an average $<\mu_{\rm PU}>$ of 50 additional pileup interactions per crossing  (LHC run 2, Phase 2)  
and (b) 3000 fb$^{-1}$ with an average $<\mu_{\rm PU}>$ of 140 additional pileup interactions per crossing (HL-LHC).

\item{\bf Multiple-lepton channels}
In table~\ref{tab:multirates14tev300PU0}, the expected signal and background event yields are tabulated, 
for  the different event categories, assuming an integrated luminosity of 300 fb$^{-1}$ of data accumulated 
at $\sqrt{s}$=14 TeV and no additional pileup events.
In tables~\ref{tab:multirates14tev300PU50} and ~\ref{tab:multirates14tev3000PU140},
 the expected signal and background yields for two different run scenarios: 
(a) LHC run 2, Phase 2, with 300  fb$^{-1}$ dataset and $<\mu_{\rm PU}>$=50, 
and (b) HL-LHC with 3000 fb$^{-1}$ dataset with $<\mu_{\rm PU}>$=140.
The background events are summed and reflect the contribution of various backgrounds in the different event categories as described in Section~\ref{multilep}.

\end{itemize}

\begin{table}[!h]\small
\begin{center}
\begin{tabular}{|l|c|c|c|c|c|c|c|c|}\hline\hline
Mass (GeV) & $e3+\mu3$,0$b$ &  $e3+\mu3$,1$b$   &    $e3+\mu3$,2$b$  &   $e3+\mu3$,3$b$   &  $e4+\mu4$,0$b$    &   $e4+\mu4$,1$b$    &   $e4+\mu4$,2$b$   &  $e4+\mu4$,3$b$    \\
\hline
\multicolumn{8}{|l|}{\bf Signal Event Yields} \\
500   &      4709           &   9809           &    11419          &  4901             &   11828              &   26439           &   34885              &   16800   \\
700   &      1295           &   3333           &     4168          &  1857             &   1520              &    4074           &    5579              &    3148   \\
900   &      329.6          &   1016           &     1246          &   586             &    249              &     757           &    1036              &     568  \\
1100  &      92.0           &    301           &      376          &   173.4           &    54.1             &     183           &     241              &     127 \\
1300  &      27.7           &    95.6          &      120.5        &   53.7            &    15.2             &     51.6          &     69.0             &     34.7 \\
1500  &      9.03           &    31.1          &       39.0        &   16.9            &     4.55            &     16.9          &     22.4             &     10.6 \\
1700  &      2.89           &    10.4          &       13.1        &   5.66            &     1.67            &     6.21          &     8.02             &     3.73  \\
1900  &      0.99           &     3.62         &       4.49        &   1.94            &     0.63            &     2.34          &     2.98             &     1.35 \\
\hline
\multicolumn{8}{|l|}{\bf Background Event Yields} \\
$t\bar{t}$  &    69251        &  84075             &  72516           &  11219               &   418164           &   375164            &  331133           &  52361   \\
Electroweak  &    146485       &  16193             &   2581           &  161                 &   1104532          &  80507              & 15173             & 1236   \\
\hline
Total  Background      &    215736	      &  100268            & 75097           &  11380               &   1522696          &   455671            & 346306            & 53597           \\
\hline
\end{tabular}
\caption{Number of expected signal and background events for 300 fb$^{-1}$ of pp collisions at 14 TeV for $<\mu_{\rm PU}>$=0 in different event categories.}
\label{tab:rates14tev300PU0}
\end{center}
\end{table}

\begin{table}[!h]\small
\begin{center}
\begin{tabular}{|l|c|c|c|c|c|c|c|c|}\hline\hline
Mass (GeV) & $e3+\mu3$,0$b$ &  $e3+\mu3$,1$b$   &    $e3+\mu3$,2$b$  &   $e3+\mu3$,3$b$   &  $e4+\mu4$,0$b$    &   $e4+\mu4$,1$b$    &   $e4+\mu4$,2$b$   &  $e4+\mu4$,3$b$    \\
\hline
\multicolumn{8}{|l|}{\bf Signal Event Yields} \\
500   &      4908           &   10400          &    12532          &  5204             &   11450              &   26134           &   33489              &   16010   \\
700   &      1293           &   3484           &     4291          &  1938             &   1568              &    4079           &    5458              &    3024   \\
900   &      340.4          &   1014           &     1273          &   587             &    246              &     773           &    1011              &     559  \\
1100  &      92.8           &    304           &      380          &  174.3           &    57.6             &     181           &     238              &     127 \\
1300  &      28.6           &    96.2          &      120.8        &  54.0            &    15.6             &     53.5          &     68.5             &     35.1 \\
1500  &      8.92           &    31.4          &      39.0        &   17.3            &     4.99            &     17.5          &     22.6             &     10.9 \\
1700  &      2.89           &    10.6          &      13.2        &   5.78            &     1.77            &     6.39          &     8.31             &     3.72  \\
1900  &      1.01           &    3.63         &       4.53        &   1.96            &     0.64            &     2.42          &     3.14             &     1.39 \\
\hline
\multicolumn{8}{|l|}{\bf Background Event Yields} \\
$t\bar{t}$  &    78239        &  96423             &  80897           &  12981               &   407214           &   353340            &  318334       &  51805   \\
Electroweak  &   186598       &  20463             &   3019           &  255                 &   1220454          &  86384              &   16872       & 1106   \\
\hline
Total Background   &    264838	      &  116886            &  83916           &  13236               &   1627668          &   439724            & 335206      & 52911           \\
\hline
\end{tabular}
\caption{Number of expected signal events for 300 fb$^{-1}$ of pp collisions at 14 TeV for $<\mu_{\rm PU}>$=50 in different event categories.}
\label{tab:rates14tev300PU50}
\end{center}
\end{table}

\begin{table}[!h]\small
\begin{center}
\begin{tabular}{|l|c|c|c|c|c|c|c|c|}\hline\hline
Mass (GeV) & $e3+\mu3$,0$b$ &  $e3+\mu3$,1$b$   &    $e3+\mu3$,2$b$  &   $e3+\mu3$,3$b$   &  $e4+\mu4$,0$b$    &   $e4+\mu4$,1$b$    &   $e4+\mu4$,2$b$   &  $e4+\mu4$,3$b$    \\
\hline
\multicolumn{8}{|l|}{\bf Signal Event Yields} \\
\hline
500   &      67927         &   143746          &    166297          &  69421             &   102273              &   235739           &   288660              &   118514   \\
700   &      16281          &   41495           &     49359          &  22218             &   15097              &    38205           &    47904              &    24971   \\
900   &      3830          &    11370           &     13987          &   6356             &    2541              &      7428           &    9617              &     4786  \\
1100  &      1050           &    3325           &     4025           &  1878              &    549               &     1856           &     2362              &     1150 \\
1300  &      313           &     1044          &      1248           &  554               &    167               &     562            &     708               &     330 \\
1500  &      100           &      332          &      408           &   176               &     52.1              &     188           &     239               &     105 \\
1700  &      32.2           &     112          &      138           &   58.5              &     19.0             &     70.2           &     88                &     37.6  \\
1900  &      10.5           &    38.5         &       47.4          &   20.1              &     7.31             &     27.3           &     34.0              &     14.1 \\
\hline
\multicolumn{8}{|l|}{\bf Background Event Yields} \\
$t\bar{t}$  &    1465927        &  1580542             &  1245725           &  189075               &   3577667           &   3000107            &  2496862       &  363147   \\
Electroweak  &   4662073       &  426720             &  65614           &  4533                 &   15093341          &  989621              &   157748       & 9760   \\
\hline
Total Background  &    6128000       &  2007262            &  1311339           &  193608               &   18671008          &   3989728            & 2654610      & 372907           \\
\end{tabular}
\caption{Number of expected signal events for 3000 fb$^{-1}$ of pp collisions at 14 TeV for $<\mu_{\rm PU}>$=140 in different event categories.}
\label{tab:rates14tev3000PU140}
\end{center}
\end{table}

\begin{table}[!h]\small
\begin{center}
\begin{tabular}{|l|c|c|c|c|c|c|c|c|}\hline\hline
Mass (GeV) & $e3+\mu3$,0$b$ &  $e3+\mu3$,1$b$   &    $e3+\mu3$,2$b$  &   $e3+\mu3$,3$b$   &  $e4+\mu4$,0$b$    &   $e4+\mu4$,1$b$    &   $e4+\mu4$,2$b$   &  $e4+\mu4$,3$b$    \\
\hline
\multicolumn{8}{|l|}{\bf Signal Event Yields} \\
\hline
700   &      148256      &   369044       &    444898       &  185027     &  78789    &   235132           &   285382              &   115089   \\
1700   &     1773          &     5922           &       7046          &     2928            &   1189              &    4036           &    4907              &    2048   \\
1900   &     810          &      2846           &        3344          &     1401           &   651              &      2153           &    2694              &     1064  \\
2100  &      410           &     1383           &        1646           &    708            &  337               &     1246           &     1534              &     618 \\
2300  &      198           &      713          &          849           &     364            & 199               &     730            &     900               &     345 \\
2500  &      103           &      371          &         451           &     192            &117              &     428           &     538               &      210 \\
2700  &      53.7           &      200          &         239           &    103           &71.3             &     265           &     327                &    127  \\
3100  &      16.0           &      62.0         &         75.3          &     33.6           &26.3             &     103           &     128              &     49.0 \\
3300  &      9.4           &       35.3         &          43.3          &    19.3           &17.0             &     64.3           &    80.9              &     30.9 \\
3500  &      5.5           &      20.5         &          25.3          &     11.3          &10.5             &     41.8           &     52.2              &     19.8 \\
\hline
\multicolumn{8}{|l|}{\bf Background Event Yields} \\
$t\bar{t}$  &    5302513        &  6372292             &  4340076           &  798877      &1650291           &   2031398            &  1437906       &  241251   \\
Electroweak  &   12170779       &  1442692             &  217466         &  14408          & 7087186          &  733038              &   114495       & 7848   \\
\hline
Total Background  &    17473292   &  7814984     &  4557542           &  813285       &  8737477  &  2764436           & 1552401      & 249099           \\
\end{tabular}
\caption{Number of expected signal events for 3000 fb$^{-1}$ of pp collisions at 33 TeV for
$<\mu_{\rm PU}>$=140 in different event categories.}
\label{tab:rates33tev3000PU140}
\end{center}
\end{table}

\begin{table}[!h]
\begin{center}
\begin{tabular}{|l|c|c|c|c|c|c|c|c|}\hline\hline
Mass (GeV) &  OS23 & OS5+ & SS & Multi-leptons ($\geq$ 3)   \\
\hline
\multicolumn{5}{|l|}{\bf Signal Event Yields} \\
500   &      691.8  & 1311.6   & 772.6 & 629.3\\ 
700   &       323.5  &  576.9 & 213.1 & 242.5\\
900   &       143.3  & 170.6 & 50.6 & 87.4\\
1100  &     54.0  & 47.8 & 12.3 & 27.6\\
1300  &     19.9 & 13.7  & 3.3 & 7.8\\
1500  &     7.0 & 4.1 & 1.0 & 2.2\\
1700  &     2.6  & 1.2 & 0.3 & 0.7\\
1900  &     9.8 & 0.4 & 0.1 & 0.2\\ \hline
{\bf Total Background }     &     197.5 & 1467.1  & 399.7 &   67.4    \\
\hline
\end{tabular}
\caption{Number of expected signal and background events for 300 fb$^{-1}$ of pp collisions at 14 TeV in different event categories
for $<\mu_{\rm PU}>$=0 pileup scenario.}
\label{tab:multirates14tev300PU0}
\end{center}
\end{table}

\begin{table}[!h]
\begin{center}
\begin{tabular}{|l|c|c|c|c|c|c|c|c|}\hline\hline
Mass (GeV) &  OS23 & OS5+ & SS & Multi-leptons ($\geq$ 3)   \\
\hline
\multicolumn{5}{|l|}{\bf Signal Event Yields} \\
500   &      629.3 &  1857.8   & 911.6 & 640.8\\ 
700   &       243.4  &   780.6 & 228.9 & 242.8\\
900   &      119.9   & 232.1 & 53.7 & 86.7\\
1100  &     46.4 &  63.8 & 14.4 & 26.5\\
1300  &     17.1 & 18.2  & 3.7 & 7.6\\
1500  &     5.8 & 5.4 & 1.2 & 2.3\\
1700  &     2.1  & 1.7 & 0.4 & 0.7\\
1900  &     8.3 & 0.6 & 0.1 & 0.2\\ \hline
{\bf Total Background}      &     235.3 & 1804.0 & 421.2 &   69.4    \\
\hline
\end{tabular}
\caption{Number of expected signal and background events for 300 fb$^{-1}$ of pp collisions at 14 TeV in different event categories for $<\mu_{\rm PU}>$=50 pileup scenario.}
\label{tab:multirates14tev300PU50}
\end{center}
\end{table}

\begin{table}[!h]
\begin{center}
\begin{tabular}{|l|c|c|c|c|c|c|c|c|}\hline\hline
Mass (GeV) &  OS23 & OS5+ & SS & Multi-leptons ($\geq$ 3)   \\
\hline
\multicolumn{5}{|l|}{\bf Signal Event Yields} \\
500   &      3622.5 &  19168.6  & 10348.1 & 6005.0\\ 
700   &      1738.9  &   8617.9 & 2716.8 & 2381.7\\
900   &       777.5  & 2875.6 & 665.8 & 881.8\\
1100  &     297.5 &  867.2 & 168.1 & 275.7\\
1300  &     109.1 & 256.5  & 46.8 & 78.9\\
1500  &     38.4 & 79.3 & 14.8 & 22.7\\
1700  &     14.7  & 25.9 & 4.8 & 6.9\\
1900  &     5.6 & 8.7 & 1.7 & 2.2\\ \hline
{\bf Total Background}      &    1378.6  & 23473.7 &  4403.2 &   691.4    \\
\hline
\end{tabular}
\caption{Number of expected signal and background events for 3000 fb$^{-1}$ of pp collisions at 14 TeV in different event categories for $<\mu_{\rm PU}>$=140 pileup scenario.}
\label{tab:multirates14tev3000PU140}
\end{center}
\end{table}

\begin{table}[!h]
\begin{center}
\begin{tabular}{|l|c|c|c|c|c|c|c|c|}\hline\hline
Mass (GeV) &  OS23 & OS5+ & SS & Multi-leptons ($\geq$ 3)   \\
\hline
\multicolumn{5}{|l|}{\bf Signal Event Yields} \\
700   &     17451.9   &  12970.1 & 17225.9 & 30396.1\\ 
1700   &   1022.0     &   1033.7 &  257.2 & 352.0 \\
1900   &  552.6      & 532.4 & 125.8 & 154.0 \\
2100  &     320.7  & 386.6  &  81.4 & 99.7 \\
2300  &   171.0   &  157.4 & 34.8 & 30.5\\
2500  &   97.5   &  88.7&  19.5 & 17.5 \\
2700  & 54.5     &  50.5 & 11.4 & 8.6\\
3100  &  20.5   & 17.8 & 4.4 & 2.3\\ 
3300 & 12.4  & 10.5 & 2.3  & 1.3\\ 
3500 & 8.0    & 6.6 & 1.4 & 0.7 \\ \hline
{\bf Total Background }     & 7154.6   & 30150.3 & 13655.9 &  6400.4   \\
\hline
\end{tabular}
\caption{Number of expected signal and background events for 3000 fb$^{-1}$ of pp collisions at 33 TeV in different event categories
for $<\mu_{\rm PU}>$=140 pileup scenario.}
\label{tab:multirates33tev300PU140}
\end{center}
\end{table}

\section{Expected significance and exclusion limits}
The signal significance and exclusion limits were computed using a Bayesian approach with the ``theta statistical analysis package''~\cite{theta-stat}. 
For the single lepton channel, the  signal significance and exclusion limits were computed using 16 
independent categories, to account for eight classifications based on $H_T$ and number of bjets, 
and two channels ($e+$jets and $\mu+$jets).
The  signal significance and exclusion limits for the multilepton channels 
were computed using a 12 binned histogram, as there are four categories (OS23, OS5+, SS, and trileptons) 
each for the three channels  ($eee$, $e\mu\mu+e e \mu$ and $\mu\mu\mu$).
The final combined limits and signal significances as a function of the expected luminosity and 
pileup scenarios were also computed. 
The mutually exclusive categories constructed in the lepton+jets channel and the multi-lepton channel 
were used to obtain the combined significance and exclusion limits. The lepton+jets templates and the 
multi-lepton templates were fed as independent models to ``theta''. 

\subsection{$\sqrt{s}$=14 TeV }
\begin{figure}[!h]
\includegraphics[width=0.32\textwidth]{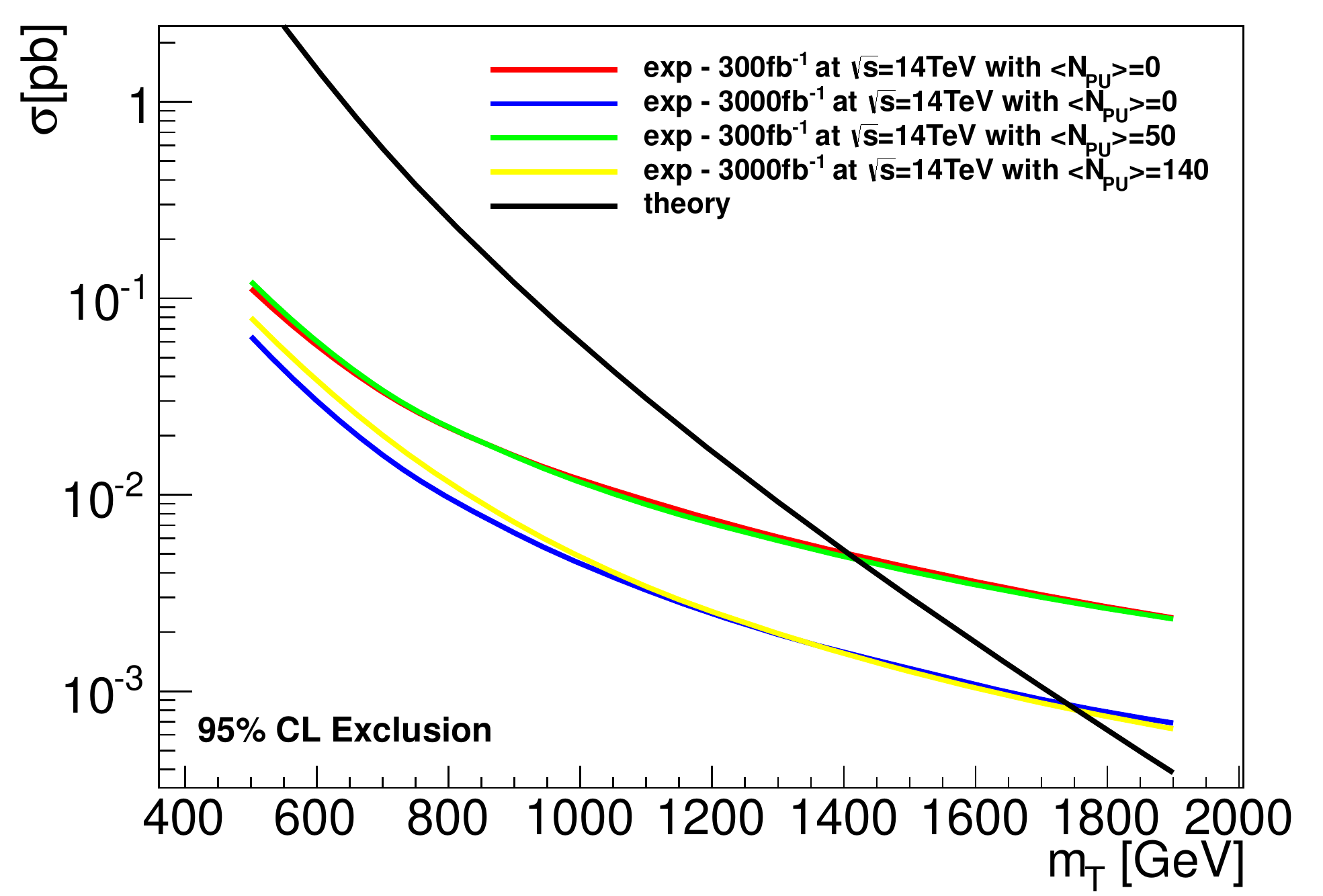}
\includegraphics[width=0.32\textwidth]{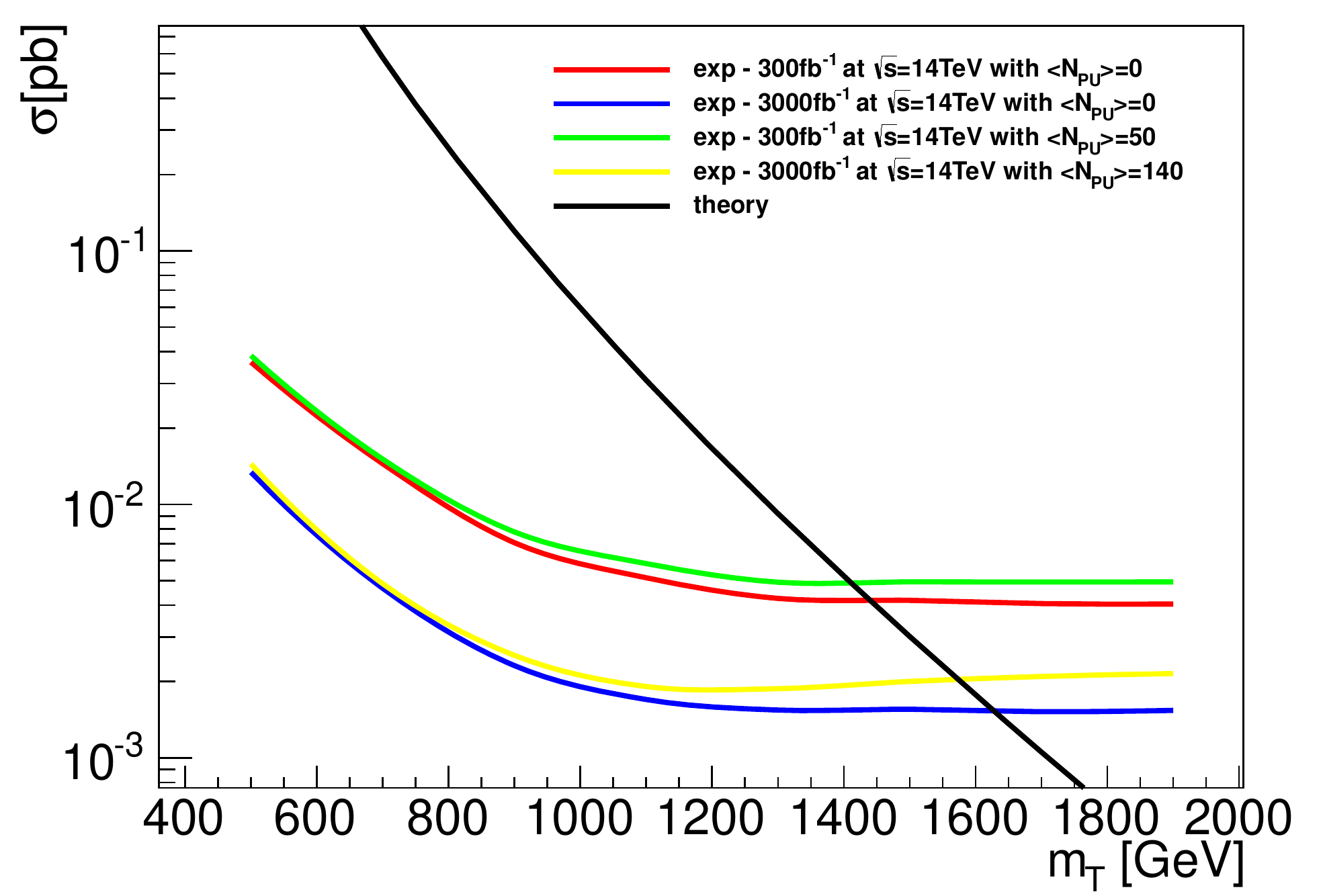}
\includegraphics[width=0.32\textwidth]{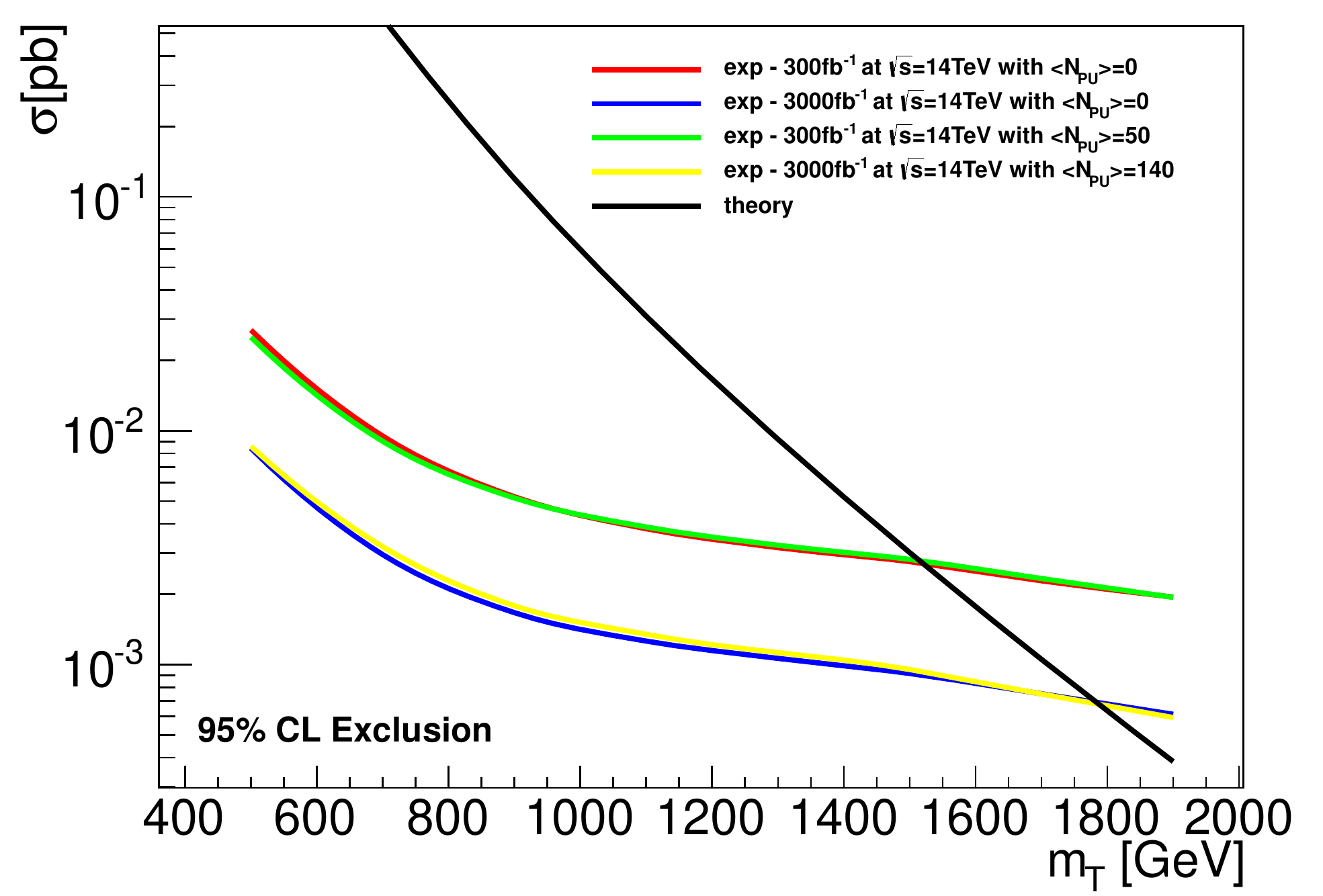}
\caption{Expected 95\% C.L. limits for $T$ quark pair production in the the $l+$ jets channel (left), multilepton channel (middle) and 
combined (right).\label{fig:explimit}}
\end{figure}

\begin{figure}[!h]
\includegraphics[width=0.32\textwidth]{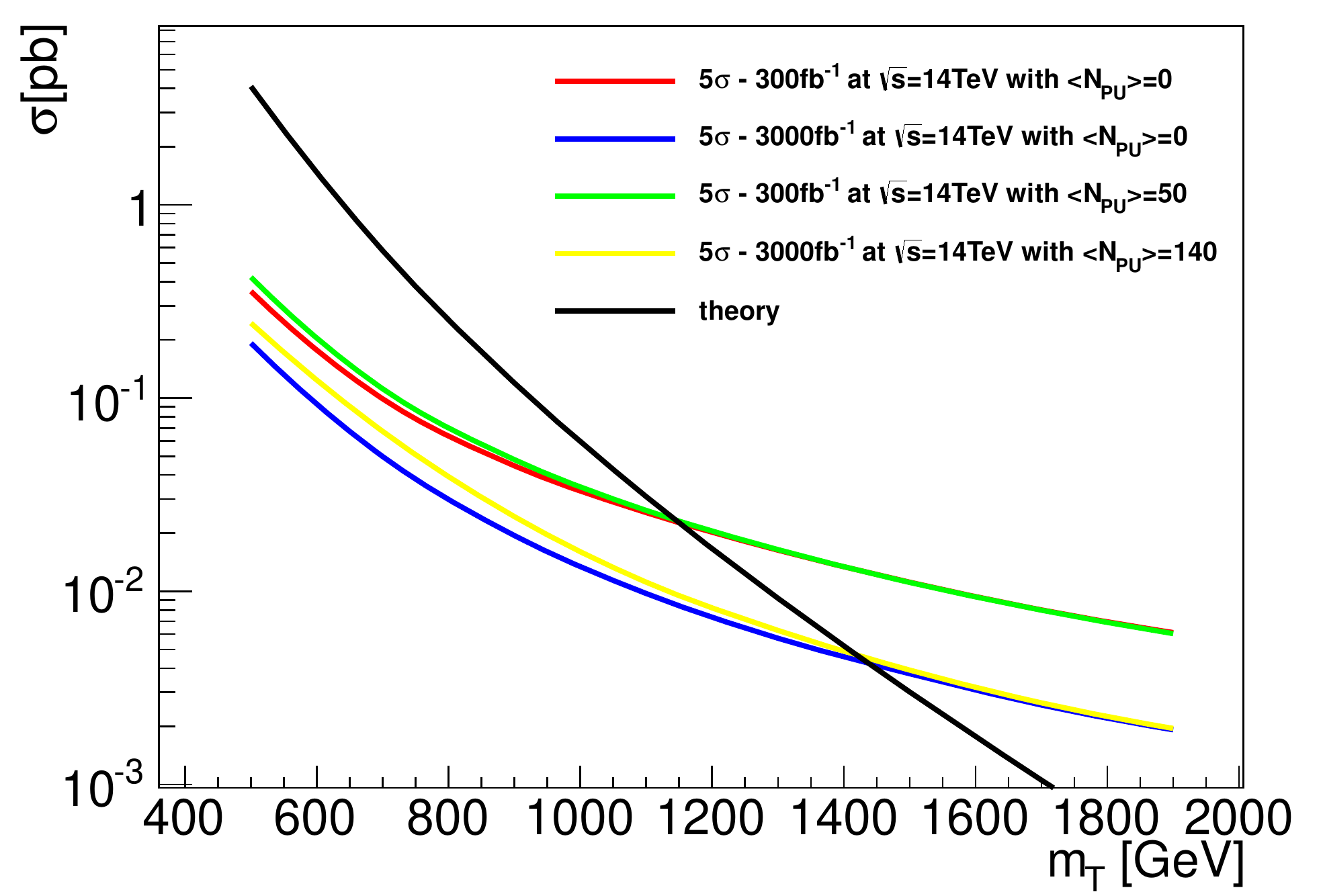}
\includegraphics[width=0.32\textwidth]{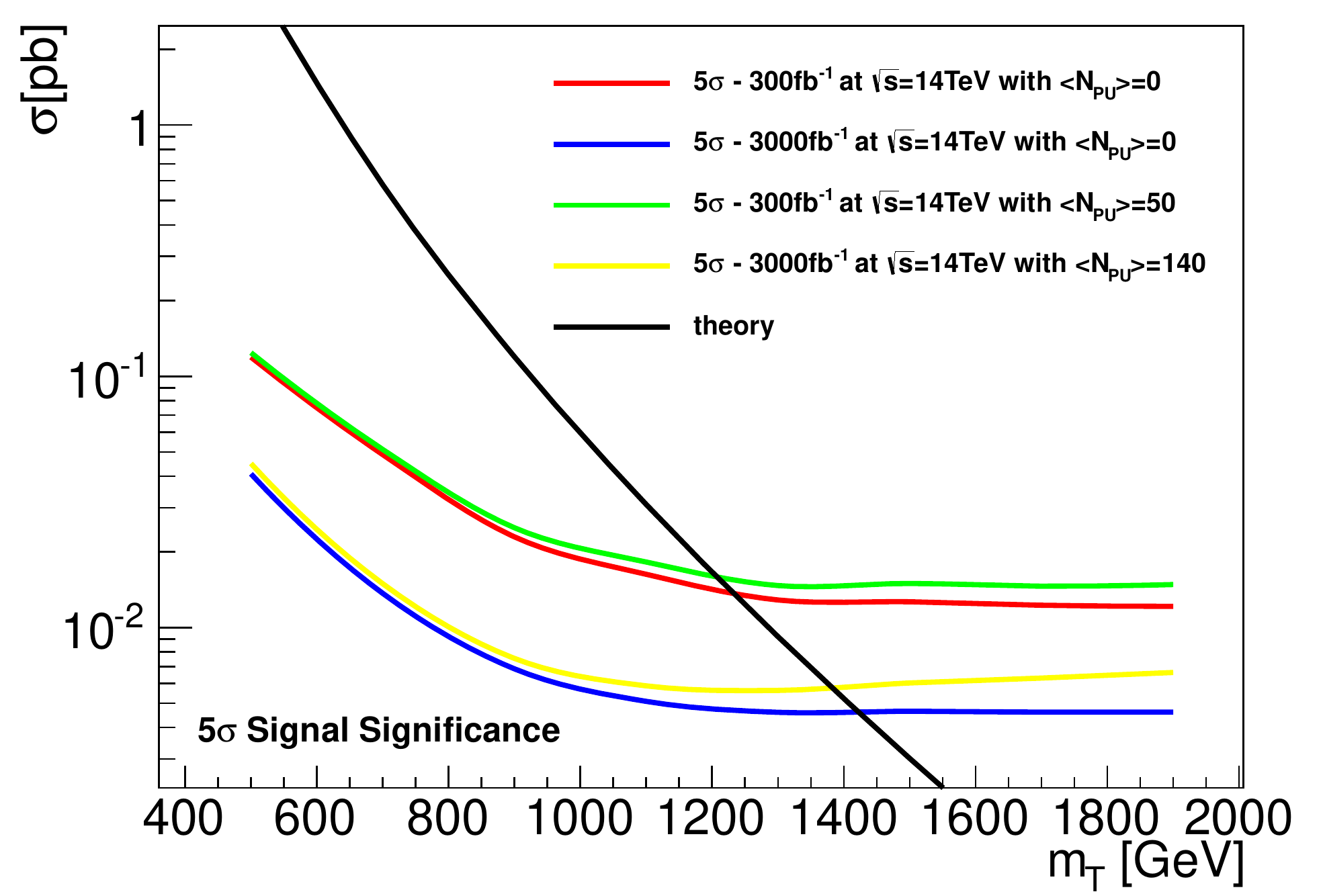}
\includegraphics[width=0.32\textwidth]{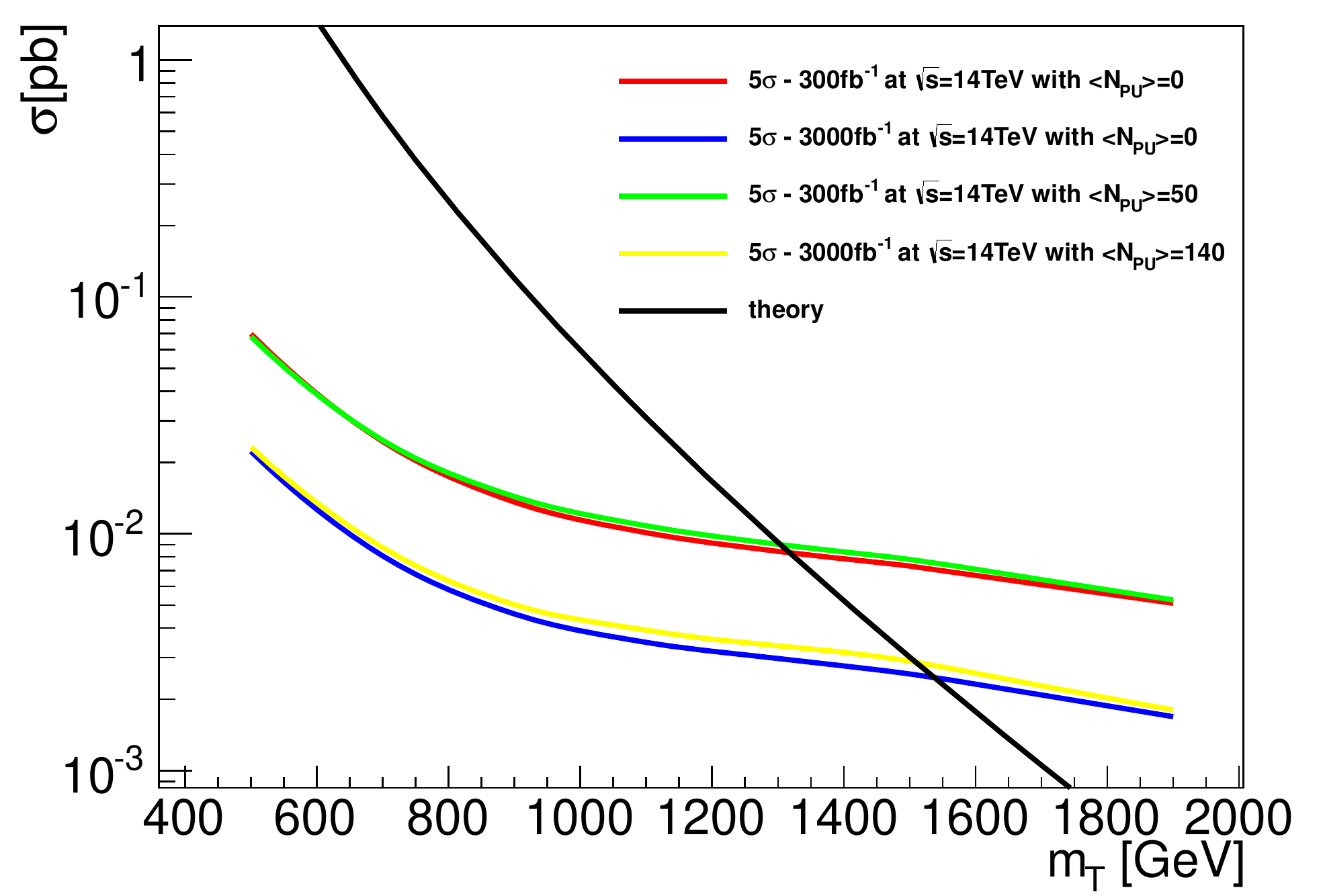}
\caption{Expected sensitivity for a 5$\sigma$ $T$ quark pair production signal 
in the the $l+$ jets channel (left), multilepton channel (middle) and 
combined (right).\label{fig:exp5sigma}}
\end{figure}

\begin{figure}[!h]
\includegraphics[width=0.32\textwidth]{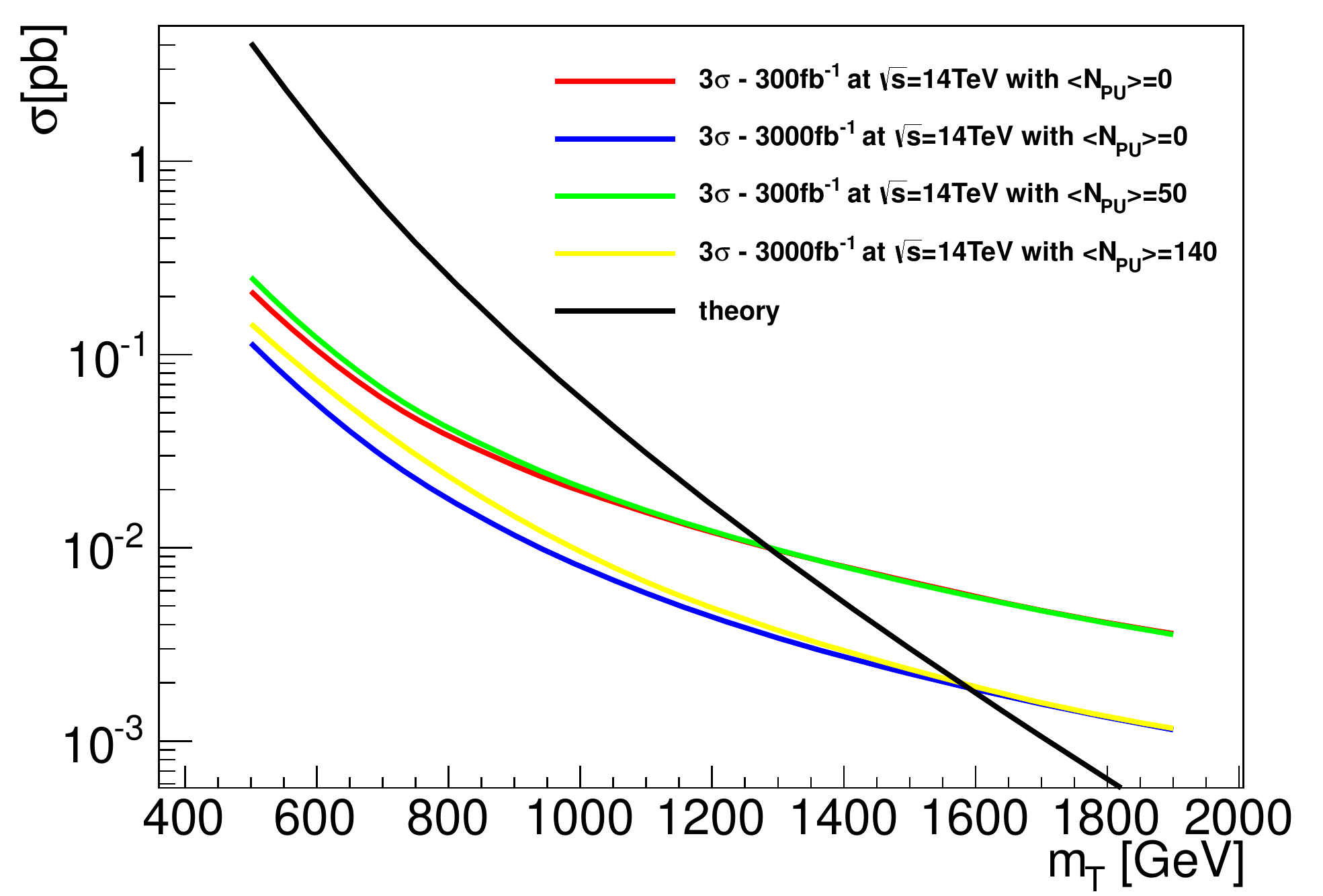}
\includegraphics[width=0.32\textwidth]{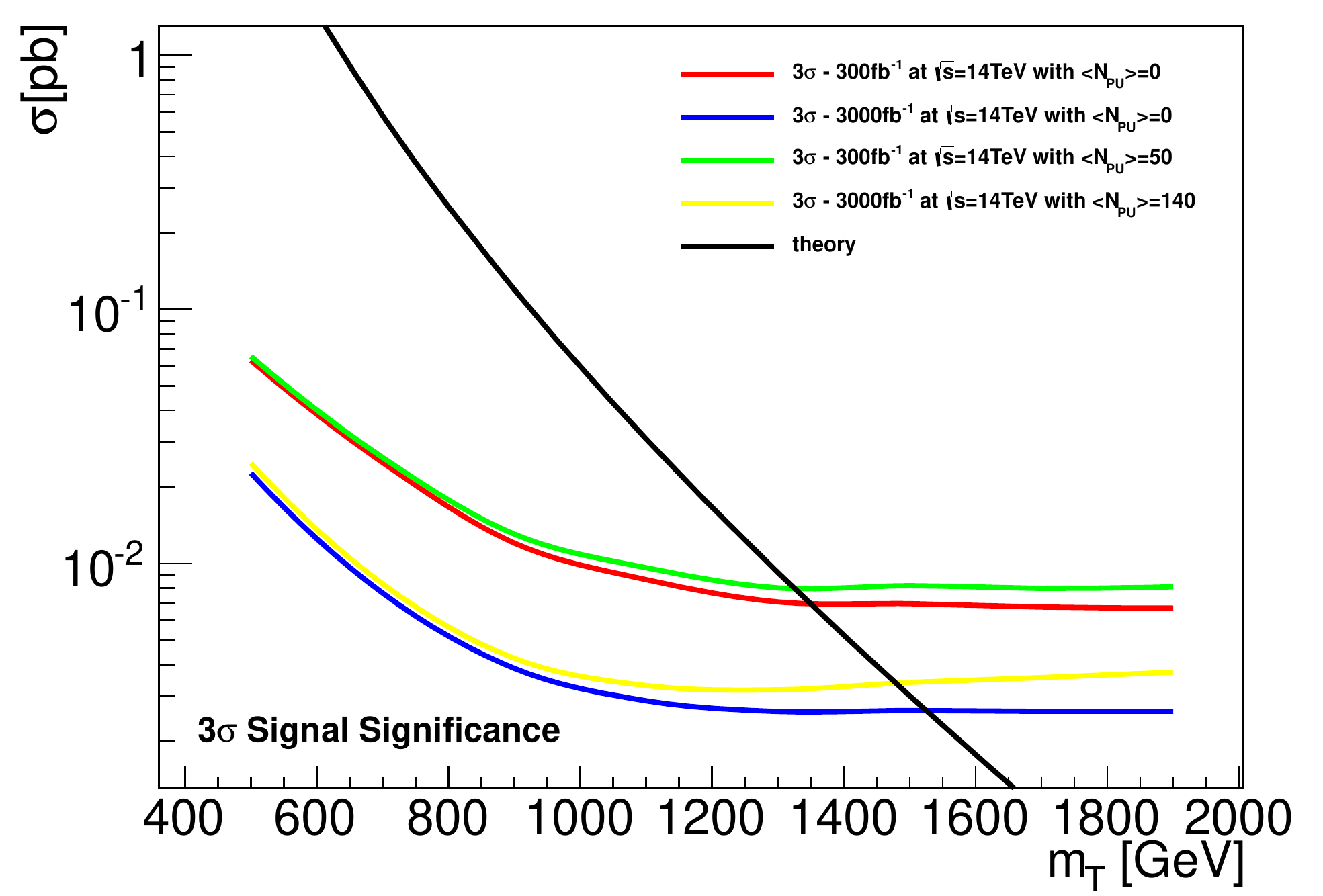}
\includegraphics[width=0.32\textwidth]{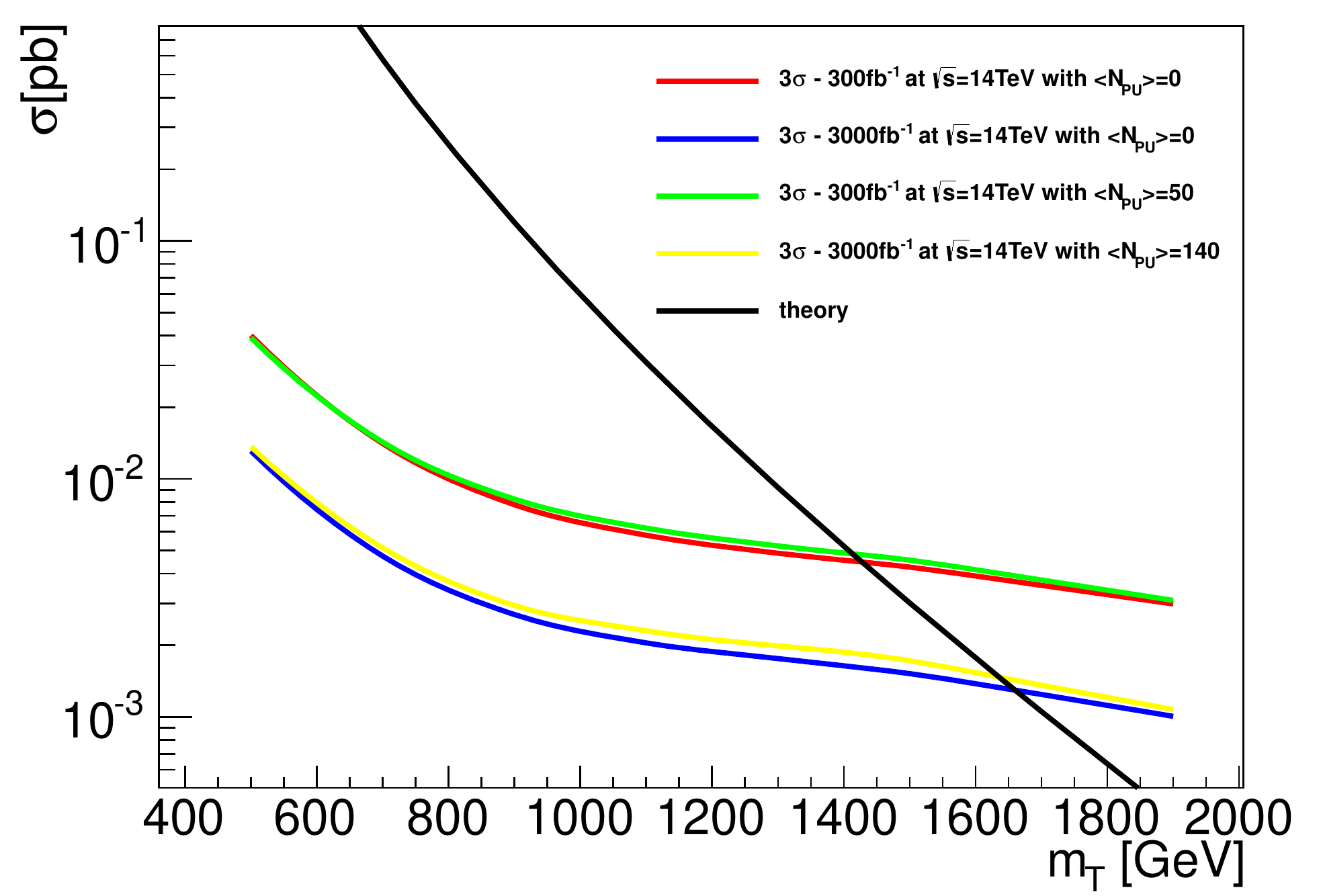}
\caption{Expected sensitivity for a 3$\sigma$ $T$ quark pair production signal 
in the the $l+$ jets channel (left), multilepton channel (middle) and 
combined (right).\label{fig:exp3sigma}}
\end{figure}

Figures~\ref{fig:explimit},~\ref{fig:exp5sigma}, and~\ref{fig:exp3sigma} show the expected 95\% C.L. limit, 
the 5$\sigma$ and 3$\sigma$ discovery reaches respectively for $\sqrt{s}$=14 TeV. 

\clearpage

\subsection{$\sqrt{s}$=33 TeV }
\begin{figure}[!h]
\includegraphics[width=0.32\textwidth]{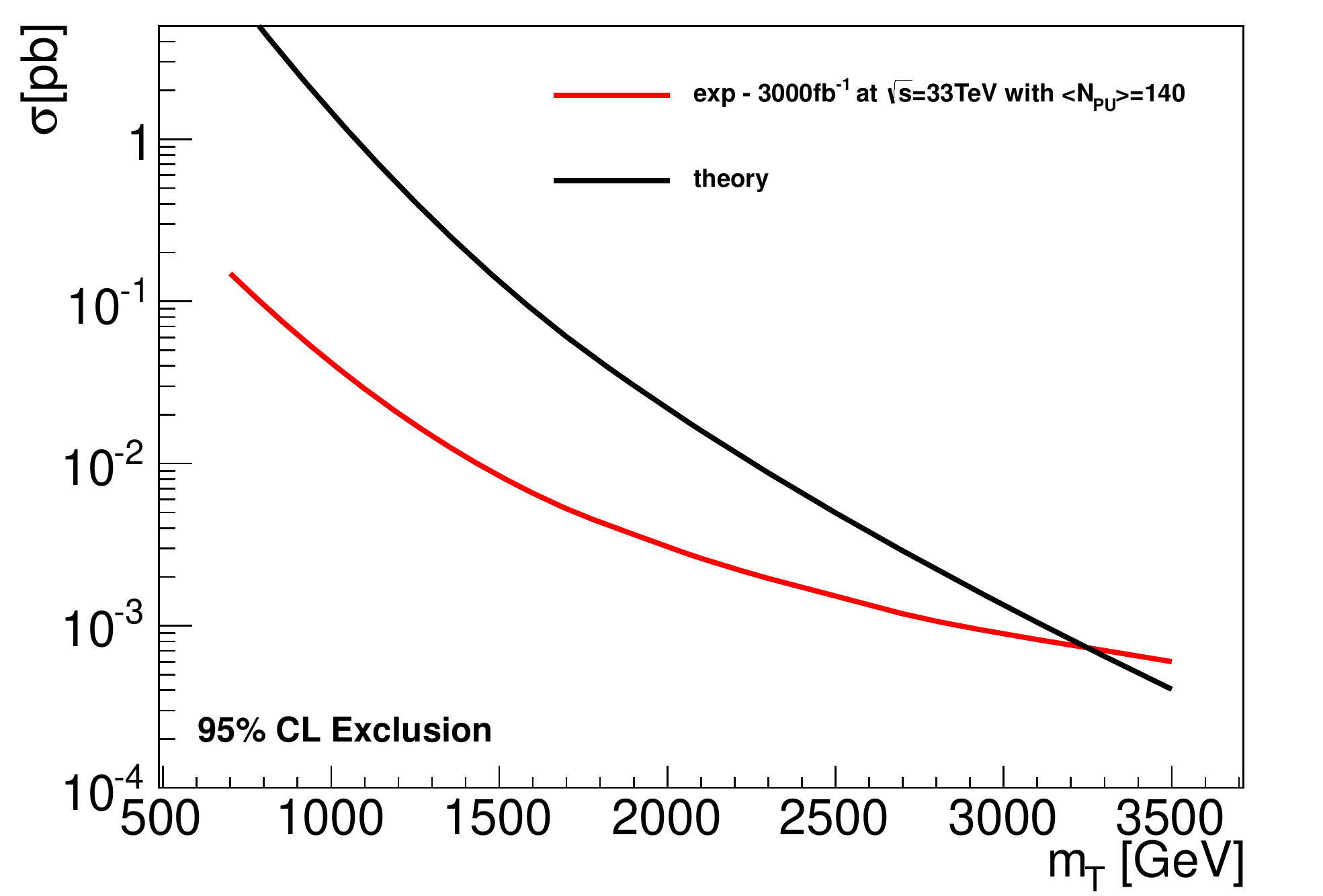}
\includegraphics[width=0.32\textwidth]{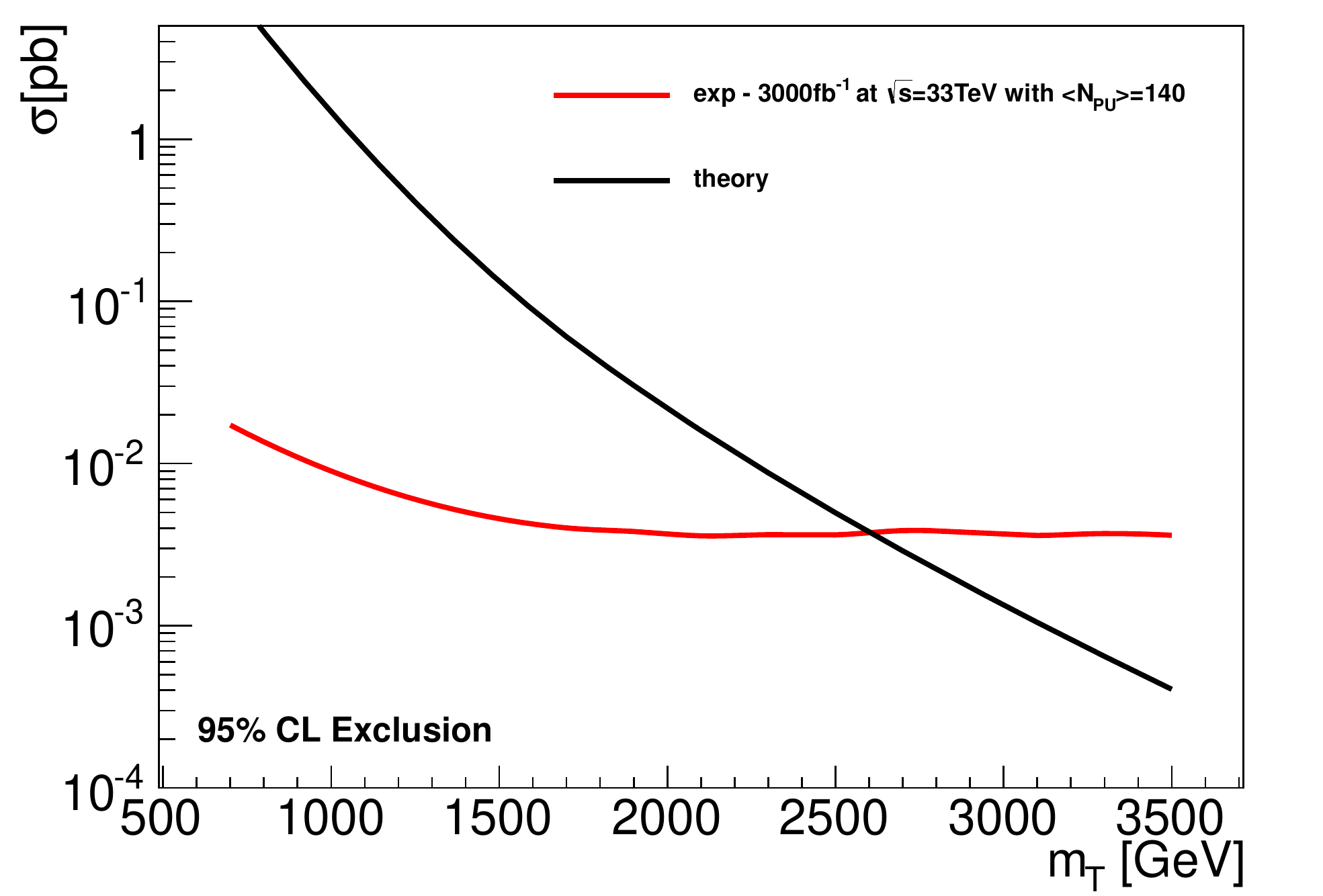}
\includegraphics[width=0.32\textwidth]{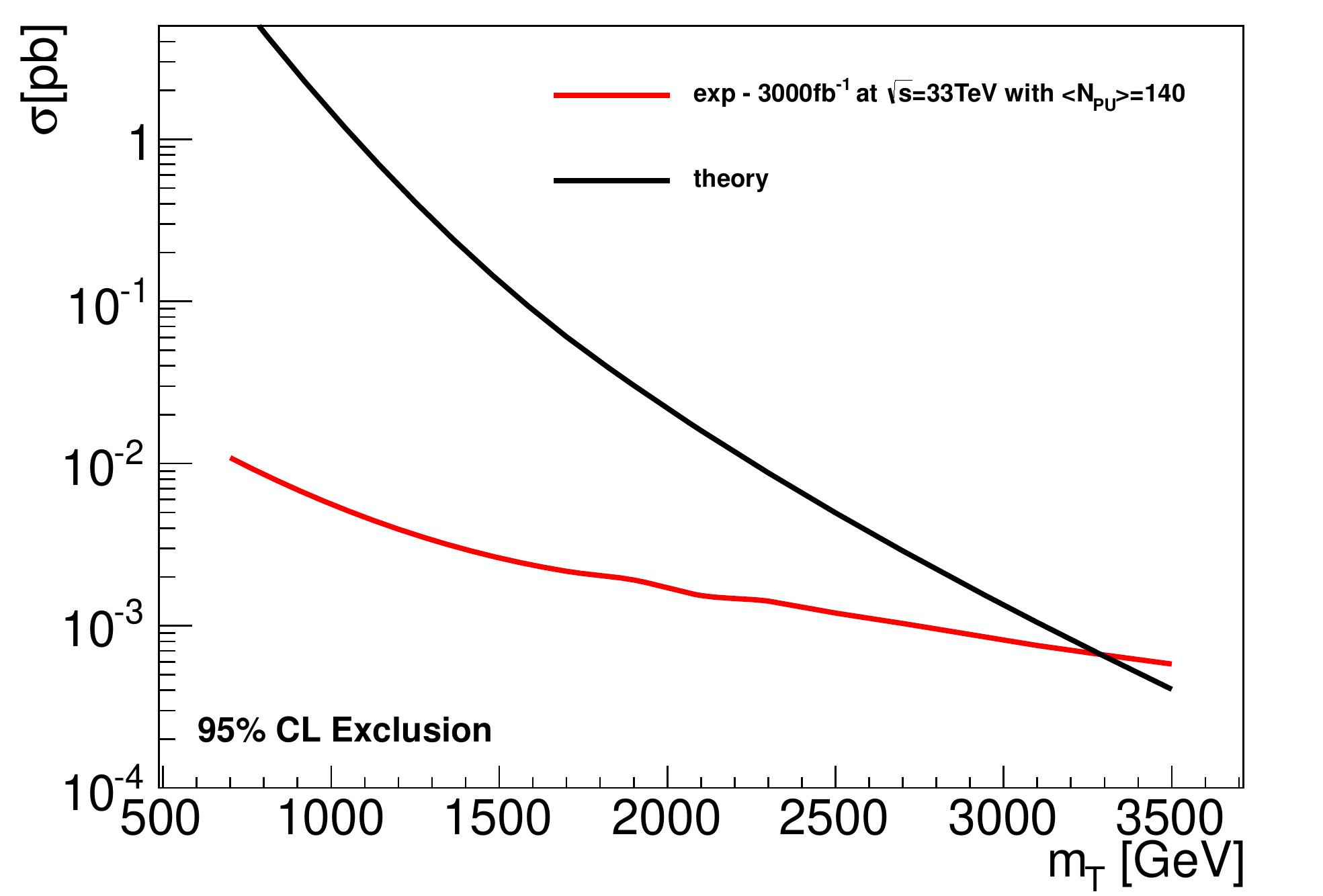}
\caption{Expected 95\% C.L. limits for $T$ quark pair production in the the $l+$ jets channel (left), multilepton channel (middle) and 
combined (right) for $\sqrt{s}$=33 TeV .\label{fig:explimit_33tev}}
\end{figure}

\begin{figure}[!h]
\includegraphics[width=0.32\textwidth]{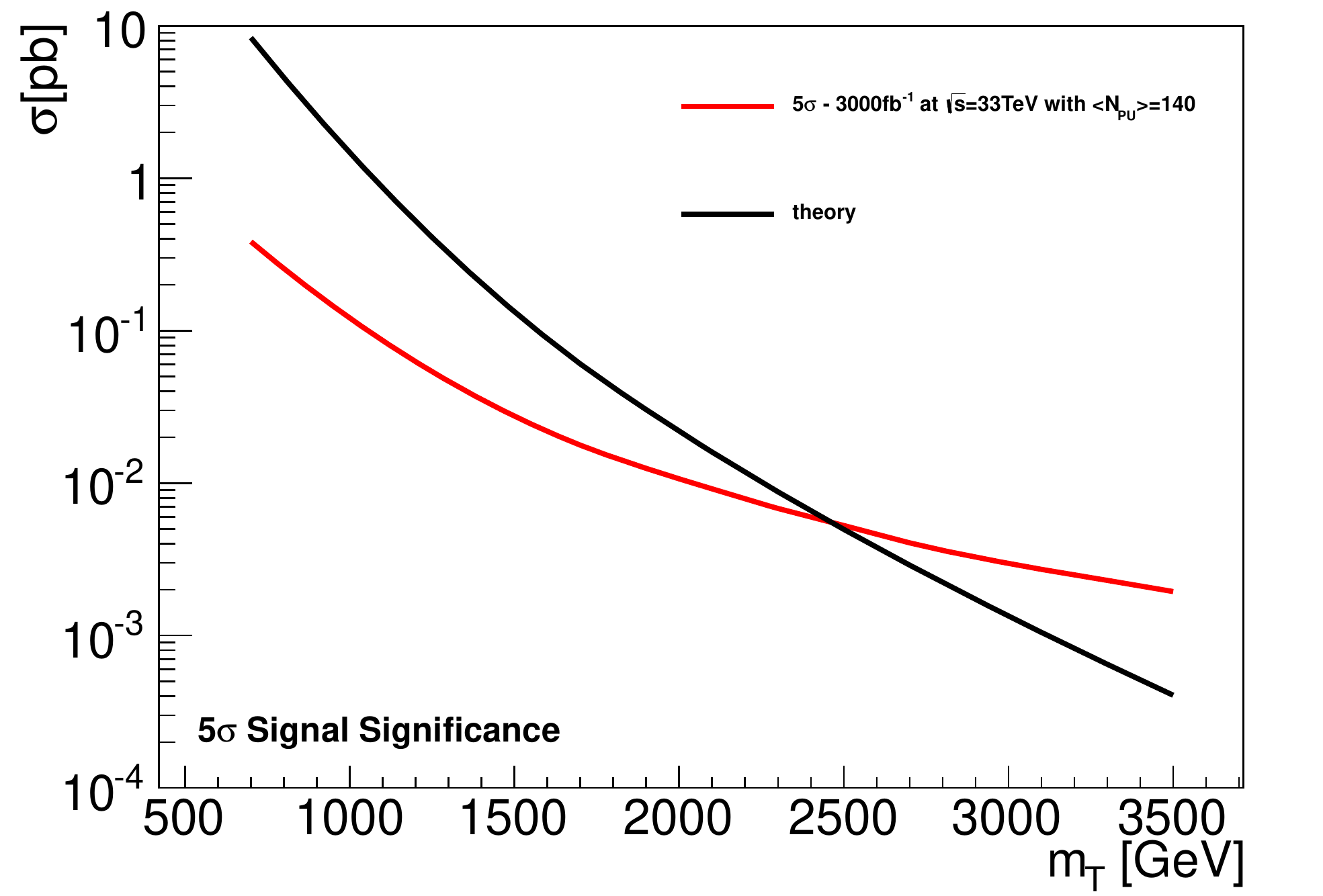}
\includegraphics[width=0.32\textwidth]{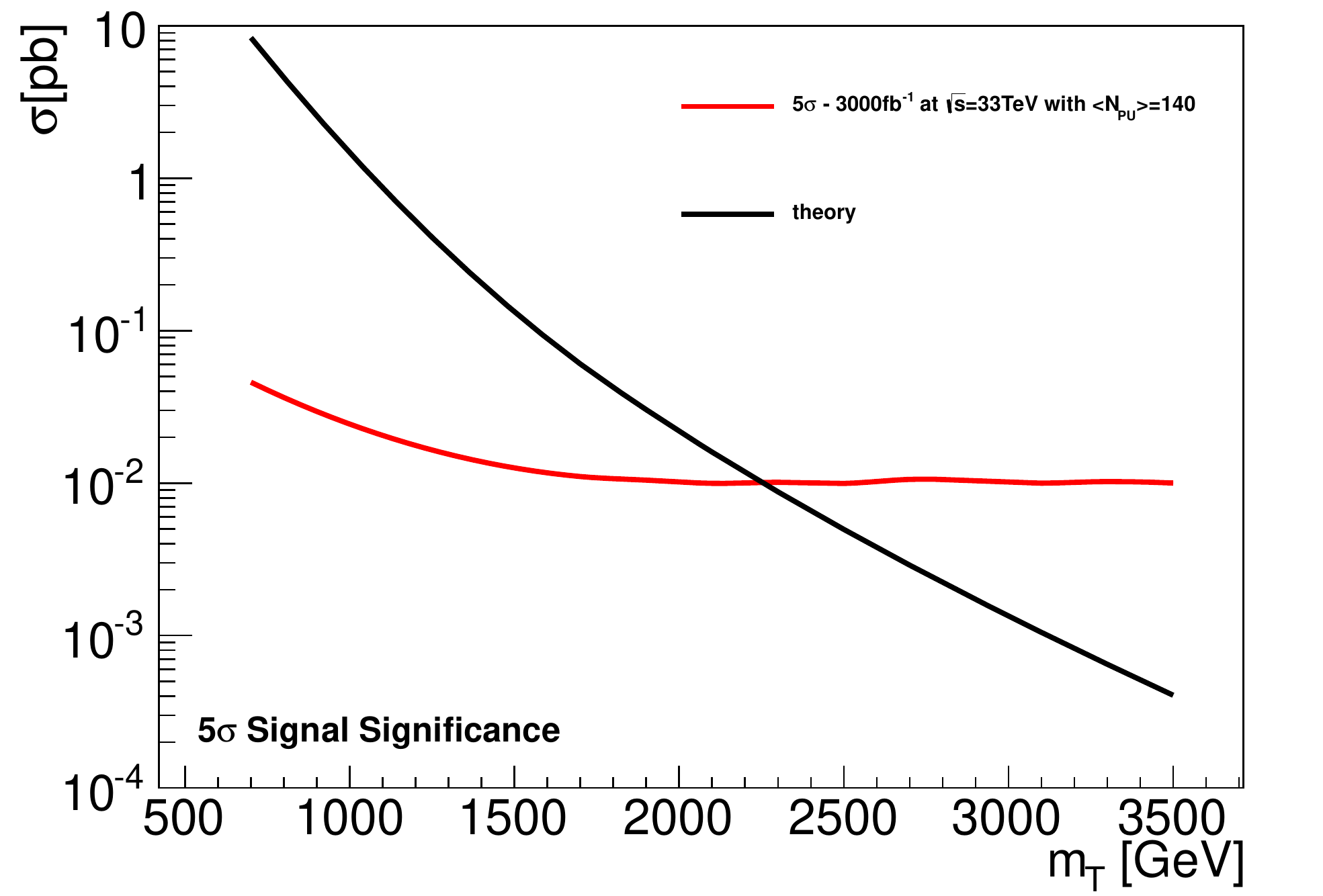}
\includegraphics[width=0.32\textwidth]{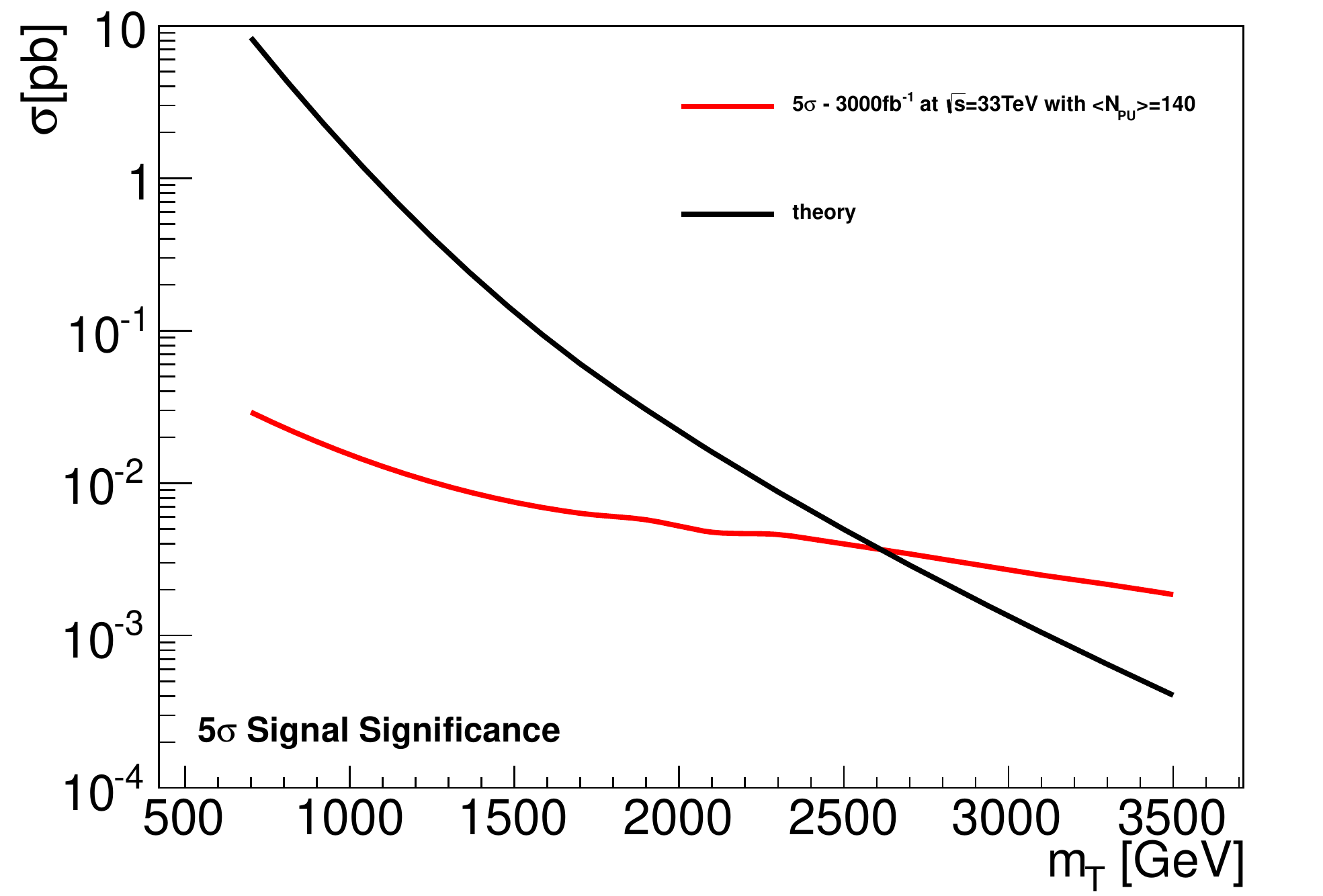}
\caption{Expected sensitivity for a 5$\sigma$ $T$ quark pair production signal 
in the the $l+$ jets channel (left), multilepton channel (middle) and 
combined (right) for $\sqrt{s}$=33 TeV.\label{fig:exp5sigma_33tev}}
\end{figure}

\begin{figure}[!h]
\includegraphics[width=0.32\textwidth]{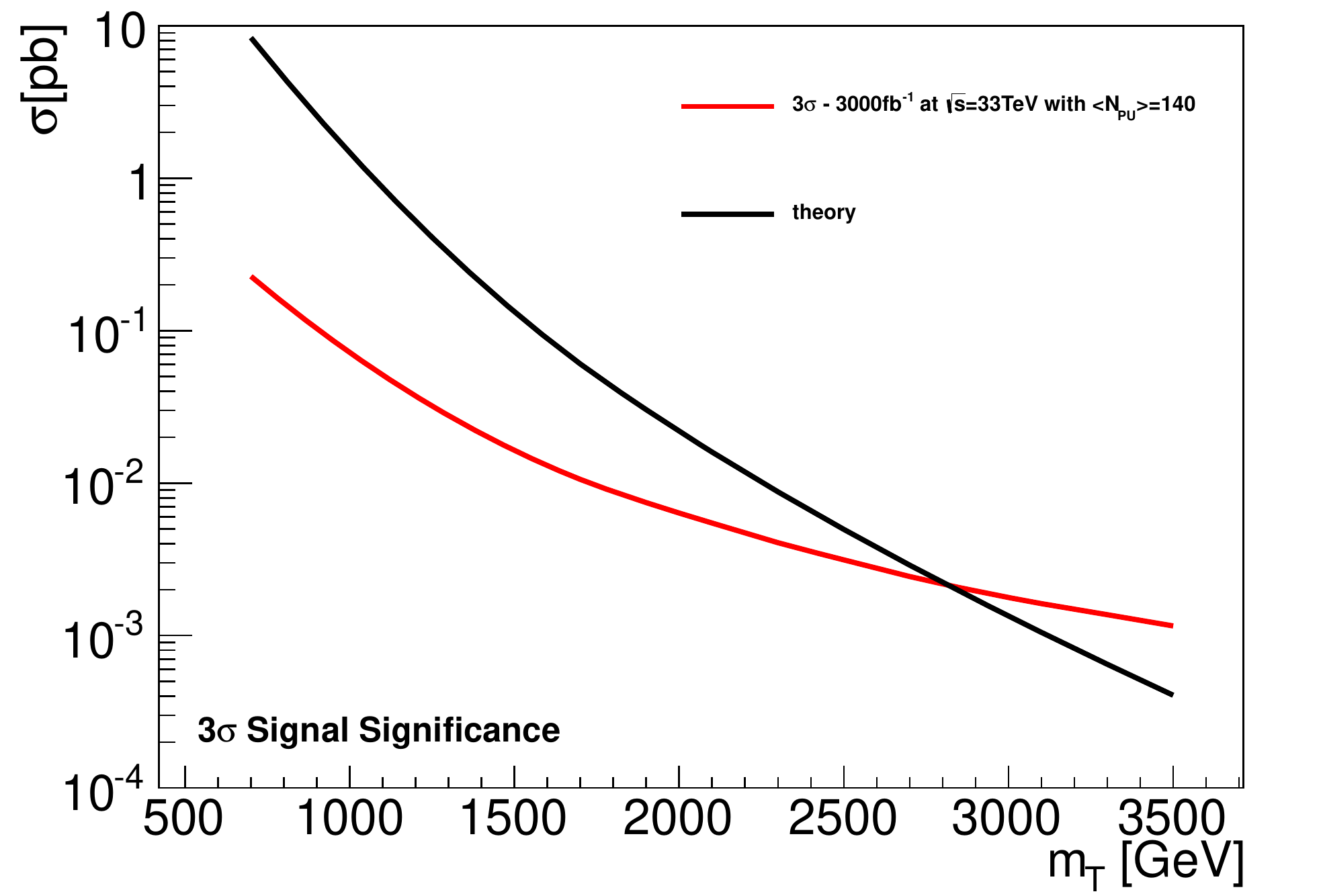}
\includegraphics[width=0.32\textwidth]{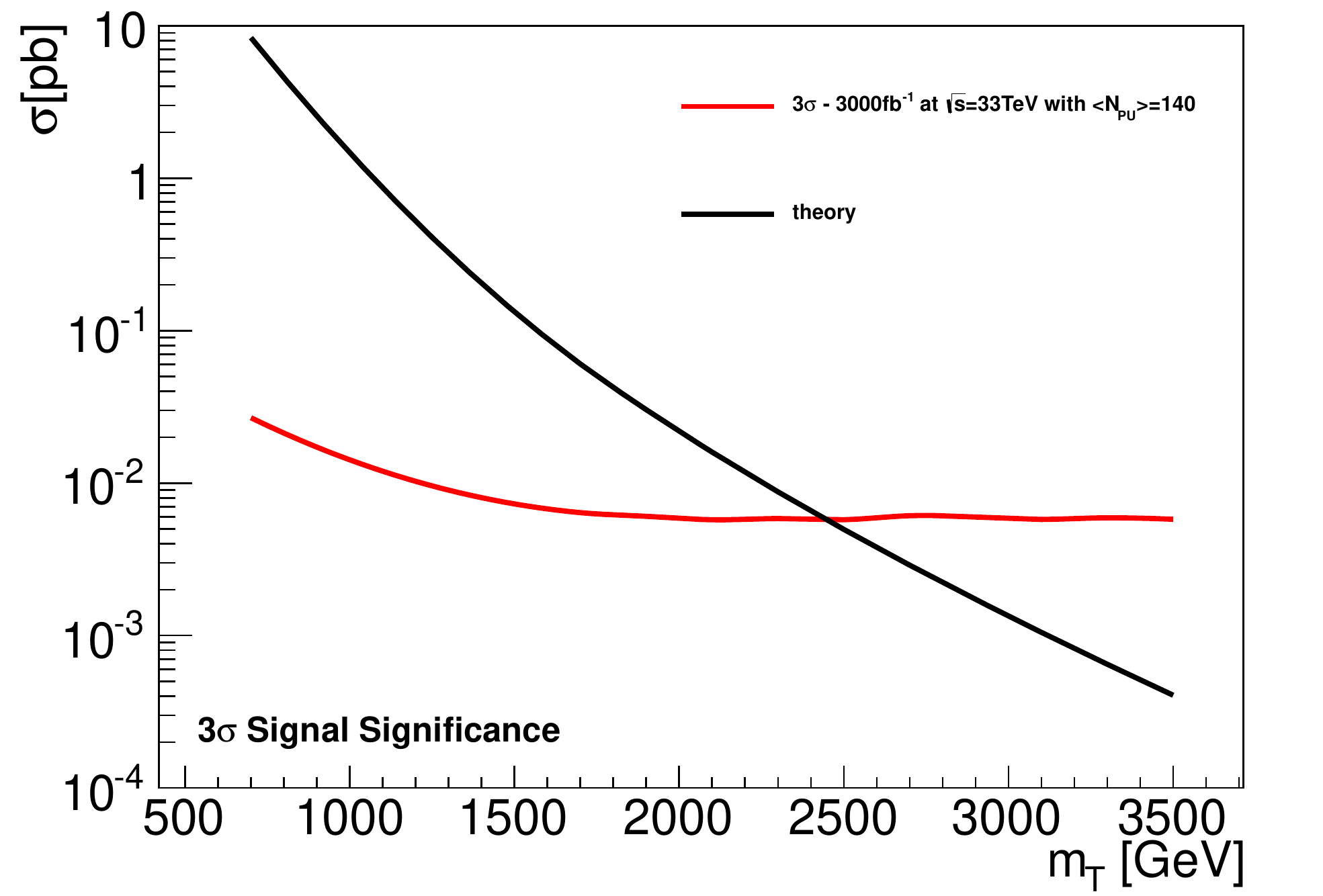}
\includegraphics[width=0.32\textwidth]{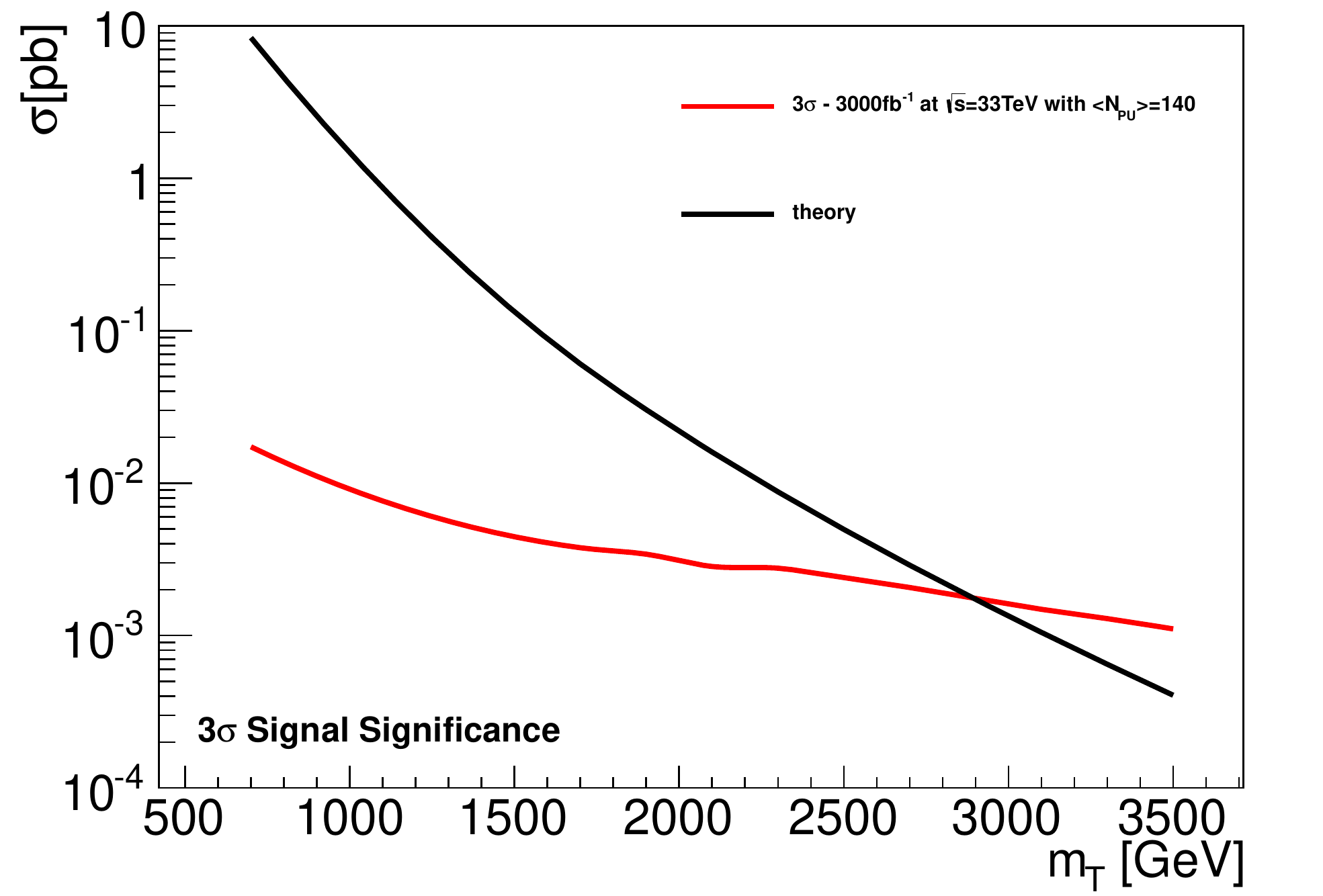}
\caption{Expected sensitivity for a 3$\sigma$ $T$ quark pair production signal 
in the the $l+$ jets channel (left), multilepton channel (middle) and 
combined (right) for $\sqrt{s}$=33 TeV.\label{fig:exp3sigma_33tev}}
\end{figure}

Figures~\ref{fig:explimit_33tev},~\ref{fig:exp5sigma_33tev} and,~\ref{fig:exp3sigma_33tev} show the expected 95\% C.L. limit, the 5$\sigma$ and 3$\sigma$ discovery reaches respectively for $\sqrt{s}$=33 TeV. 

In table~\ref{tab:topDoubleT}, the expected mass sensitivity is summaried.
The  expected 95\% C.L. limit, the 5$\sigma$ and 3$\sigma$ discovery reaches 
for both  $\sqrt{s}$=14 and 33 TeV LHC runs are listed.

\begin{table}[t]
\centering
\begin{tabular}{|c|c|c|c|c|c|}
\hline
Collider      &	Luminosity   & Pileup & 3$\sigma$ evidence & 5$\sigma$ discovery & 95\% CL \\
\hline		\hline
\multicolumn{6}{|l|}{\bf Single lepton+jets}\\
LHC 14 TeV   &	300 fb$^{-1}$	&     0  &   1280~GeV & 1150~GeV&	1390~GeV \\ \hline
LHC 14 TeV   &	300 fb$^{-1}$	&    50  &   1260~GeV & 1140~GeV&	1410~GeV     \\ \hline
LHC 14 TeV   &	3 ab$^{-1}$	&   140  &   1560~GeV & 1430~GeV&	1730~GeV \\ \hline
\multicolumn{6}{|l|}{\bf Multi-lepton analyses}\\
LHC 14 TeV   &	300 fb$^{-1}$	&     0   & 1350~GeV&	1225~GeV &   1450~GeV\\ \hline
LHC 14 TeV   &	300 fb$^{-1}$	&    50  & 1340~GeV&	1200~GeV  &   1420~GeV    \\ \hline
LHC 14 TeV   &	3 ab$^{-1}$	&   140  & 1475~GeV& 1375~GeV &   1575~GeV \\ \hline
\multicolumn{6}{|l|}{\bf Combination of single lepton+jets and multi-lepton analyses}\\
LHC 14 TeV   &	300 fb$^{-1}$	&     0  &   1410~GeV & 1300~GeV&	1525~GeV \\ \hline
LHC 14 TeV   &	300 fb$^{-1}$	&    50  &   1415~GeV & 1300~GeV&	1525~GeV     \\ \hline
LHC 14 TeV   &	3 ab$^{-1}$	&   140  &   1620~GeV & 1525~GeV&	1780~GeV \\ \hline\hline

\multicolumn{6}{|l|}{\bf Single lepton+jets}\\
LHC 33 TeV   &  3 ab$^{-1}$     &   140  &   2800~GeV & 2420~GeV&       3200~GeV \\ \hline
\multicolumn{6}{|l|}{\bf Multi-lepton analyses}\\
LHC 33 TeV   &  3 ab$^{-1}$     &   140  &  2450 ~GeV & 2250~GeV& 2600      ~GeV \\ \hline
\multicolumn{6}{|l|}{\bf Combination of single lepton+jets and multi-lepton analyses}\\
LHC 33 TeV   &  3 ab$^{-1}$     &   140  &  2900 ~GeV & 2600~GeV&  3400     ~GeV \\ \hline

\end{tabular} \hspace{-0.138cm} 
\vspace{0.3cm}
\caption{Expected mass sensitivity for a top-partner $T$ pair production in 
the lepton~$+$~jets and multi-lepton signatures for all three decay modes $bW$, $ t H$ and $ Z t $. }
\label{tab:topDoubleT}
\end{table}%

\clearpage

\section{Distributions ($\sqrt{s}$=14 TeV)}

\begin{figure}[h]
\includegraphics[width=0.45\textwidth]{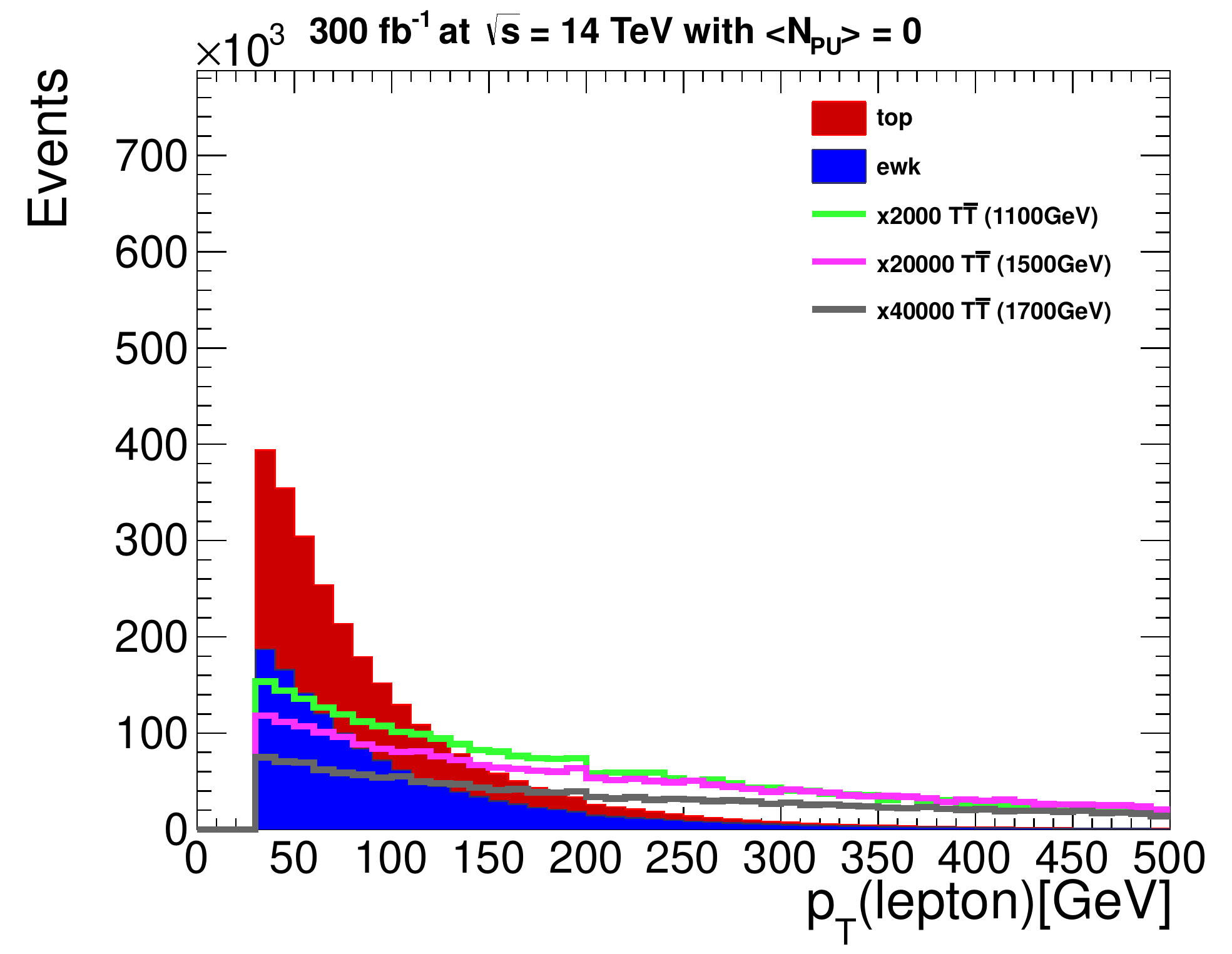}
\includegraphics[width=0.45\textwidth]{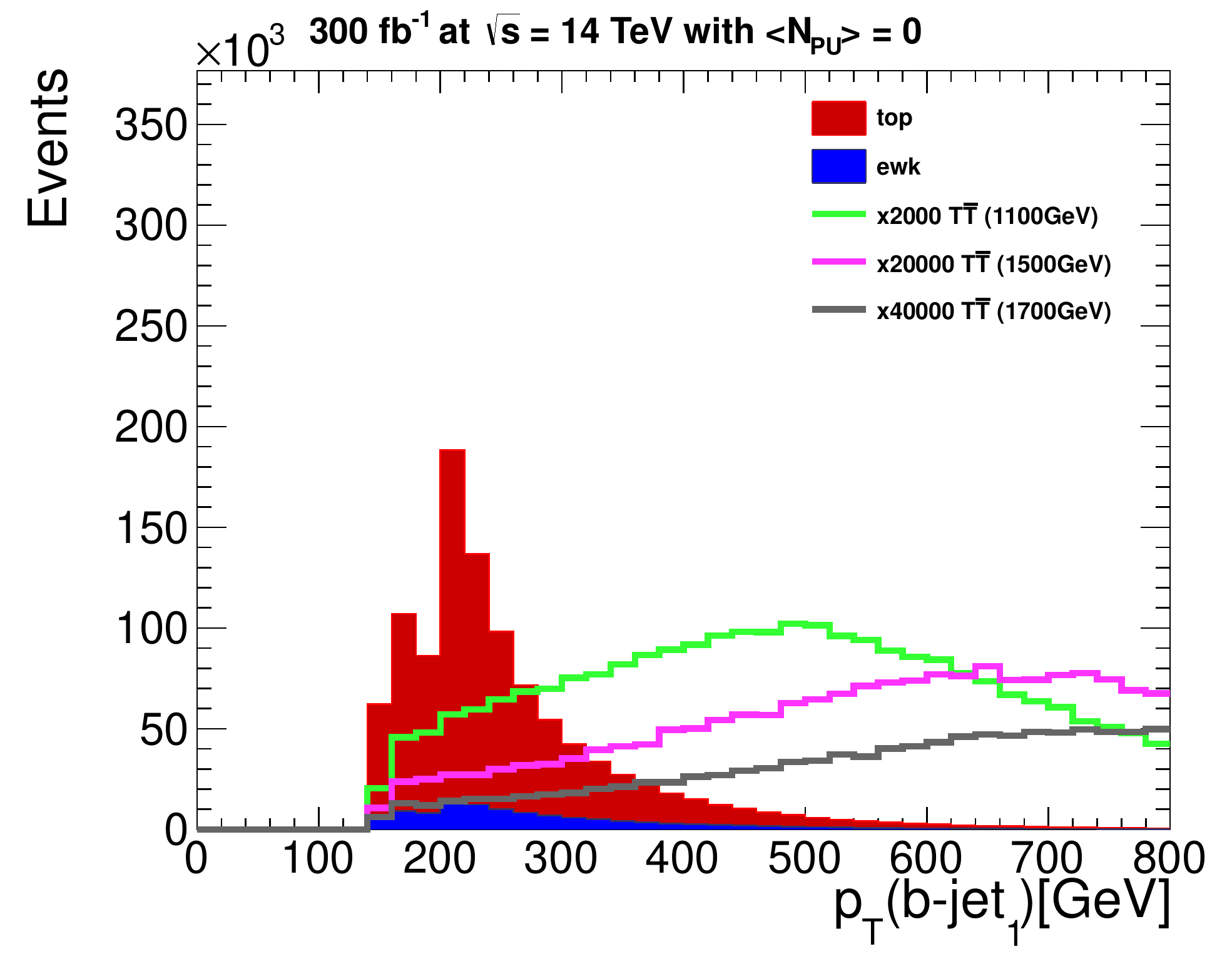}\\
\includegraphics[width=0.45\textwidth]{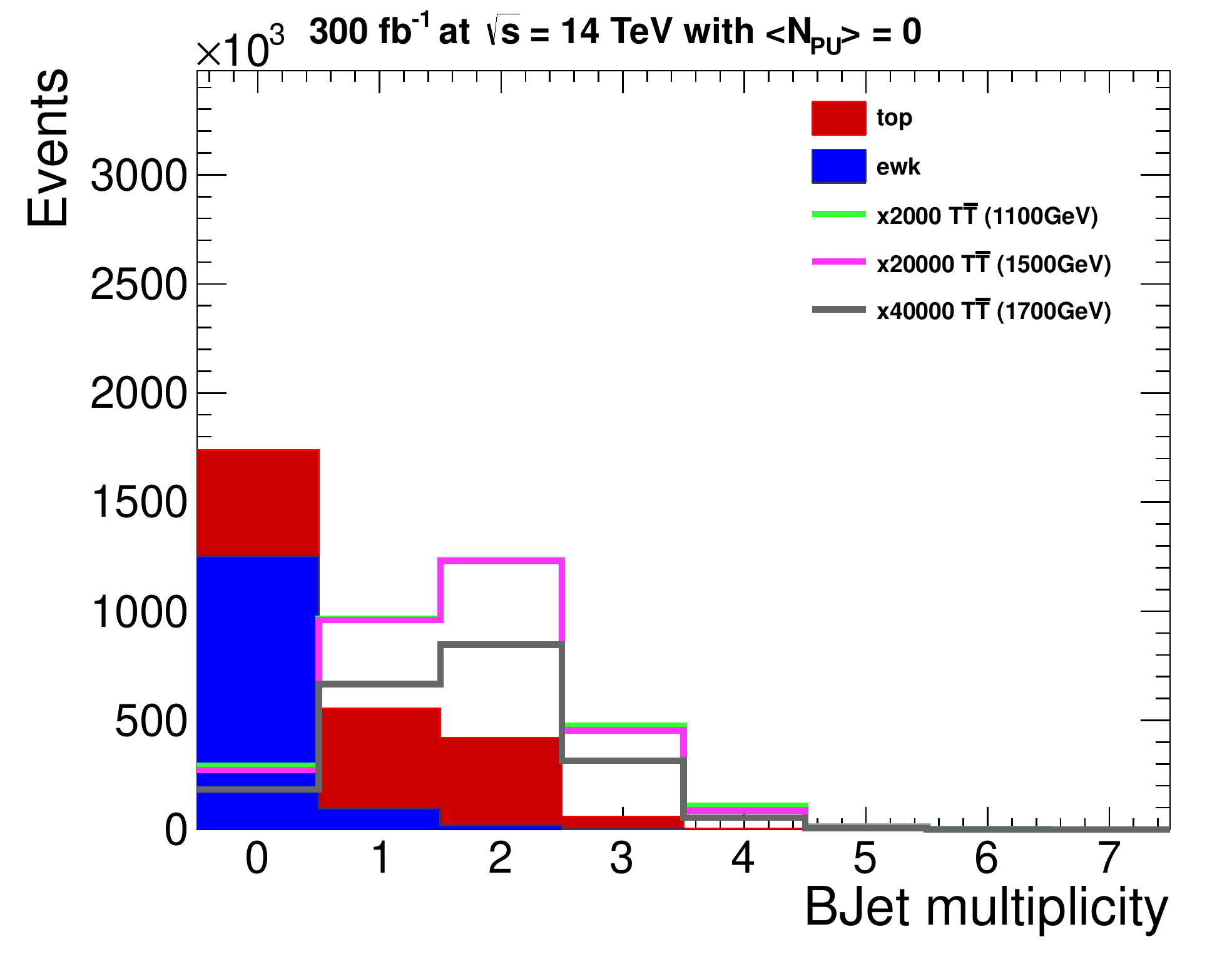}
\includegraphics[width=0.45\textwidth]{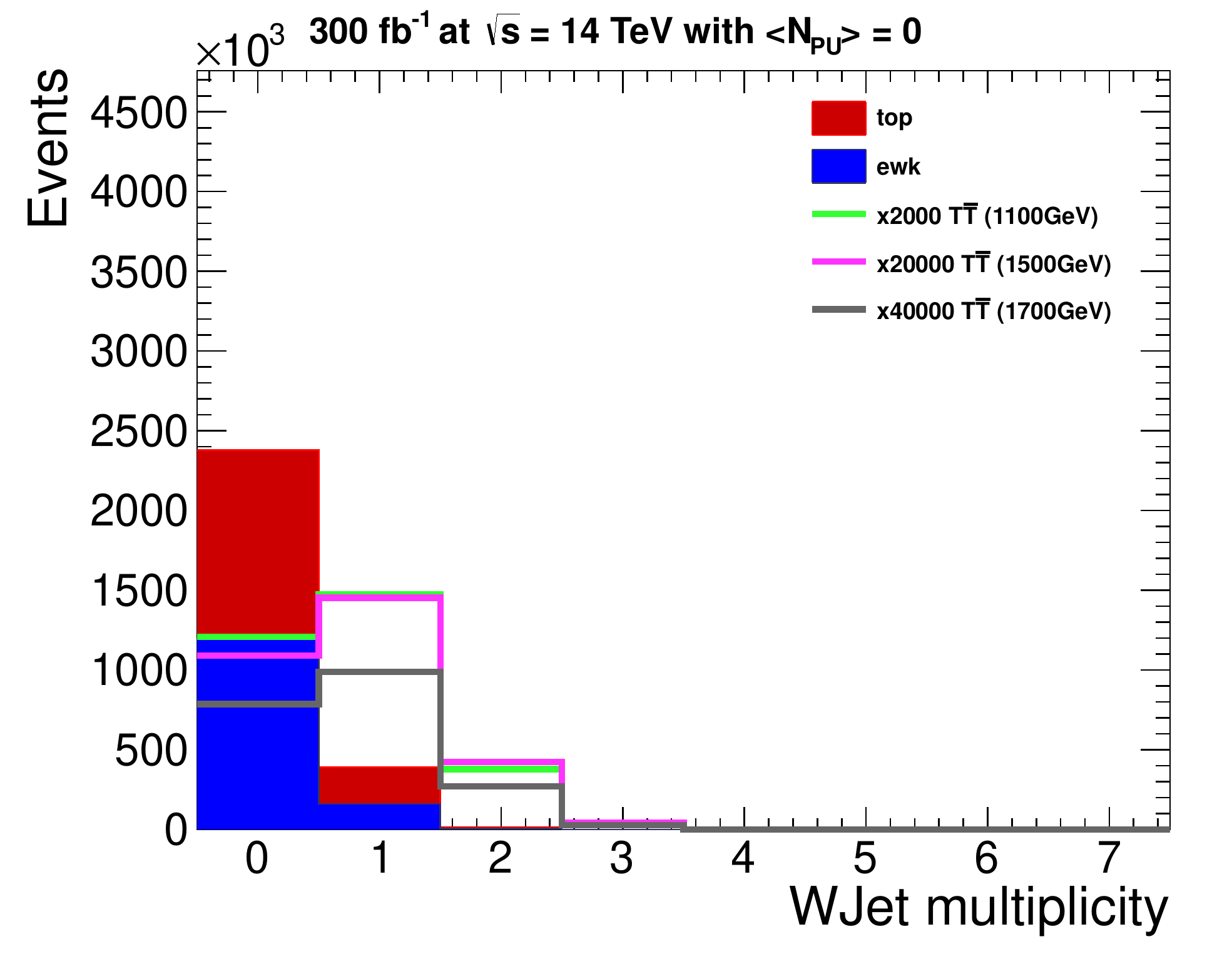}
\caption{Distributions of leading electron, leading b-jet, b-jet multiplicity and W-jet multiplicity in the $l+$jets channel .}
\end{figure}

\begin{figure}[h]
\includegraphics[width=0.45\textwidth]{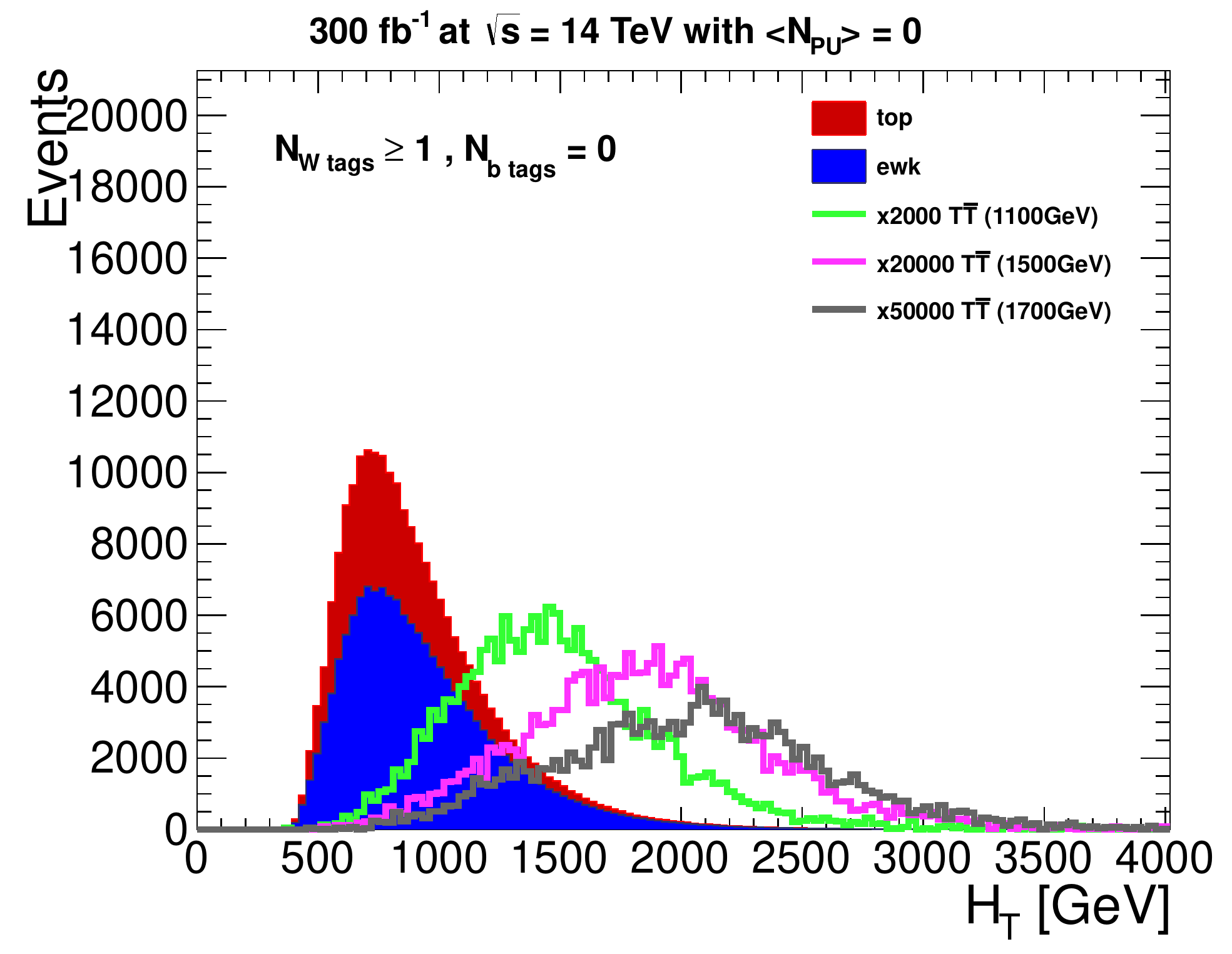}
\includegraphics[width=0.45\textwidth]{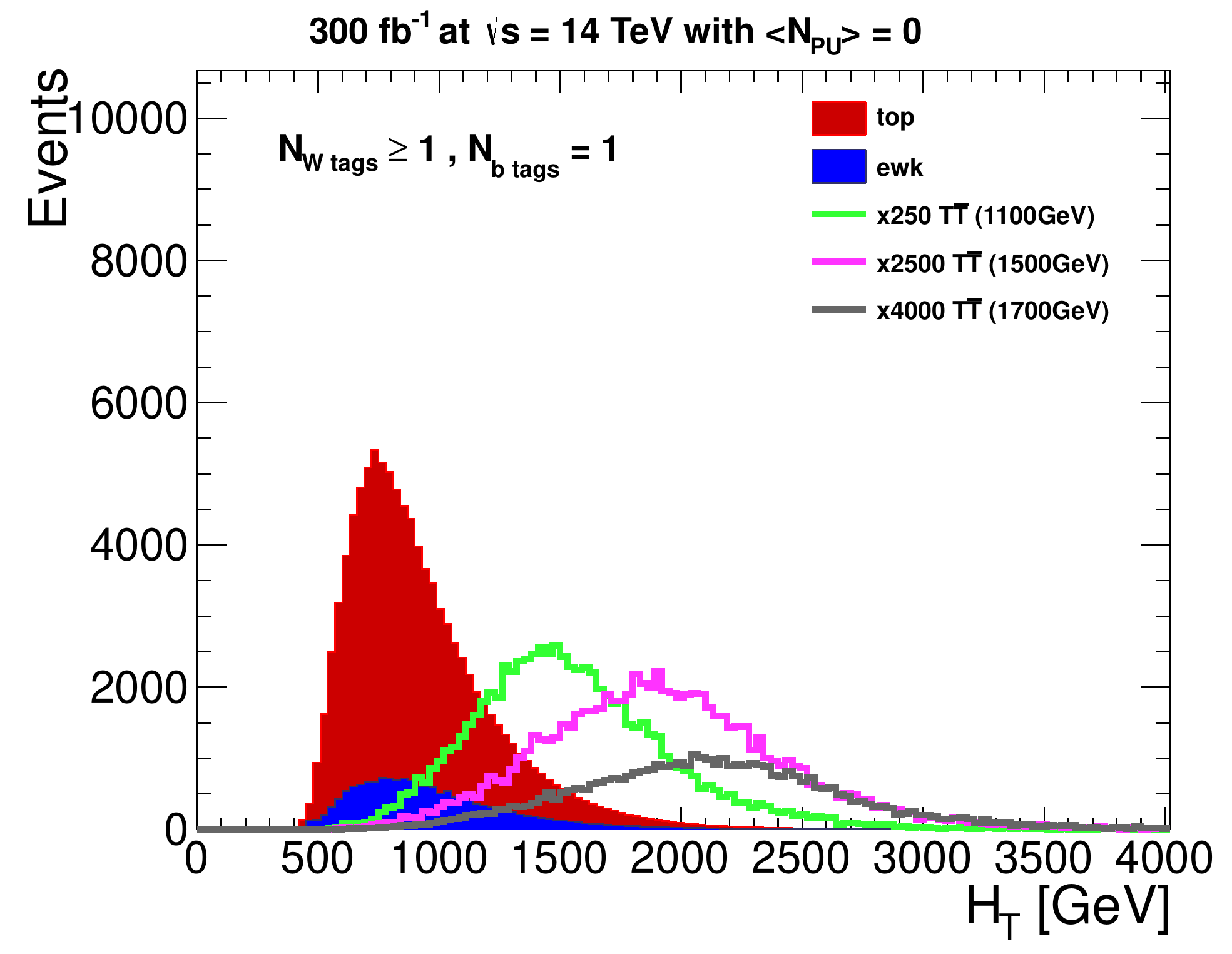}\\
\includegraphics[width=0.45\textwidth]{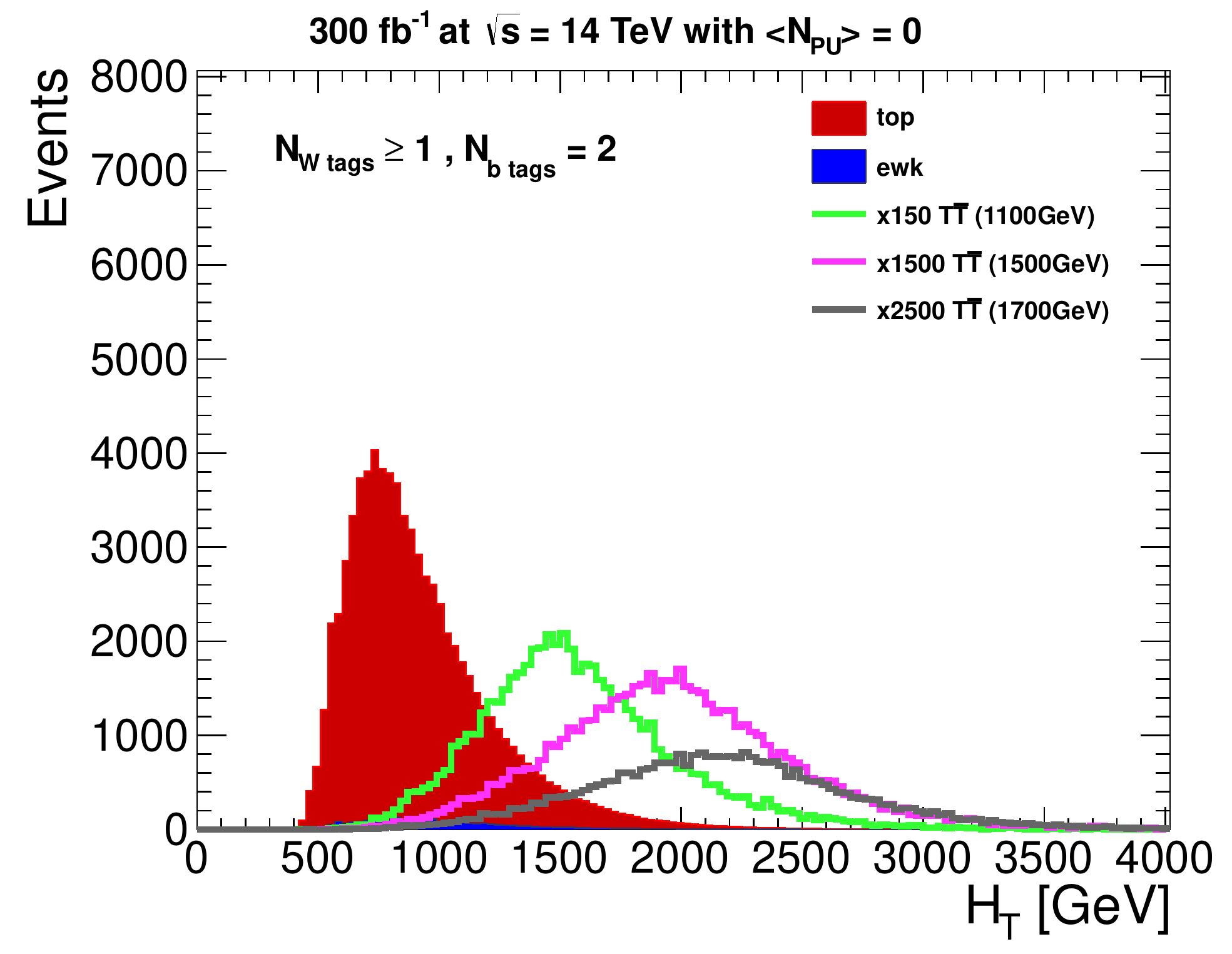}
\includegraphics[width=0.45\textwidth]{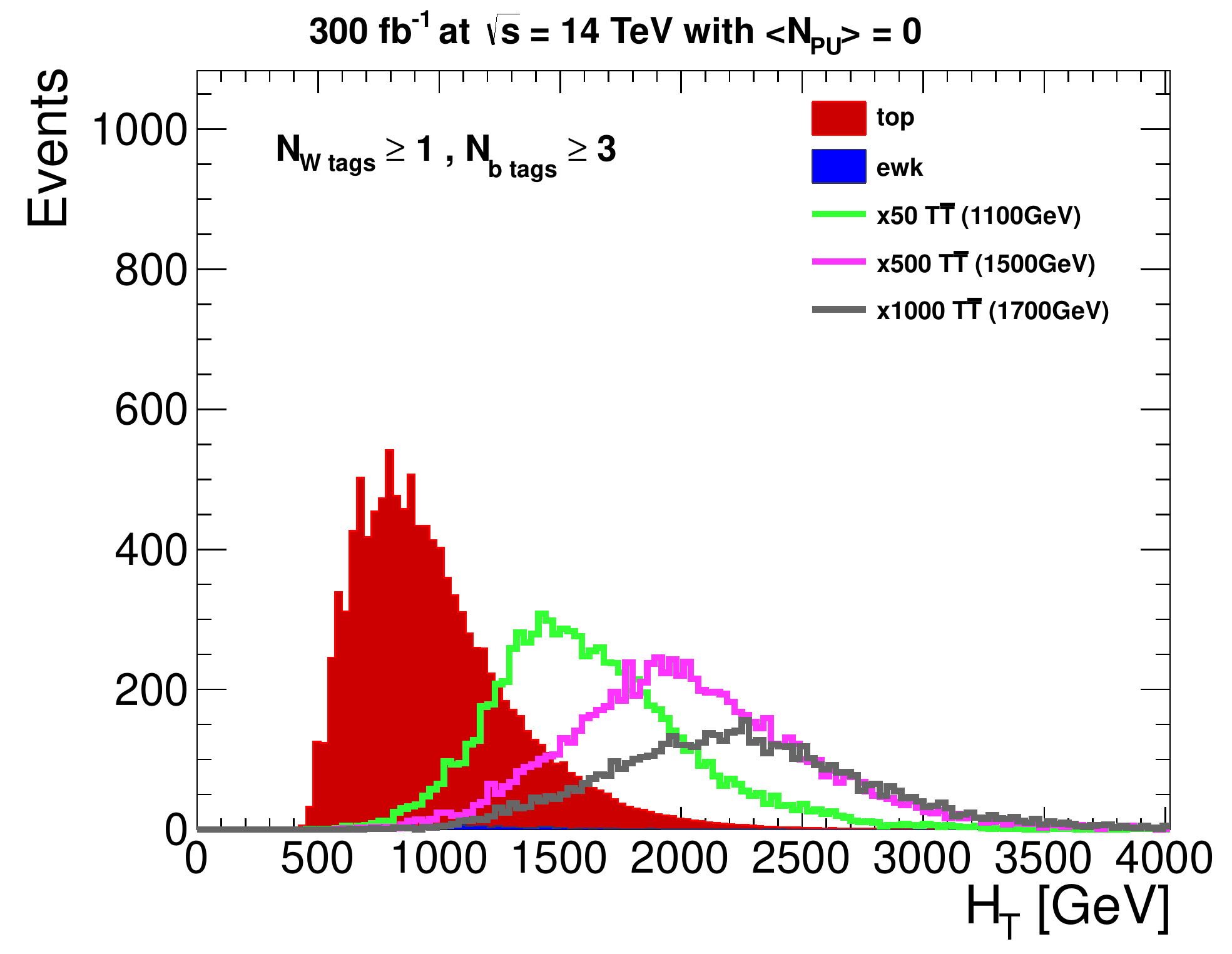}
\caption{Distributions of $H_T$ in different event categories with $l+\geq$3 jets with 1 $W-$jet and 0 b-jet (top left), 1 b-jet (top right), 2 b-jet (bottom left) and at least 3 b-jets (bottom right) .}
\end{figure}

\begin{figure}[h]
\includegraphics[width=0.45\textwidth]{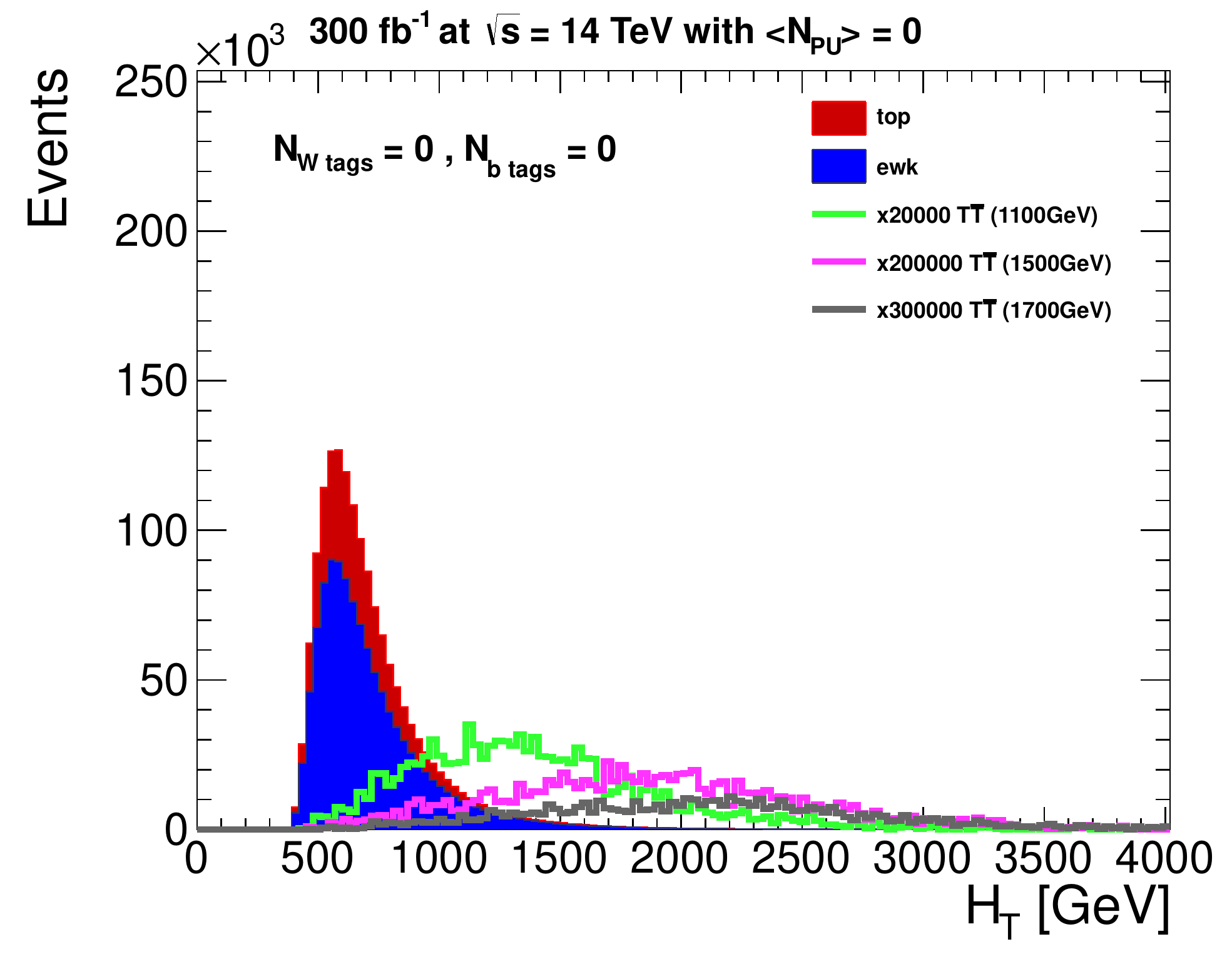}
\includegraphics[width=0.45\textwidth]{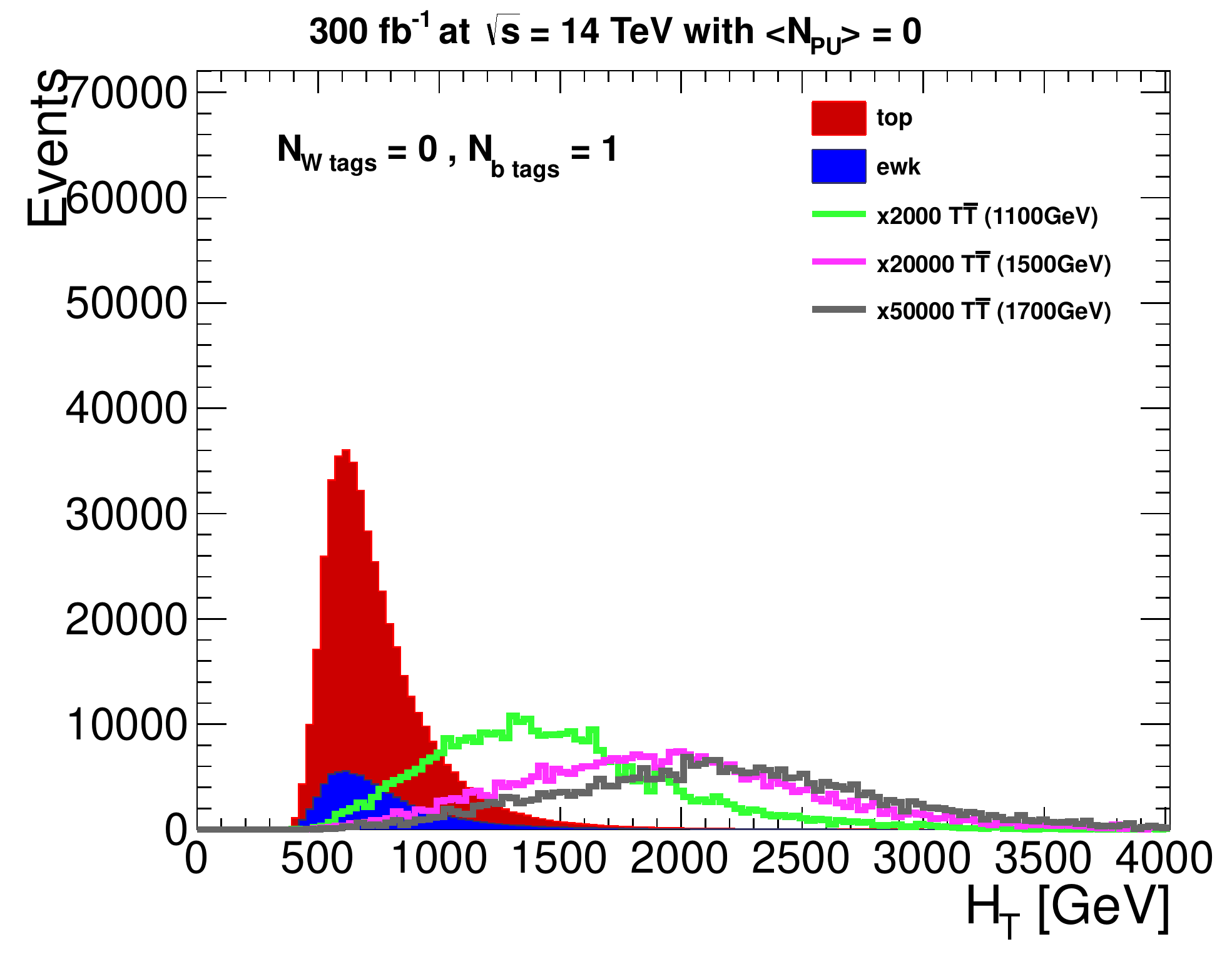}\\
\includegraphics[width=0.45\textwidth]{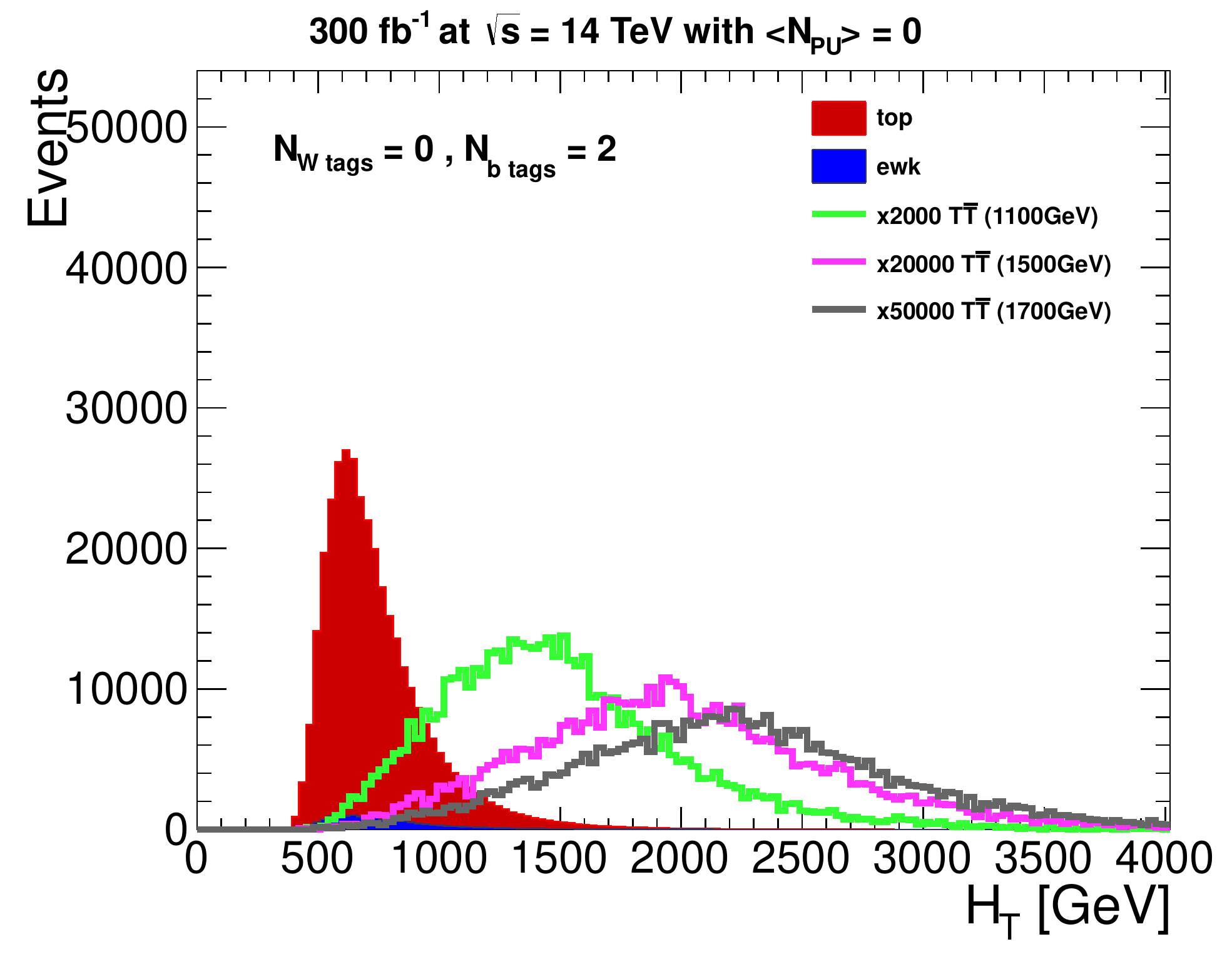}
\includegraphics[width=0.45\textwidth]{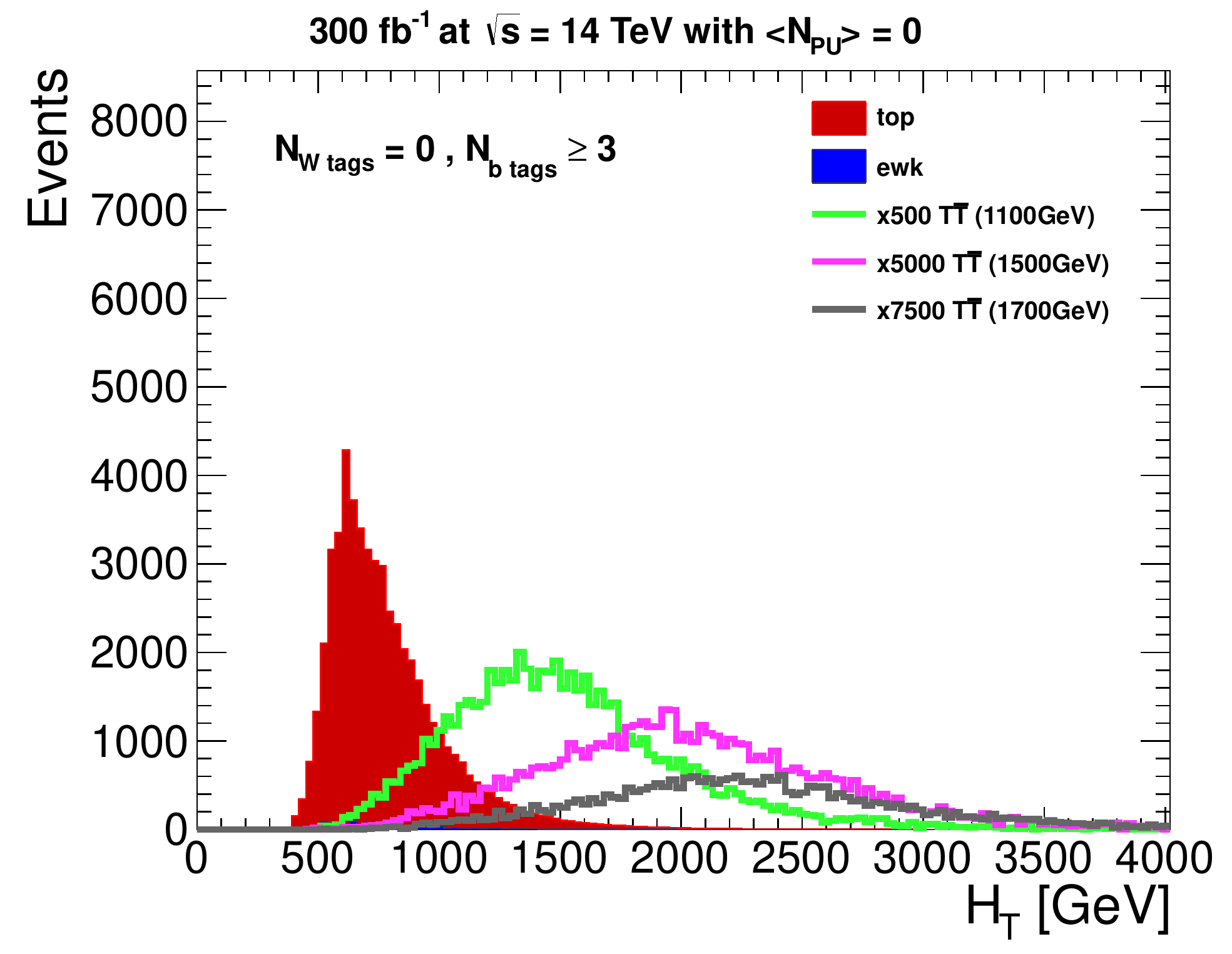}
\caption{Distributions of $H_T$ in different event categories with $l+\geq$4 jets with no $W-$jet and  0 b-jet (top left), 1 b-jet (top right), 2 b-jet (bottom left) and at least 3 b-jets (bottom right) .}
\end{figure}

\begin{figure}[!h]
\includegraphics[width=0.4\textwidth]{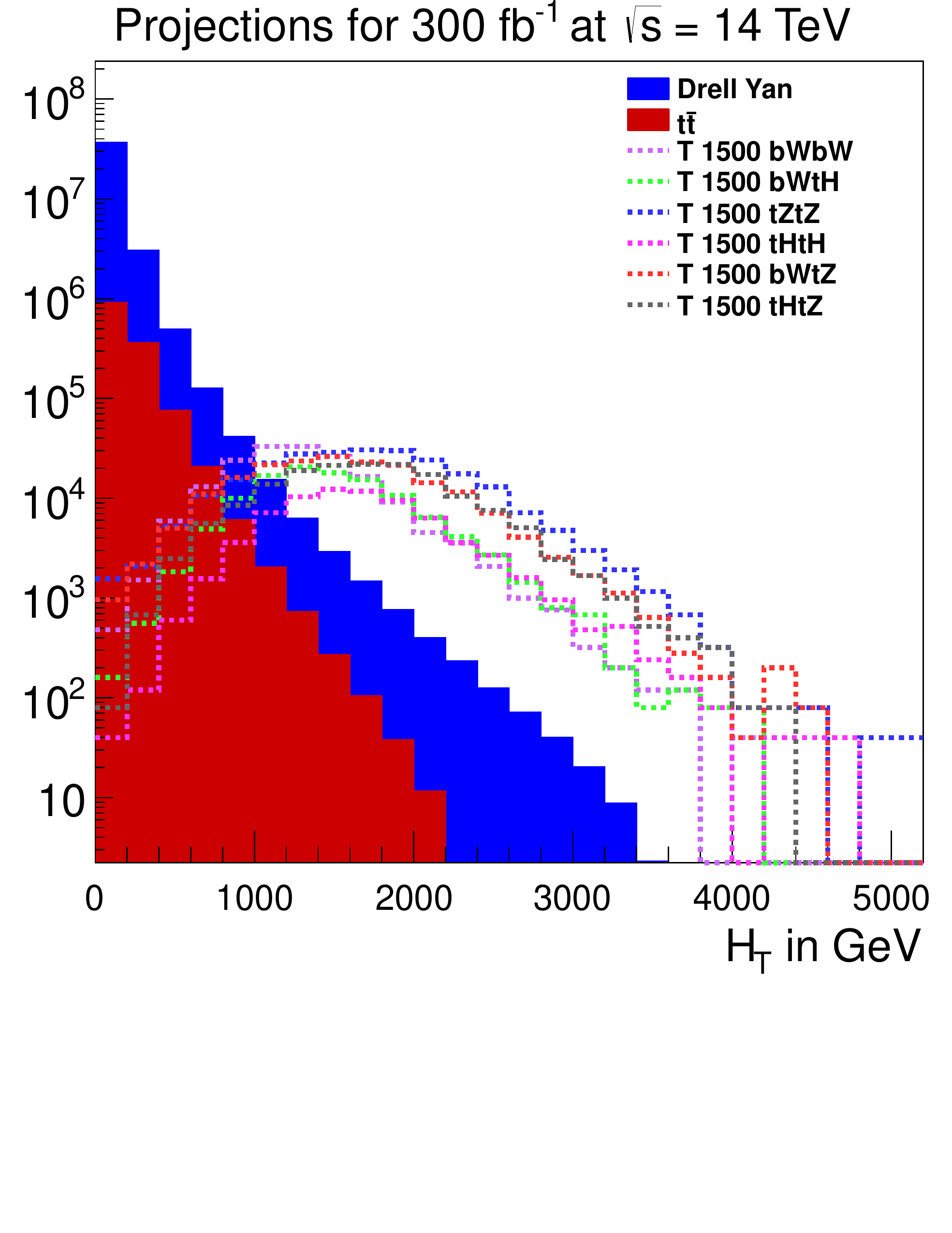}
\includegraphics[width=0.4\textwidth]{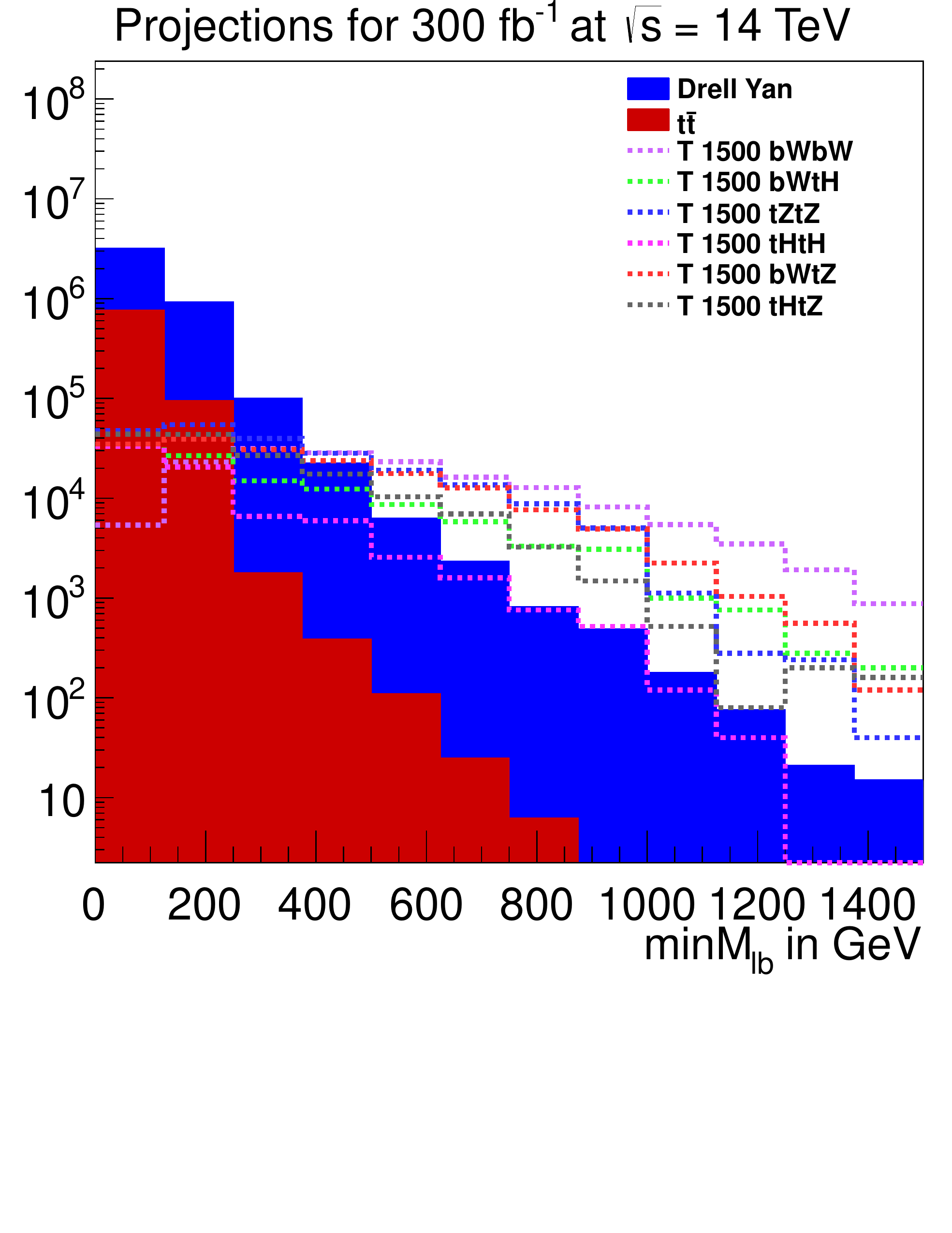}\\
\includegraphics[width=0.4\textwidth]{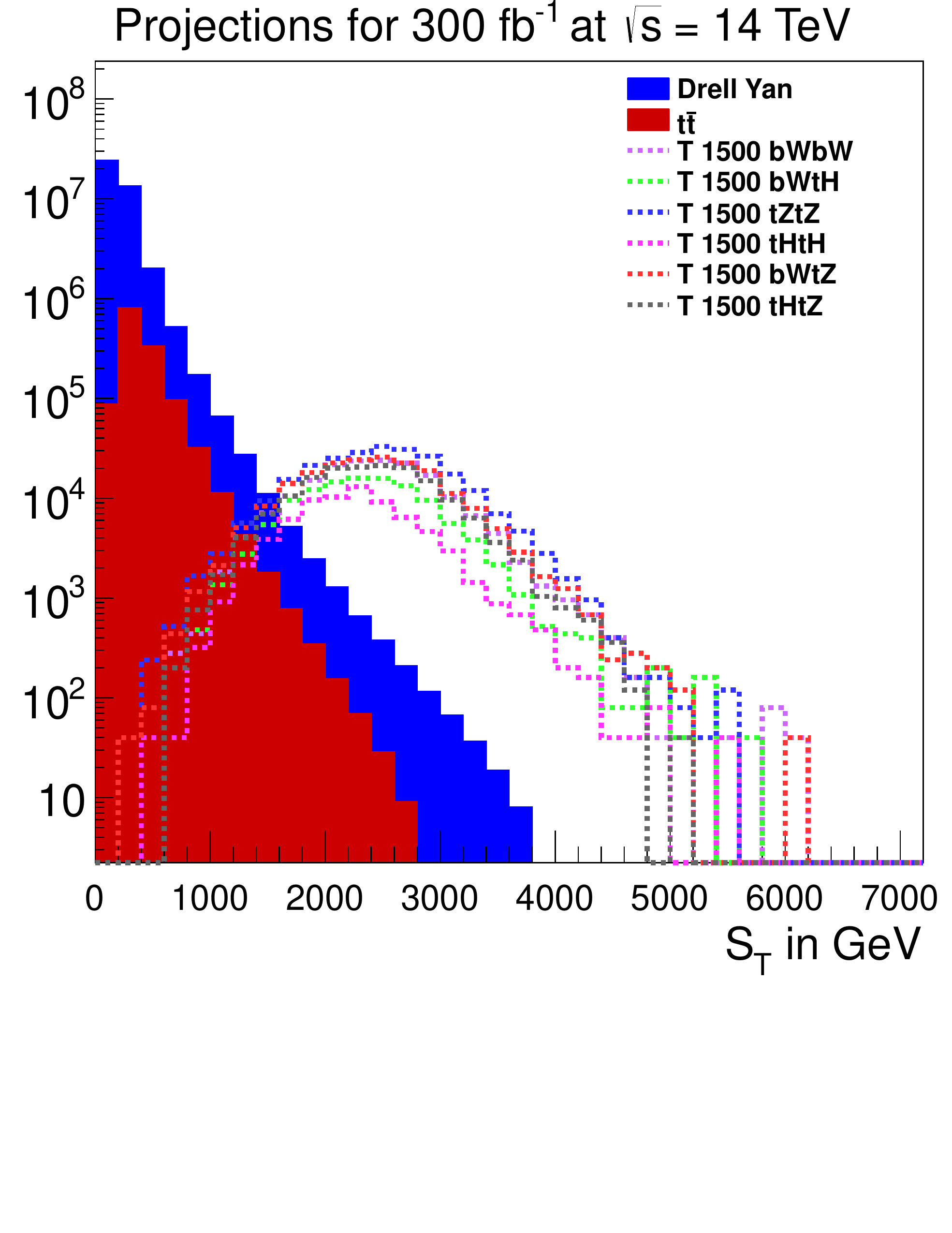}
\includegraphics[width=0.4\textwidth]{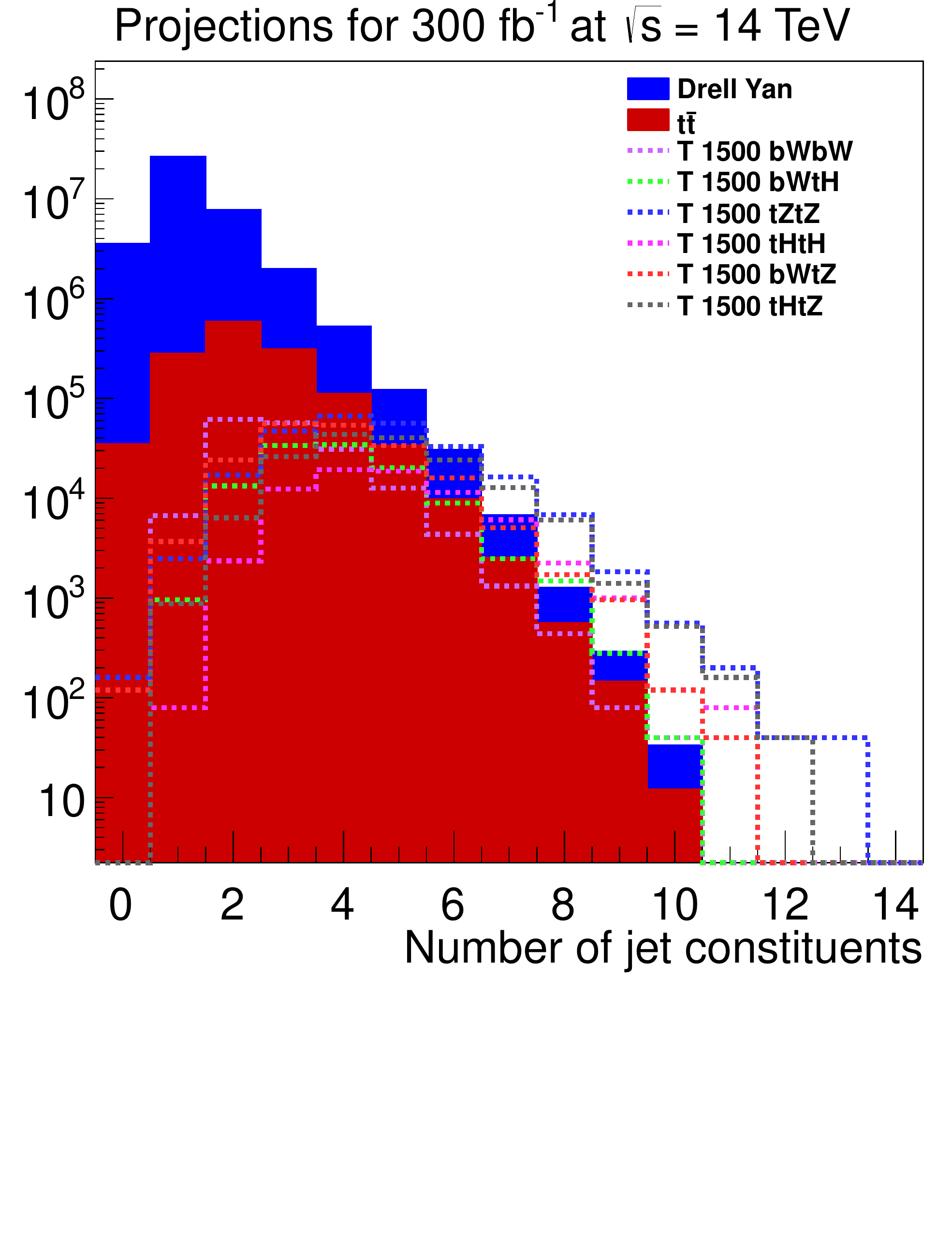}
\caption{Distributions of  H$_{T}$, minM$_{lb}$, S$_{T}$ and the number of jet constituents for the OS category. The signal is scaled by 5000.}
\end{figure}

\begin{figure}[!h]
\includegraphics[width=0.3\textwidth]{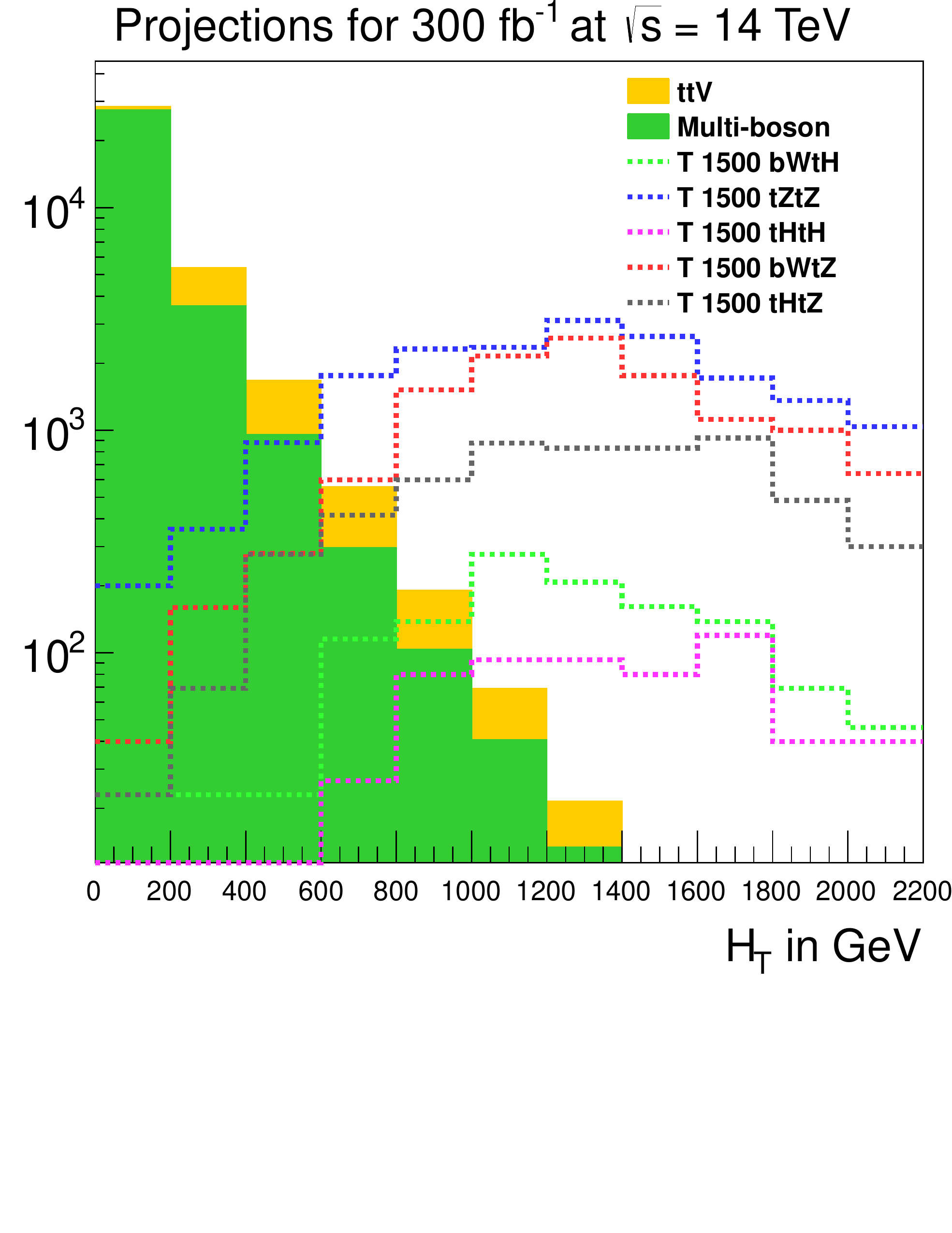}
\includegraphics[width=0.3\textwidth]{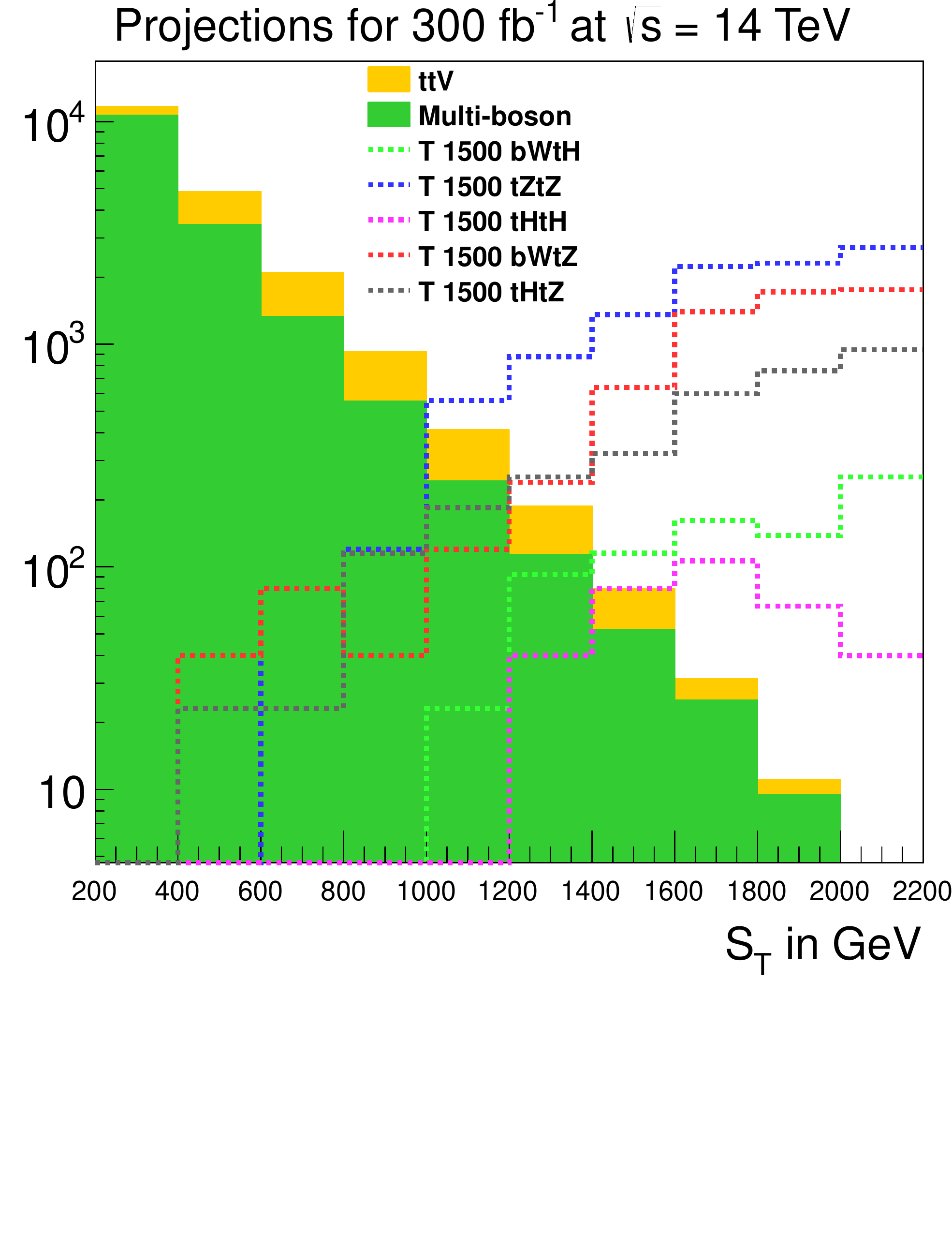}
\includegraphics[width=0.3\textwidth]{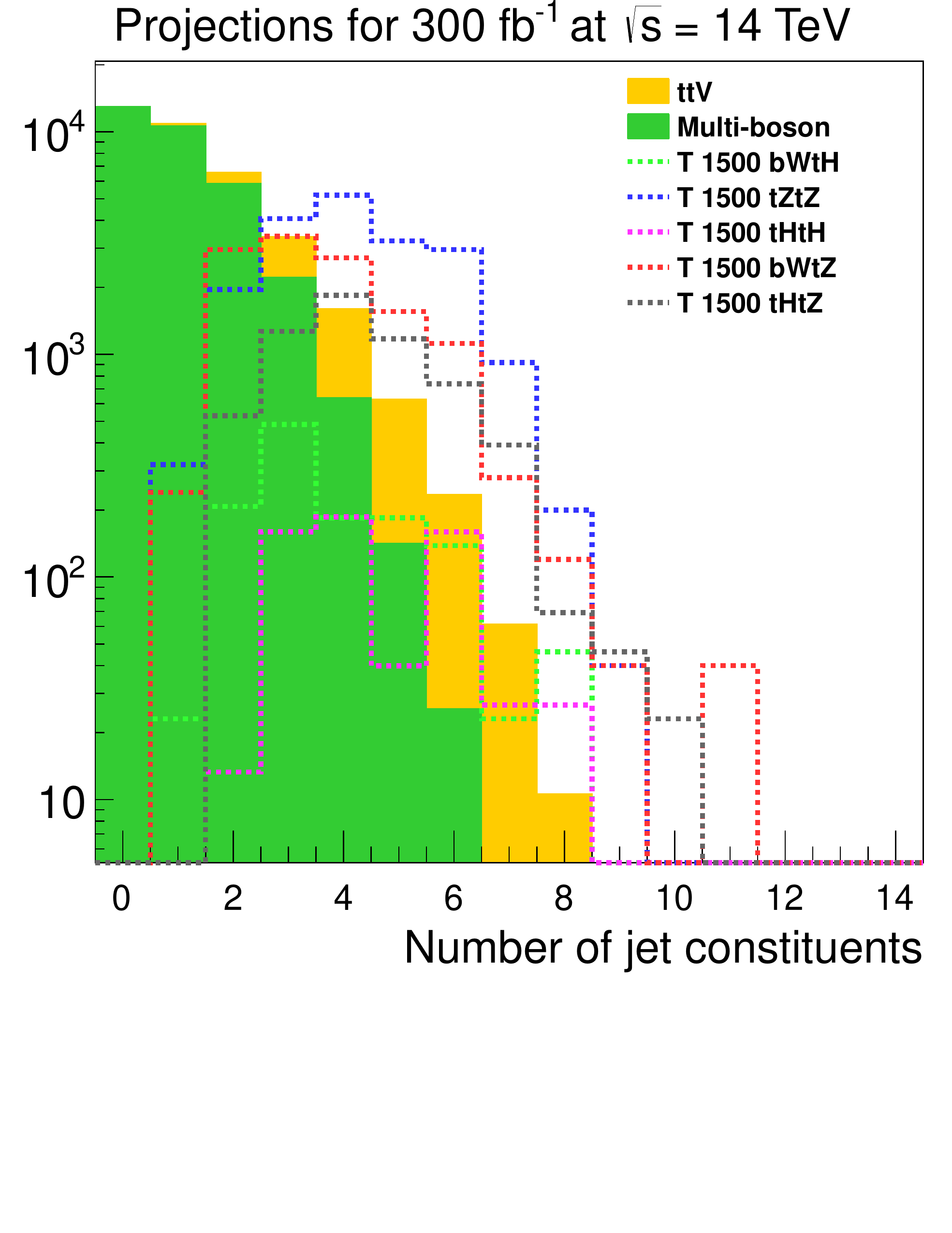}
\caption{Distributions of  H$_{T}$, S$_{T}$ and the number of jet constituents for the SS category. The signal is scaled by 5000.}
\end{figure}

\begin{figure}[!h]
\includegraphics[width=0.3\textwidth]{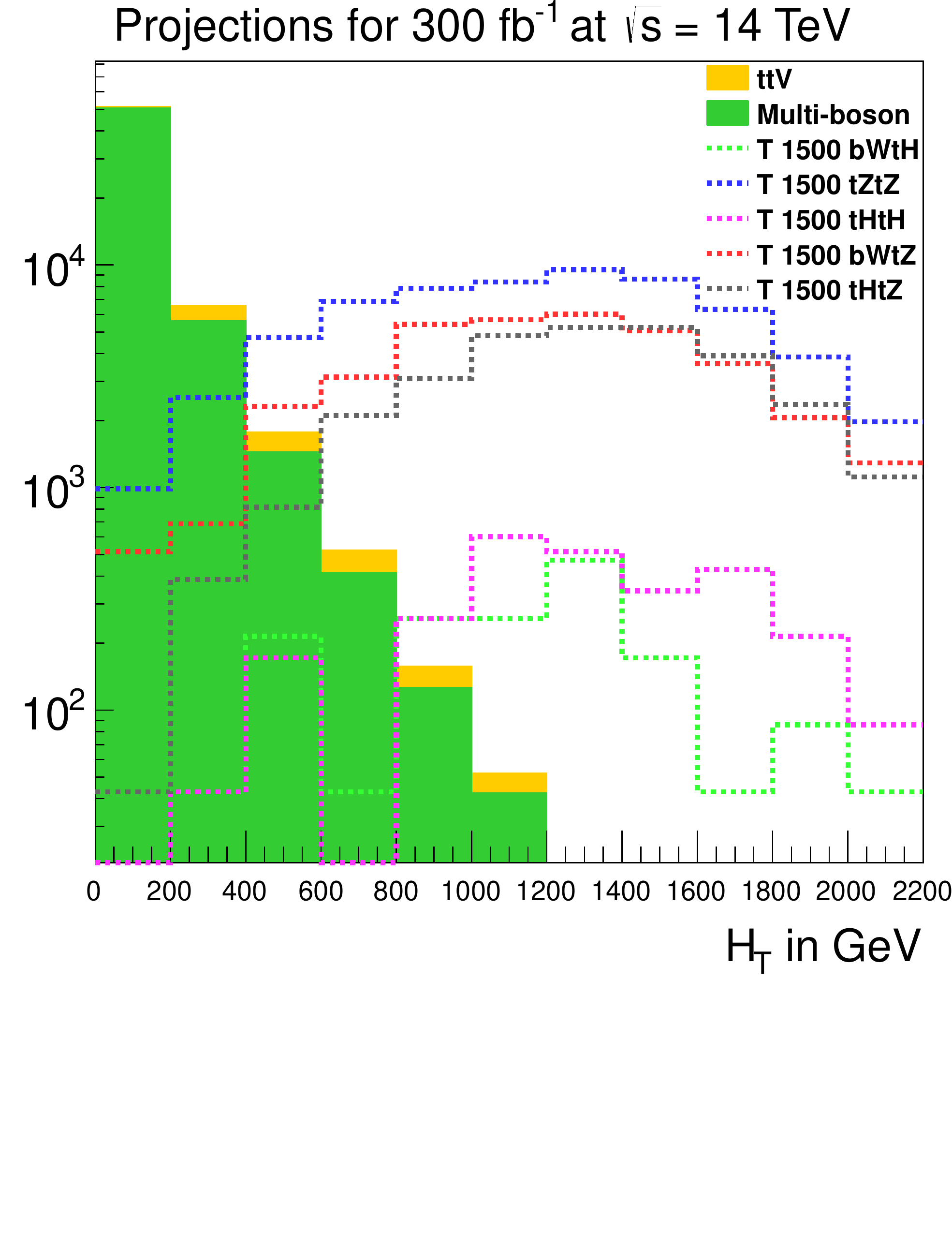}
\includegraphics[width=0.3\textwidth]{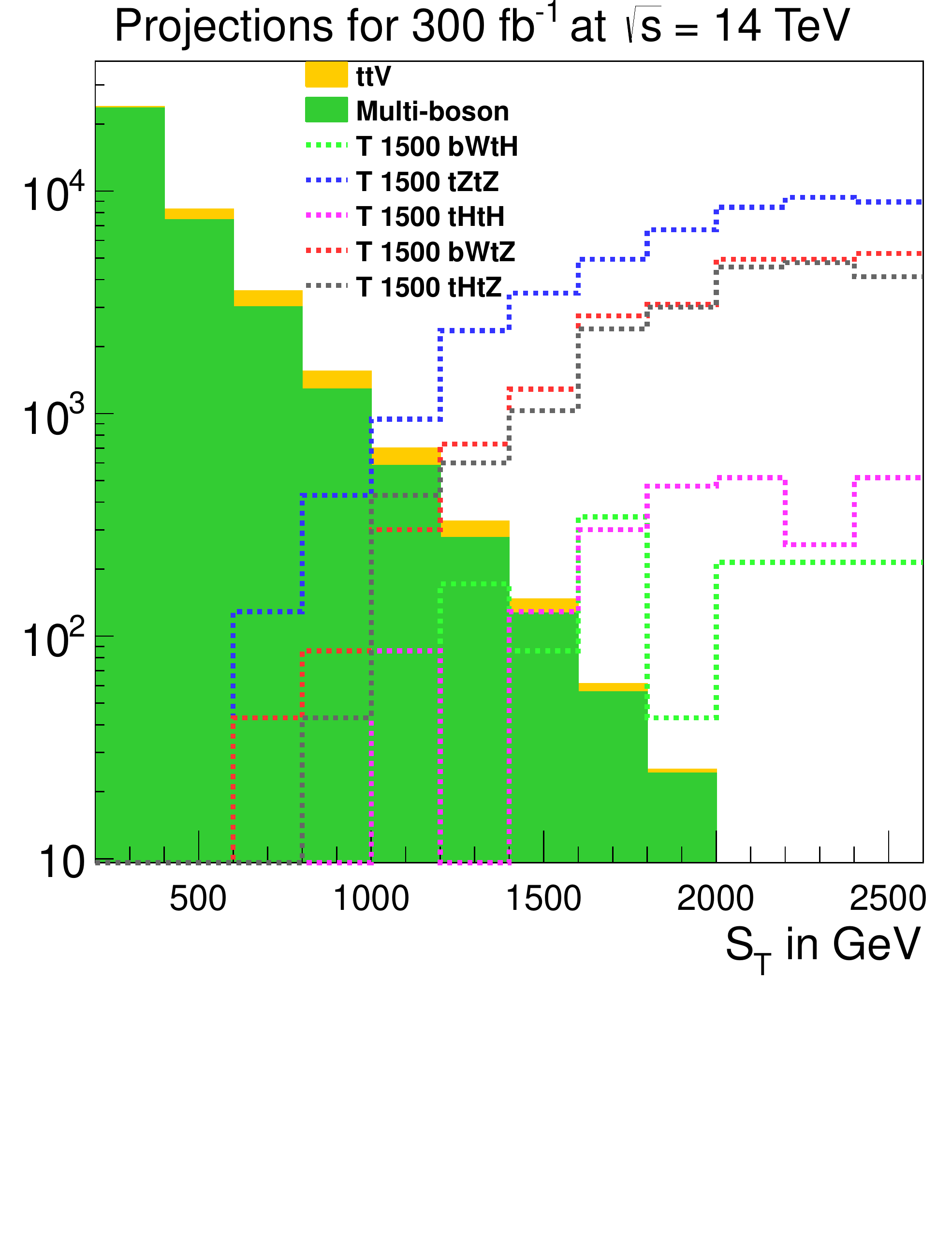}
\includegraphics[width=0.3\textwidth]{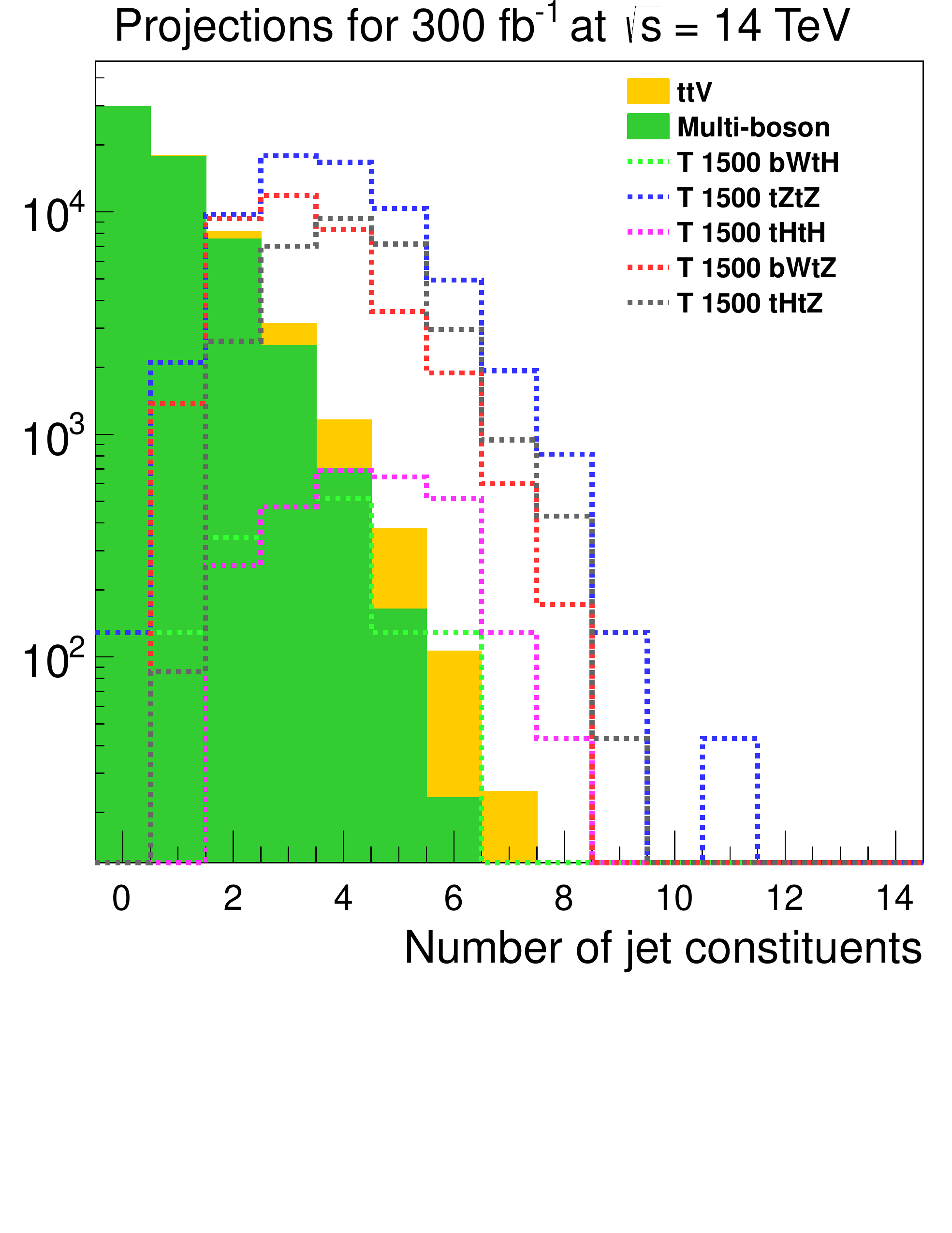}
\caption{Distributions for  H$_{T}$, S$_{T}$ and the number of jet constituents for events with $\geq$ 3 leptons. The signal is scaled by 5000.}
\end{figure}

\clearpage

\section{Distributions ($\sqrt{s}$=33 TeV)}

\begin{figure}[h]
\includegraphics[width=0.45\textwidth]{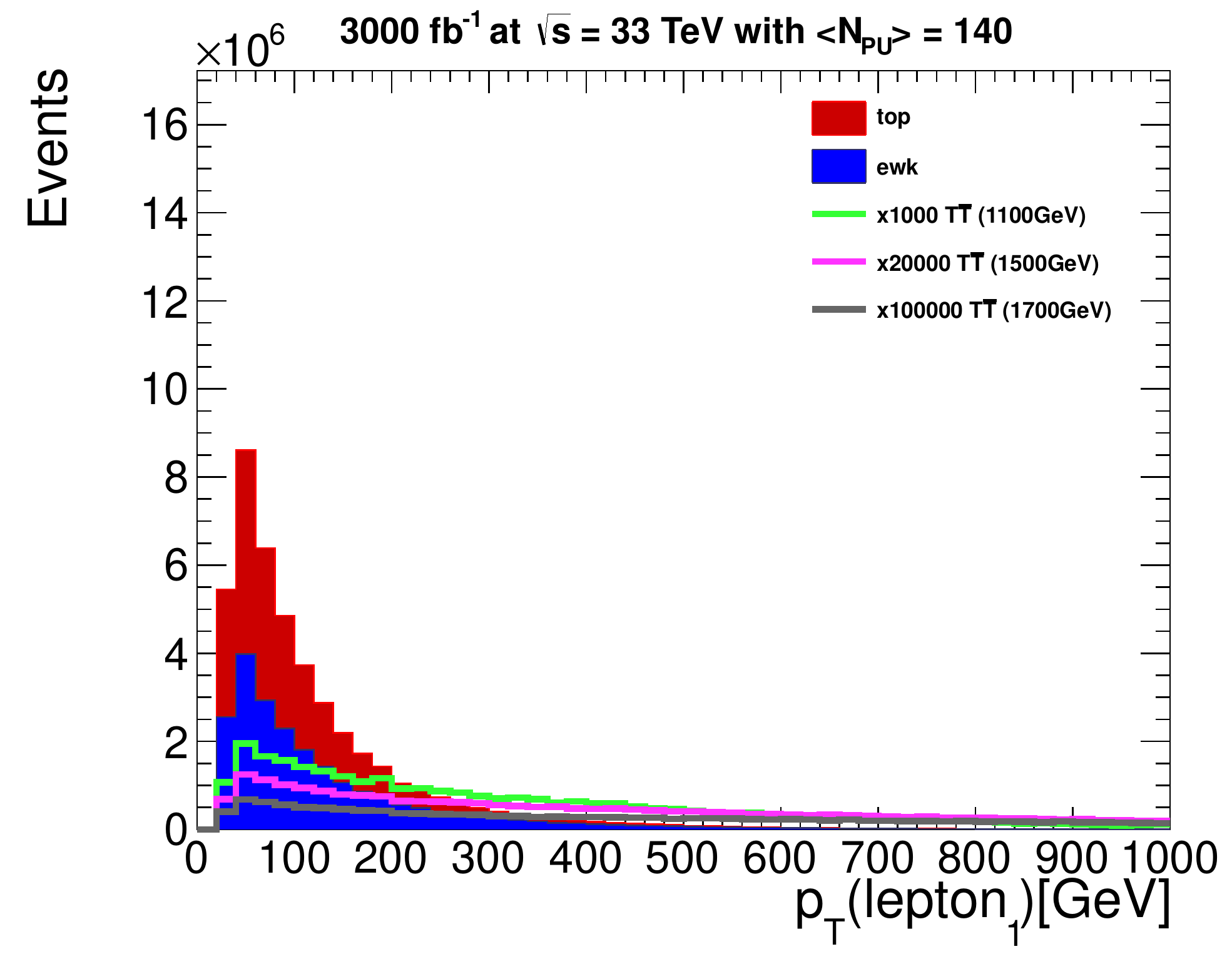}
\includegraphics[width=0.45\textwidth]{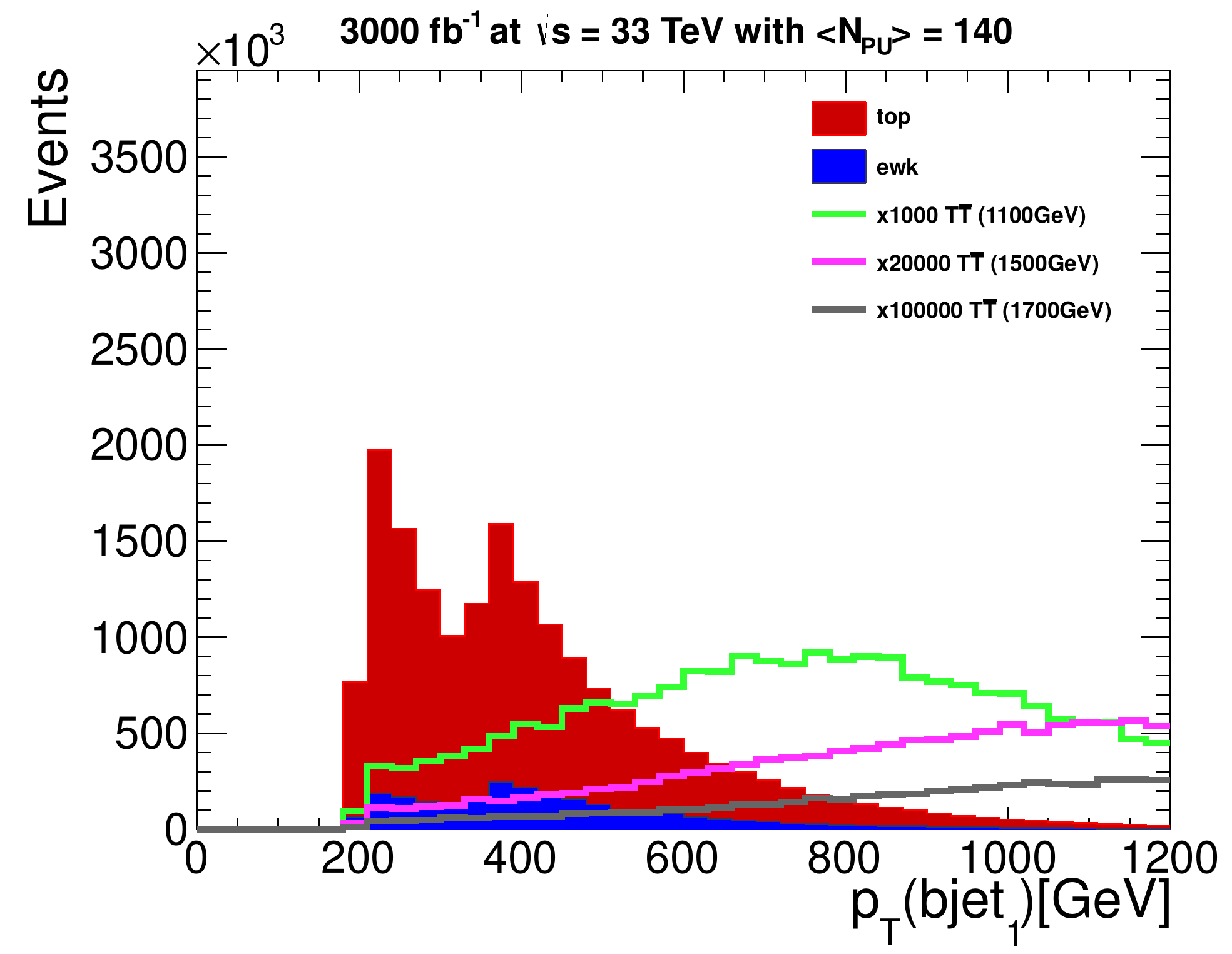}\\
\includegraphics[width=0.45\textwidth]{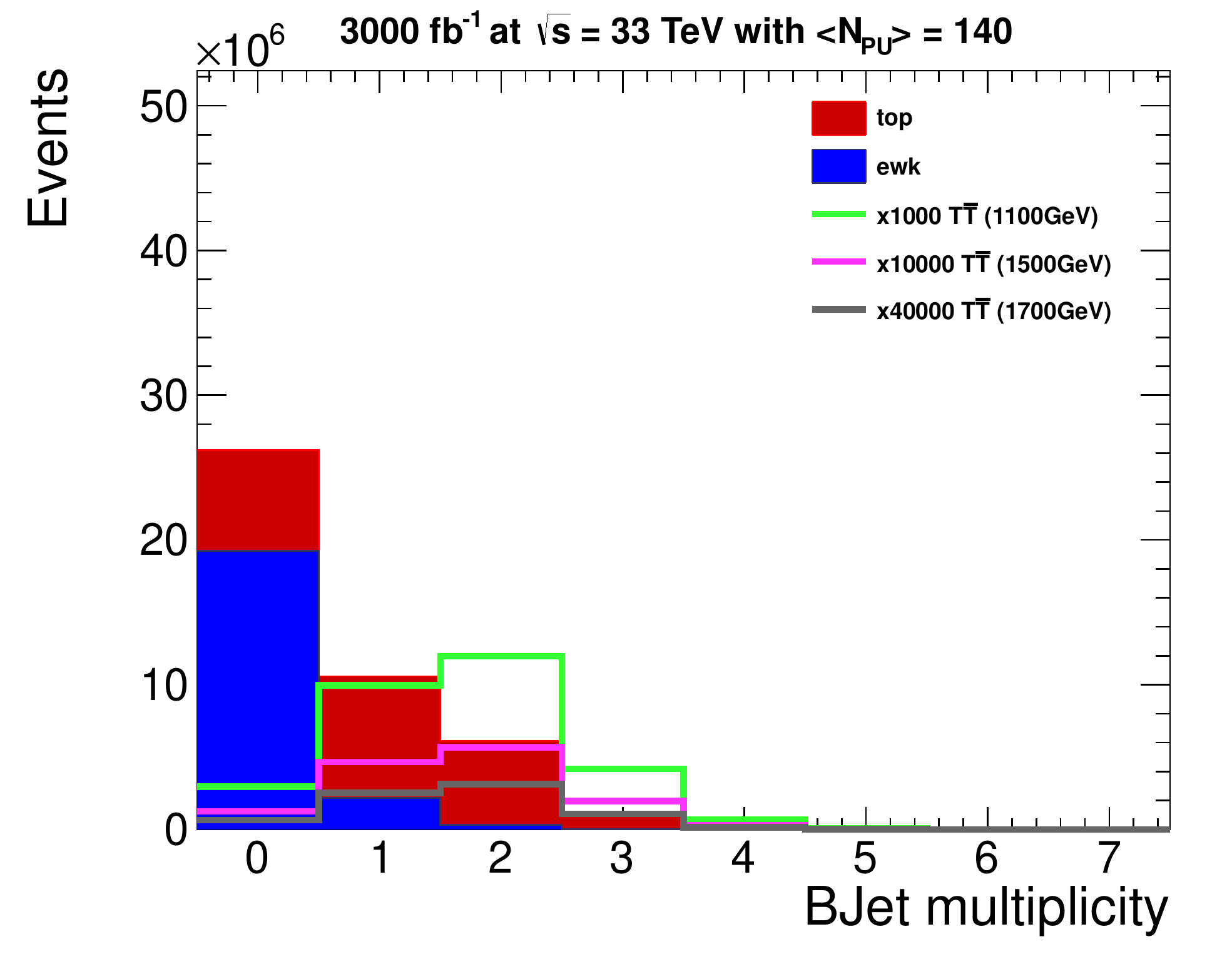}
\includegraphics[width=0.45\textwidth]{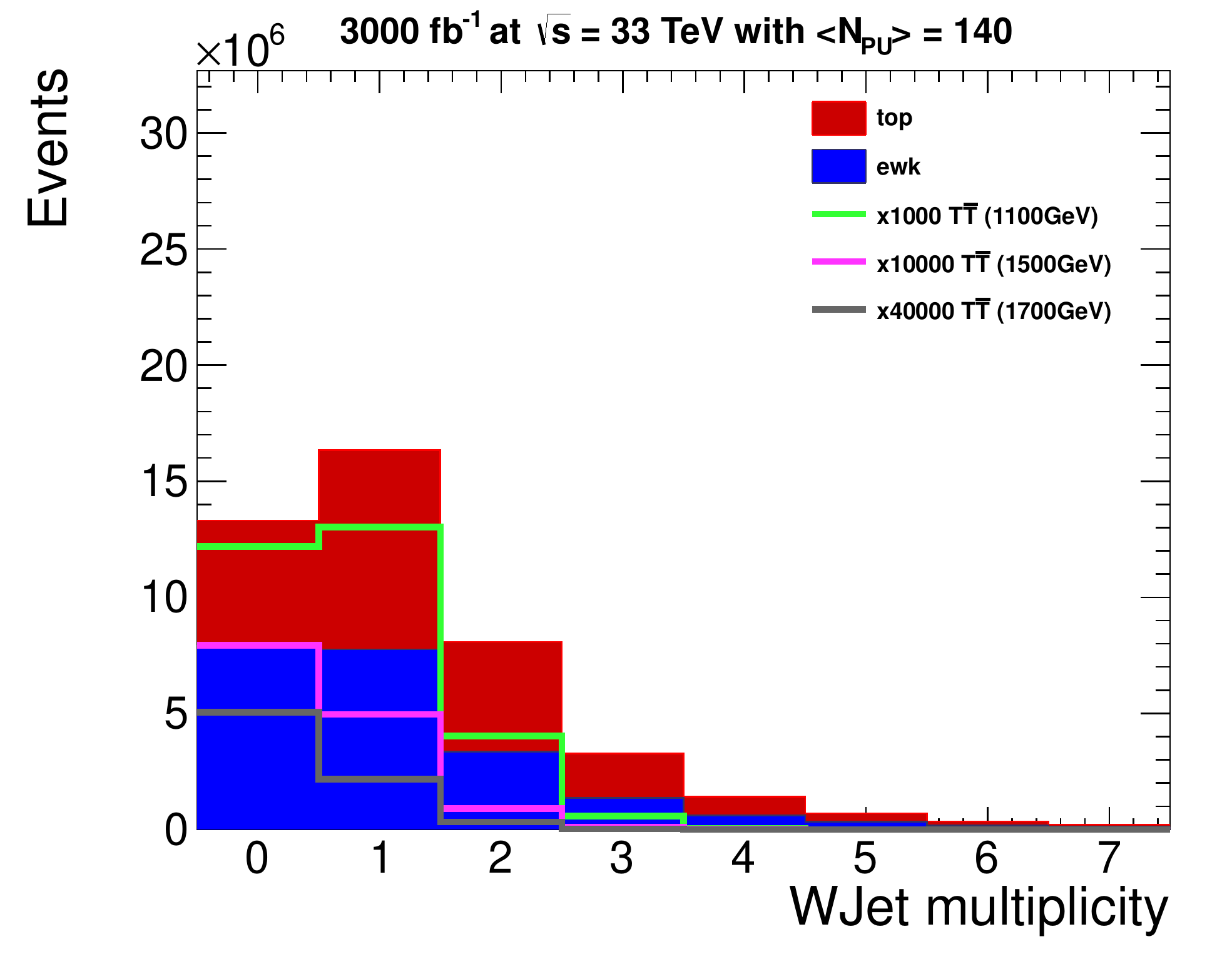}
\caption{Distributions of leading electron, leading b-jet, b-jet multiplicity and W-jet multiplicity in the $l+$jets channel .}
\end{figure}

\begin{figure}[h]
\includegraphics[width=0.45\textwidth]{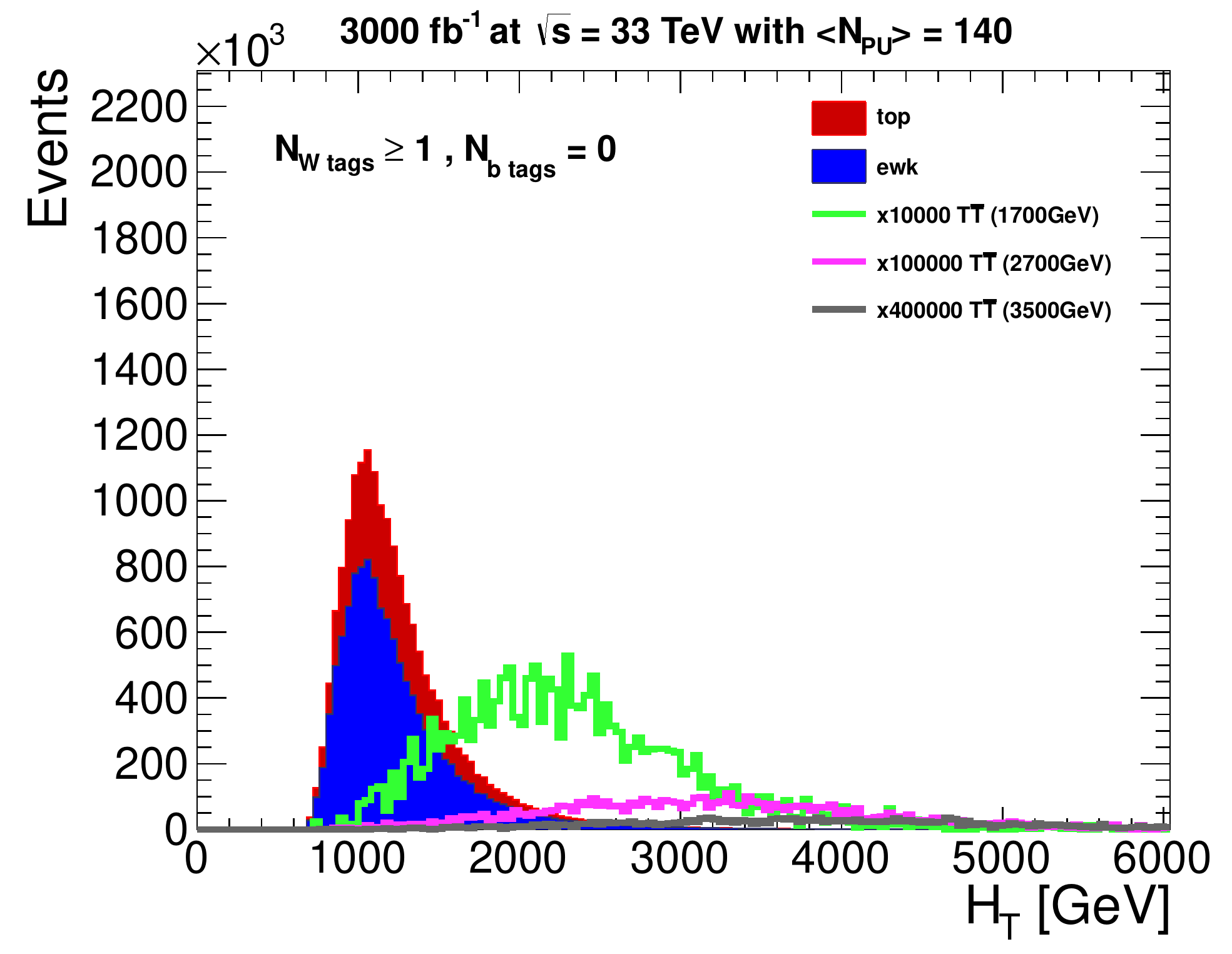}
\includegraphics[width=0.45\textwidth]{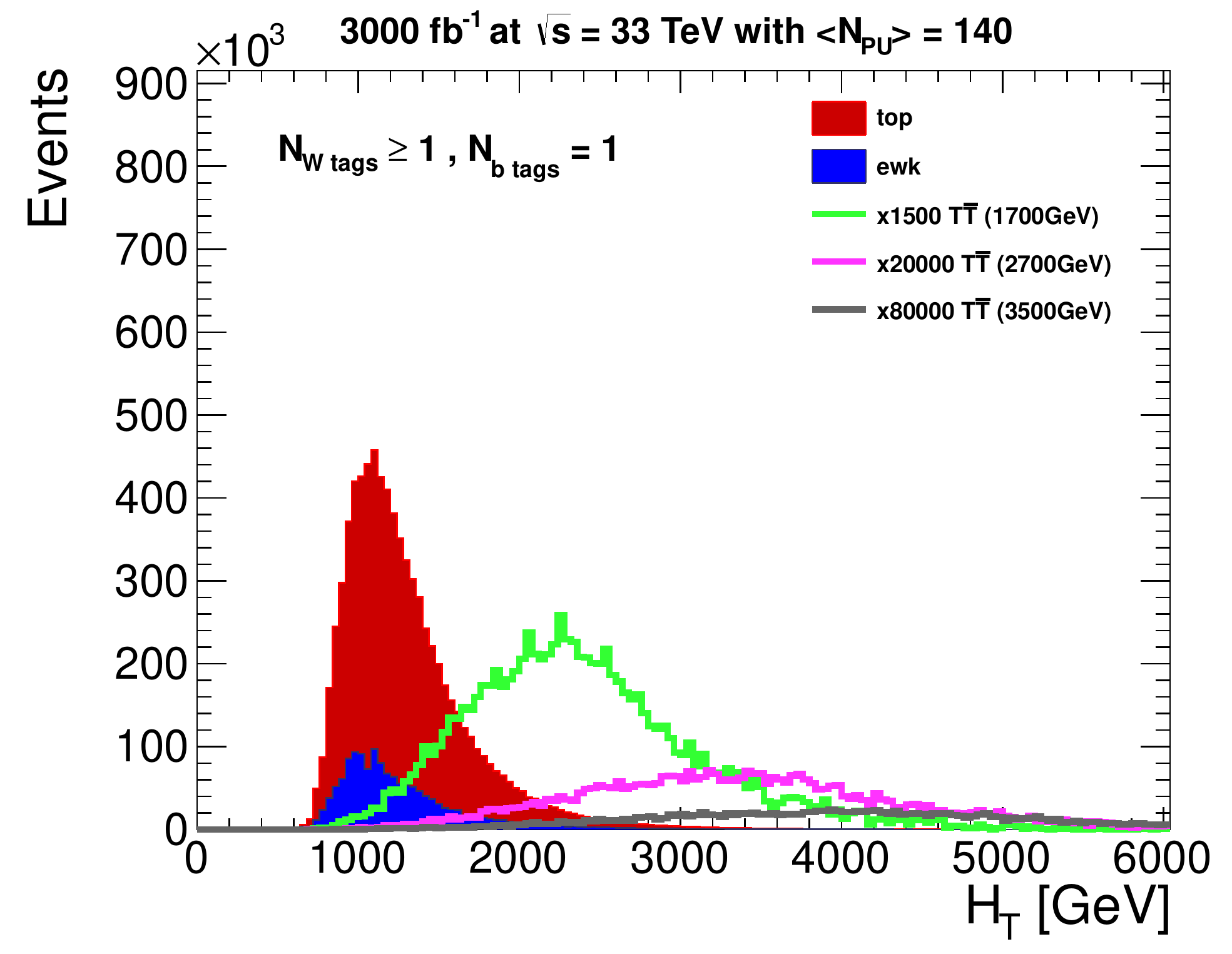}\\
\includegraphics[width=0.45\textwidth]{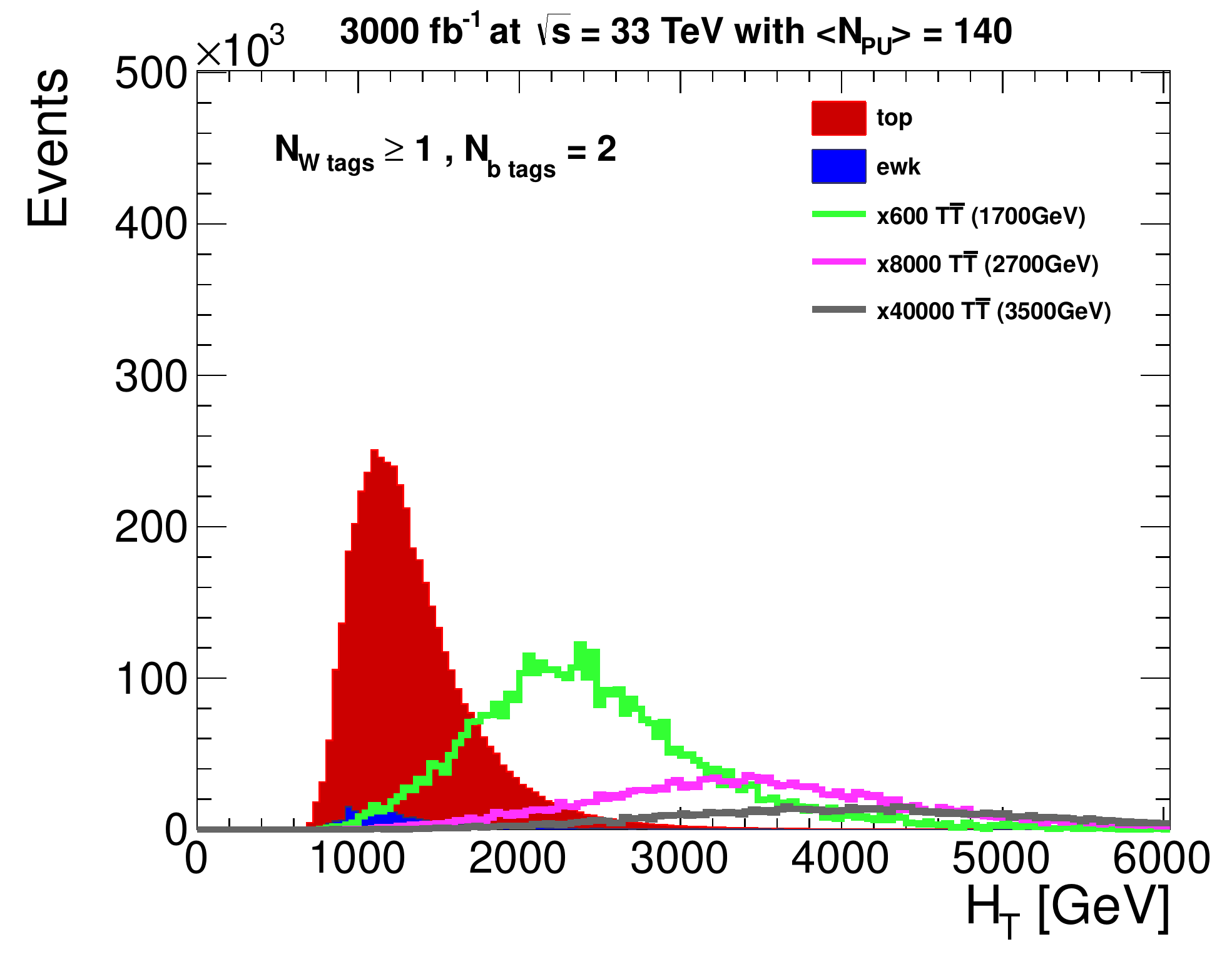}
\includegraphics[width=0.45\textwidth]{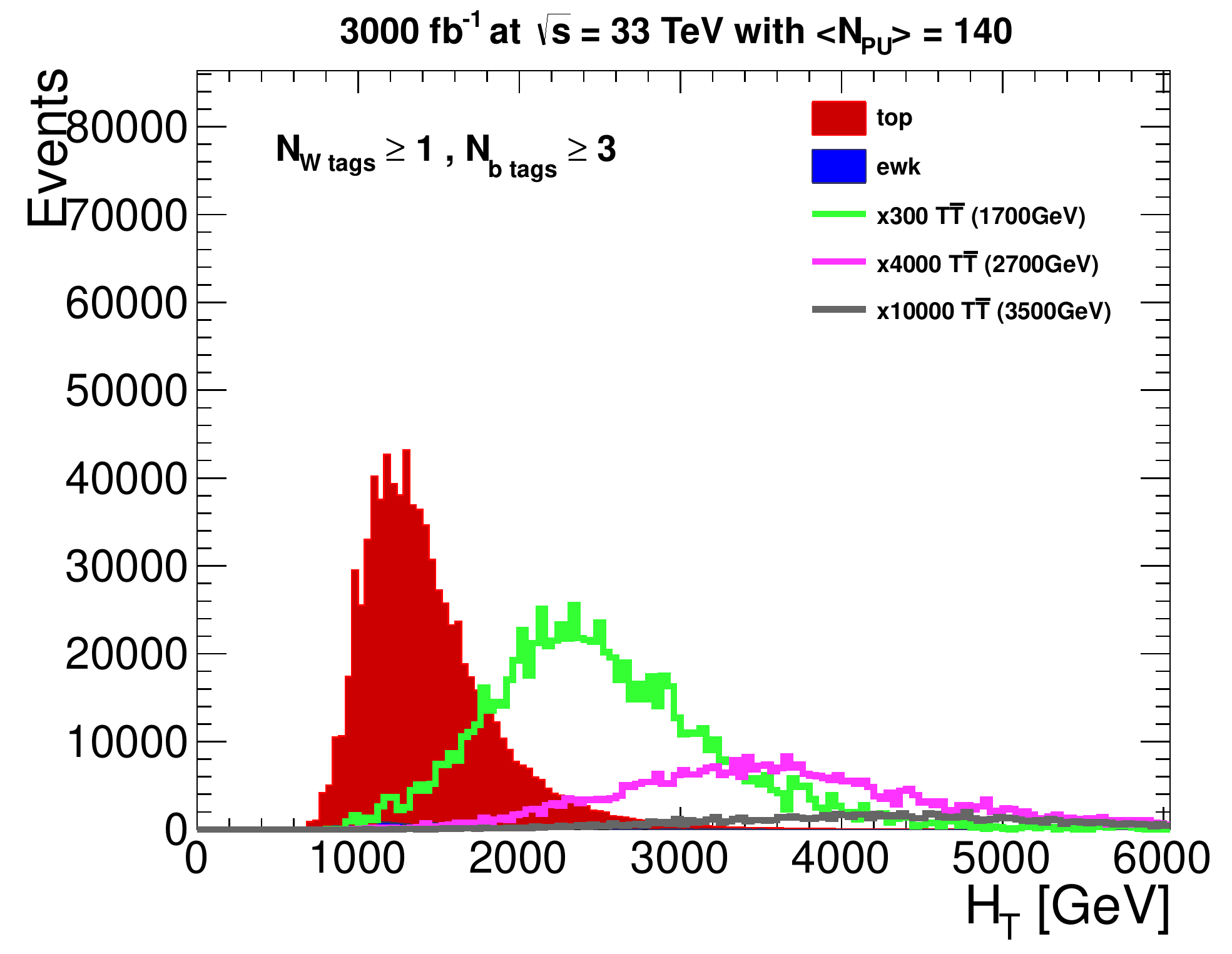}
\caption{Distributions of $H_T$ in different event categories with $l+\geq$3 jets with 1 $W-$jet and 0 b-jet (top left), 1 b-jet (top right), 2 b-jet (bottom left) and at least 3 b-jets (bottom right) .}
\end{figure}

\begin{figure}[h]
\includegraphics[width=0.45\textwidth]{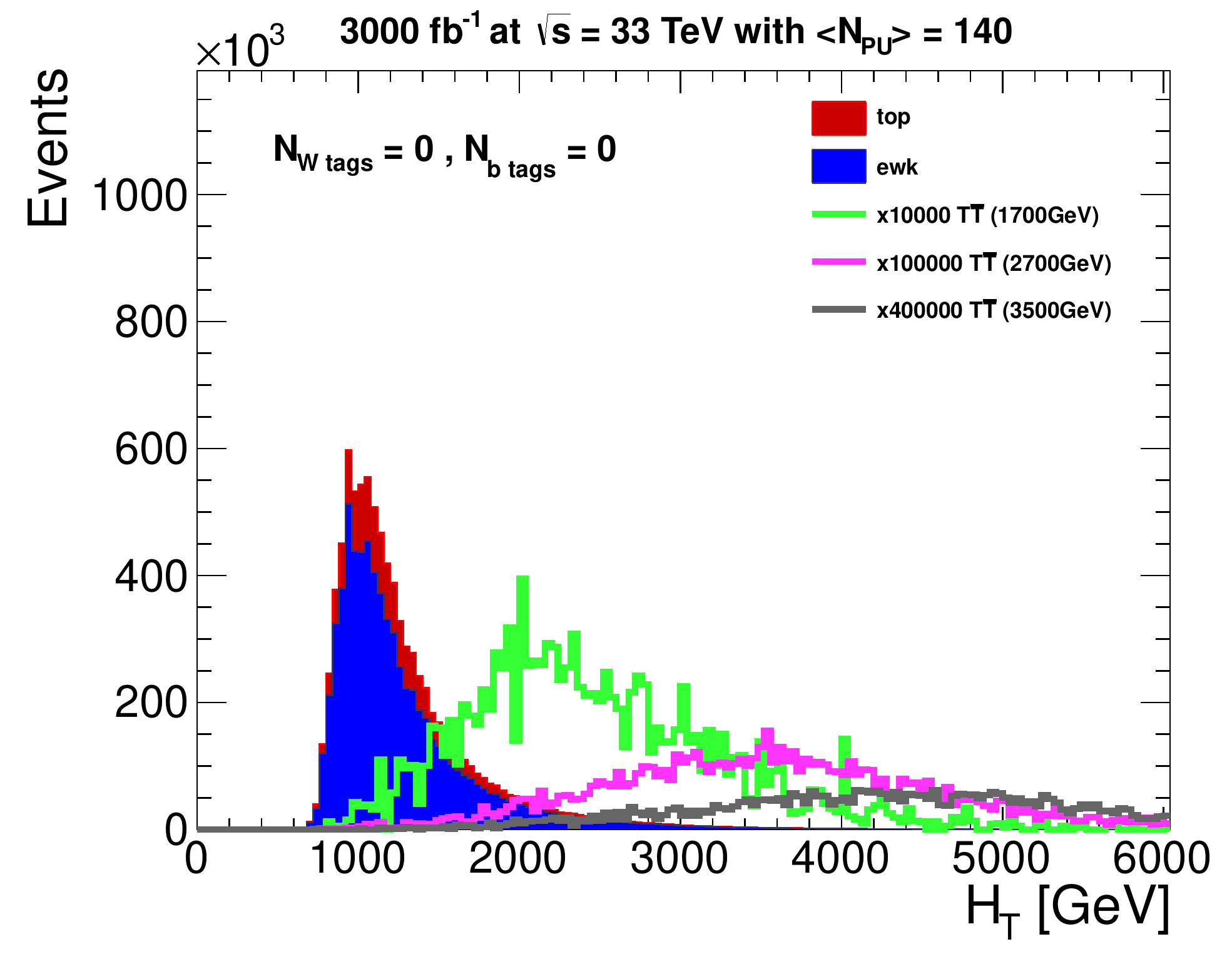}
\includegraphics[width=0.45\textwidth]{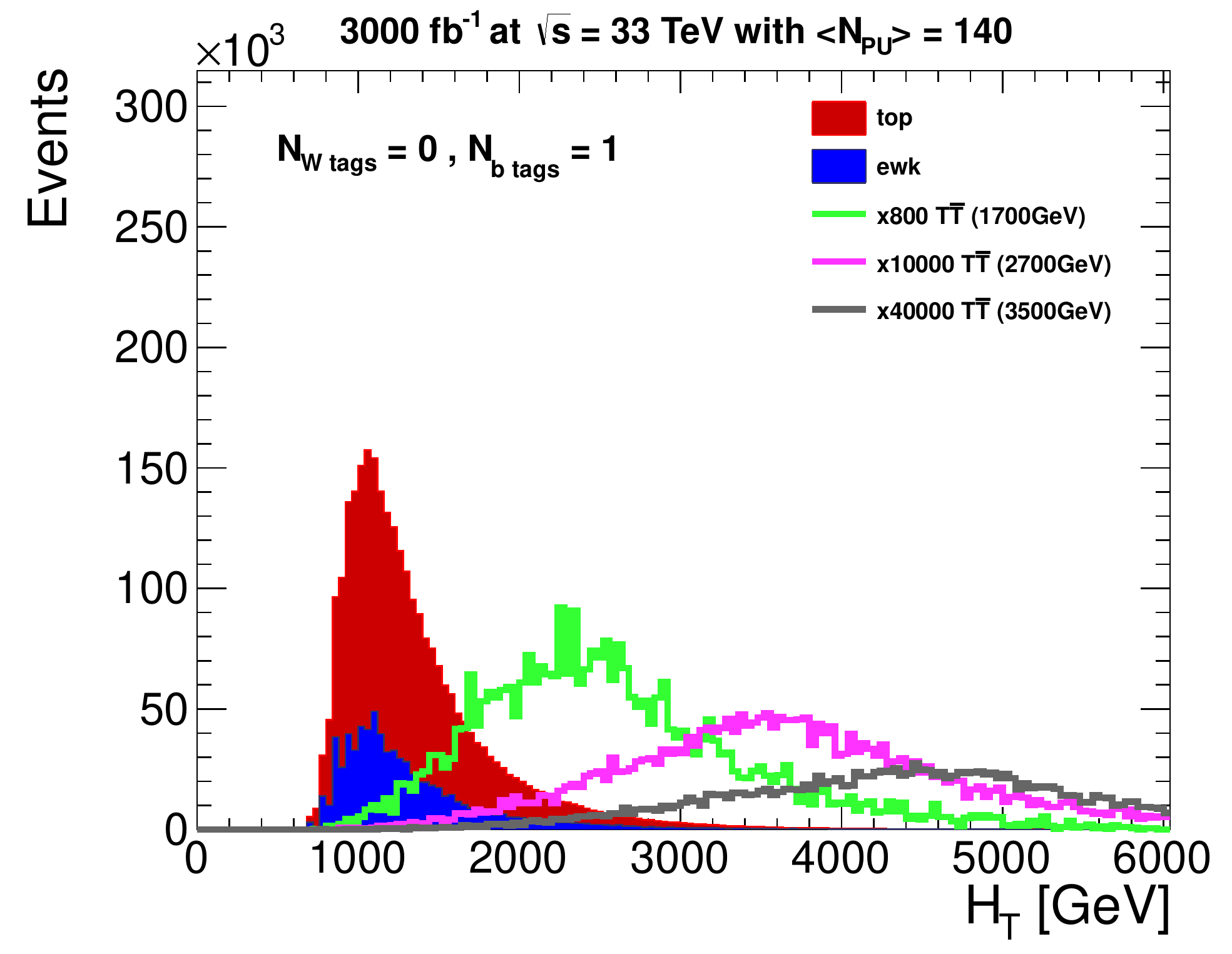}\\
\includegraphics[width=0.45\textwidth]{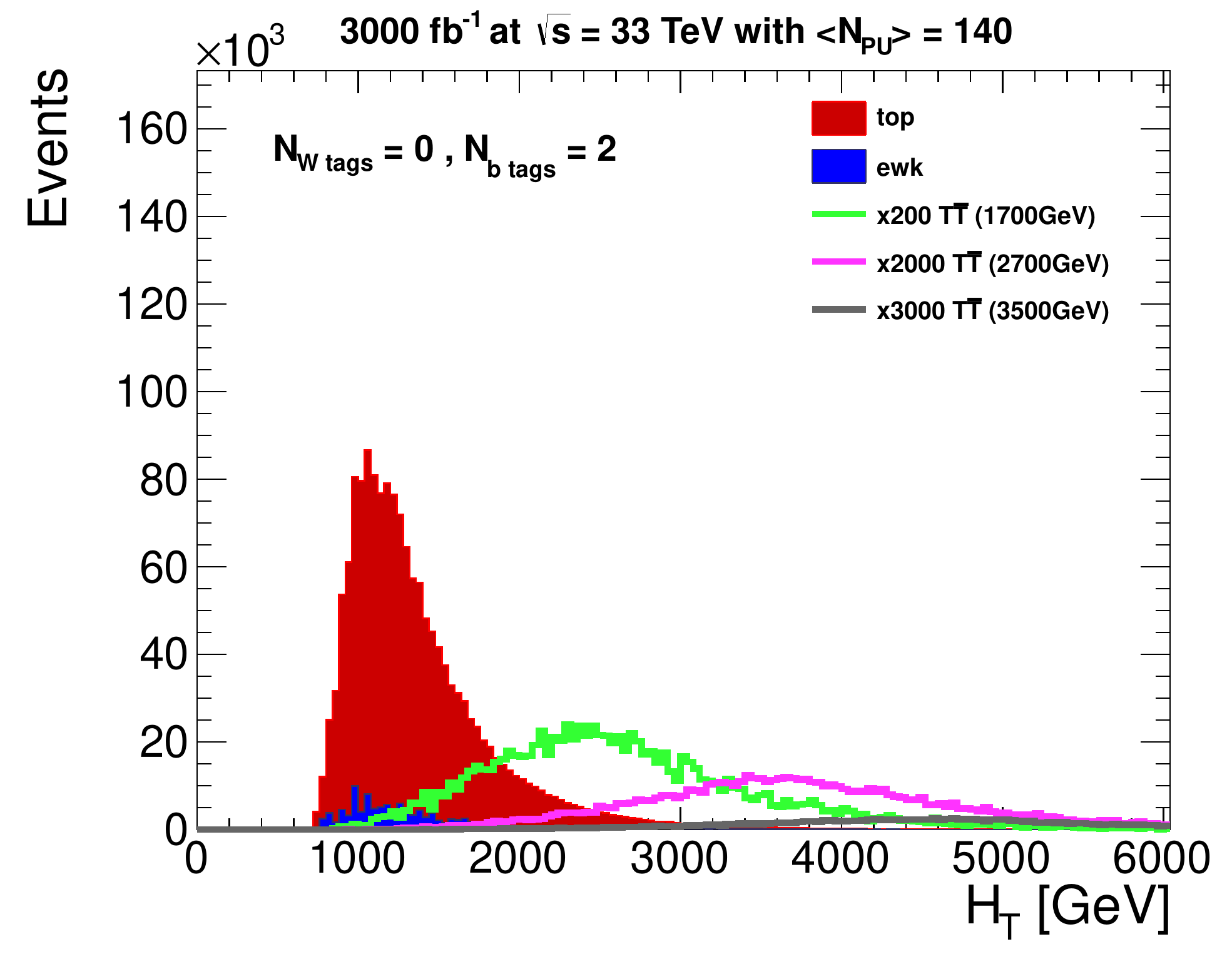}
\includegraphics[width=0.45\textwidth]{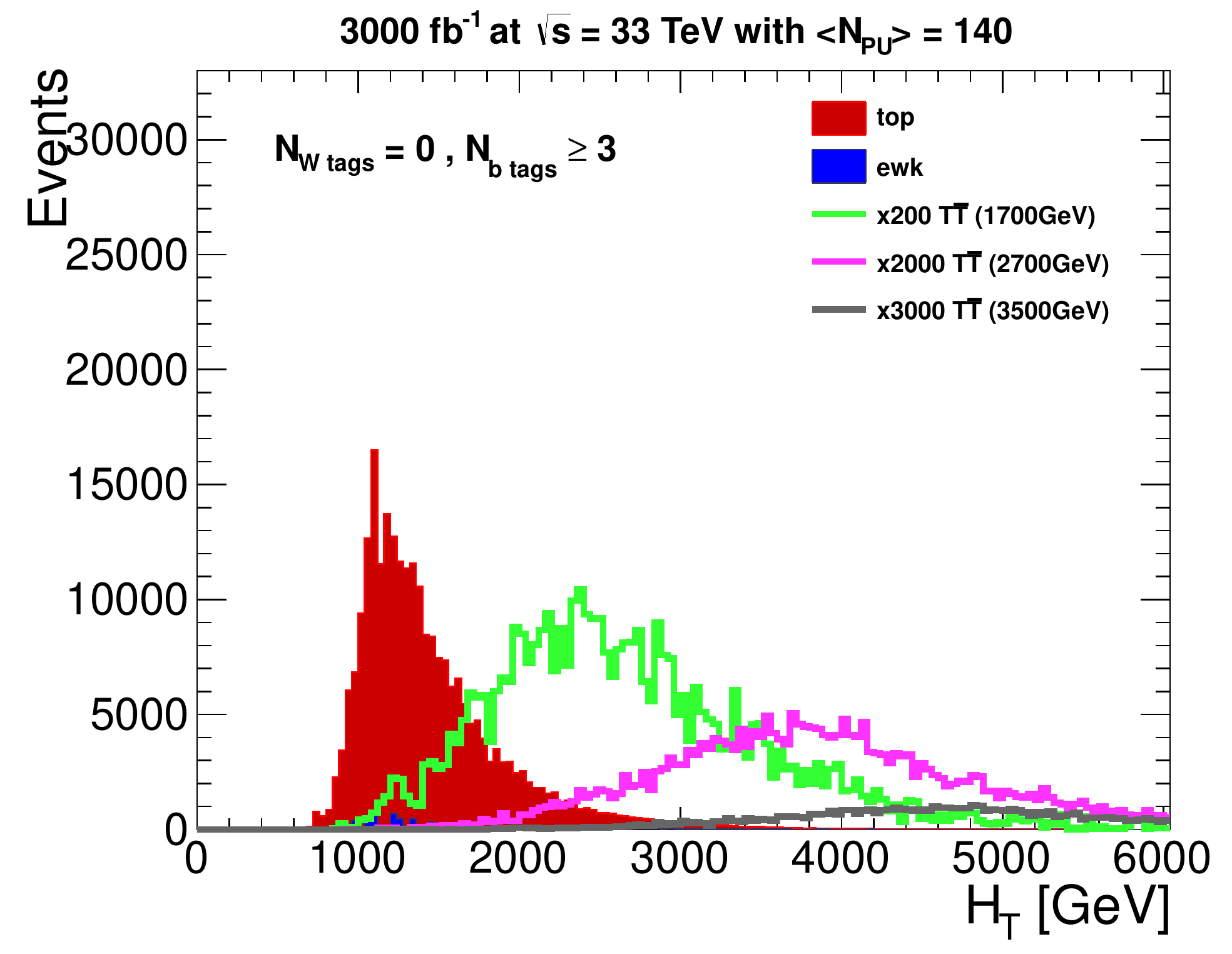}
\caption{Distributions of $H_T$ in different event categories with $l+\geq$4 jets with no $W-$jet and 0 b-jet (top left), 1 b-jet (top right), 2 b-jet (bottom left) and at least 3 b-jets (bottom right) .}
\end{figure}

\begin{figure}[!h]
\includegraphics[width=0.4\textwidth]{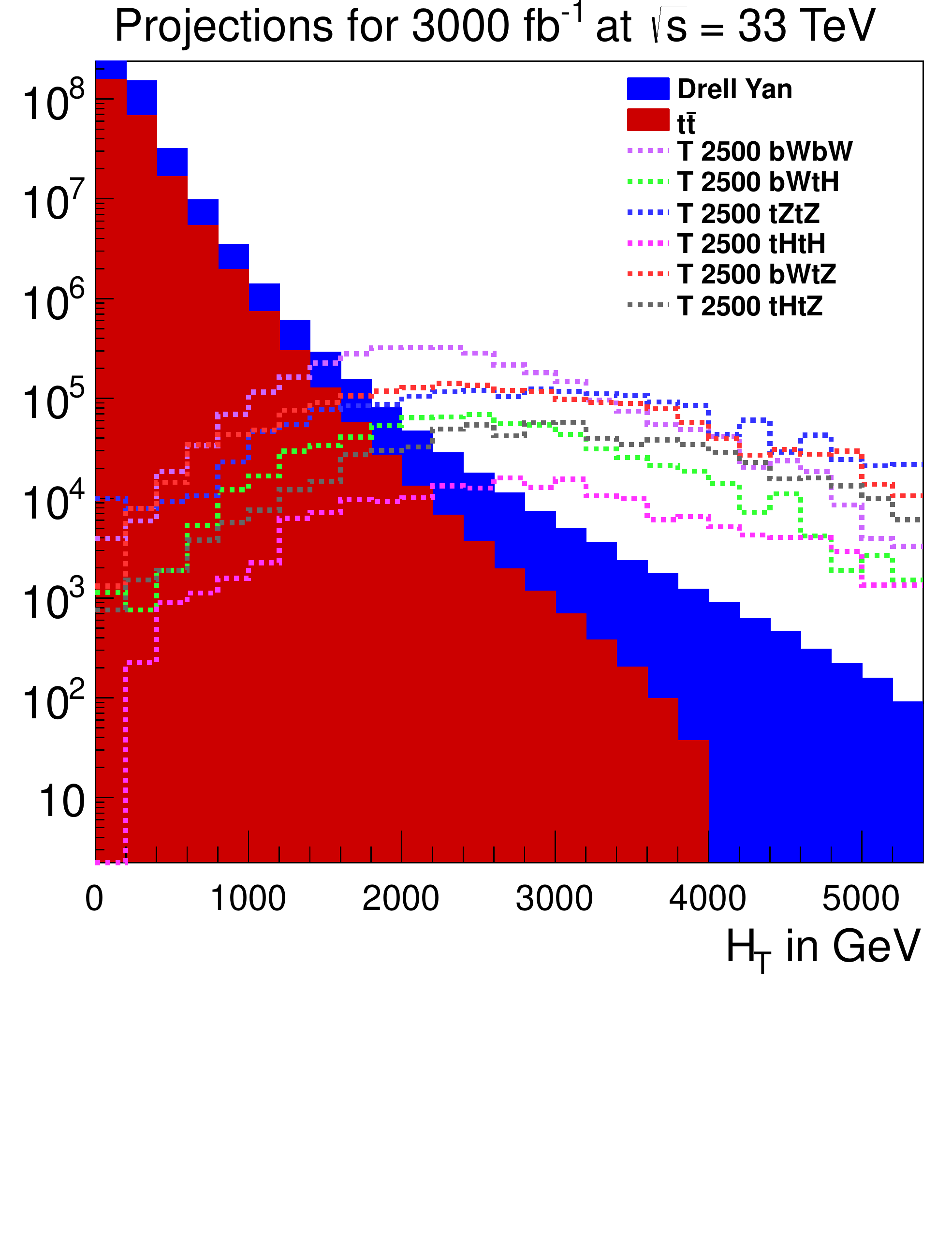}
\includegraphics[width=0.4\textwidth]{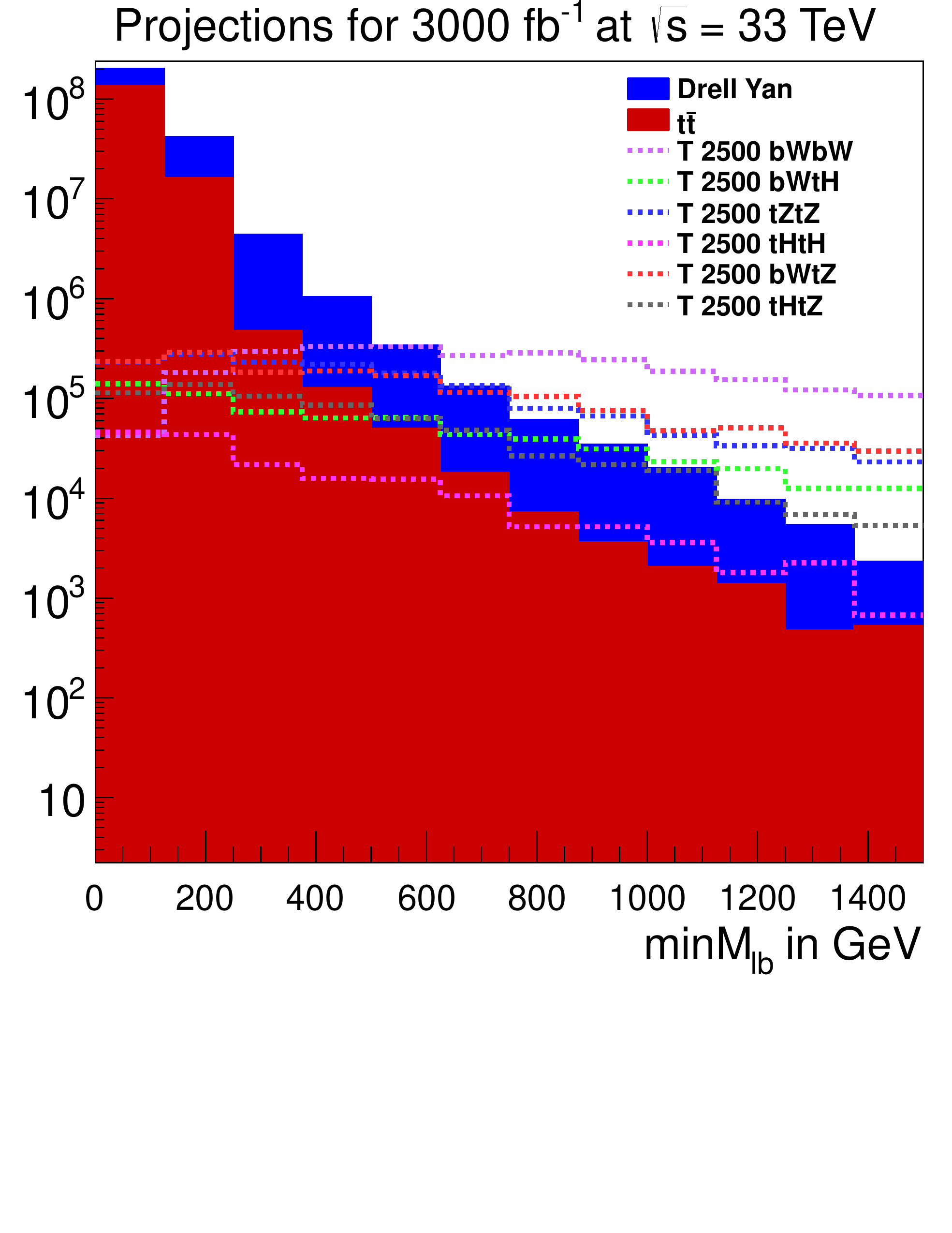}\\
\includegraphics[width=0.4\textwidth]{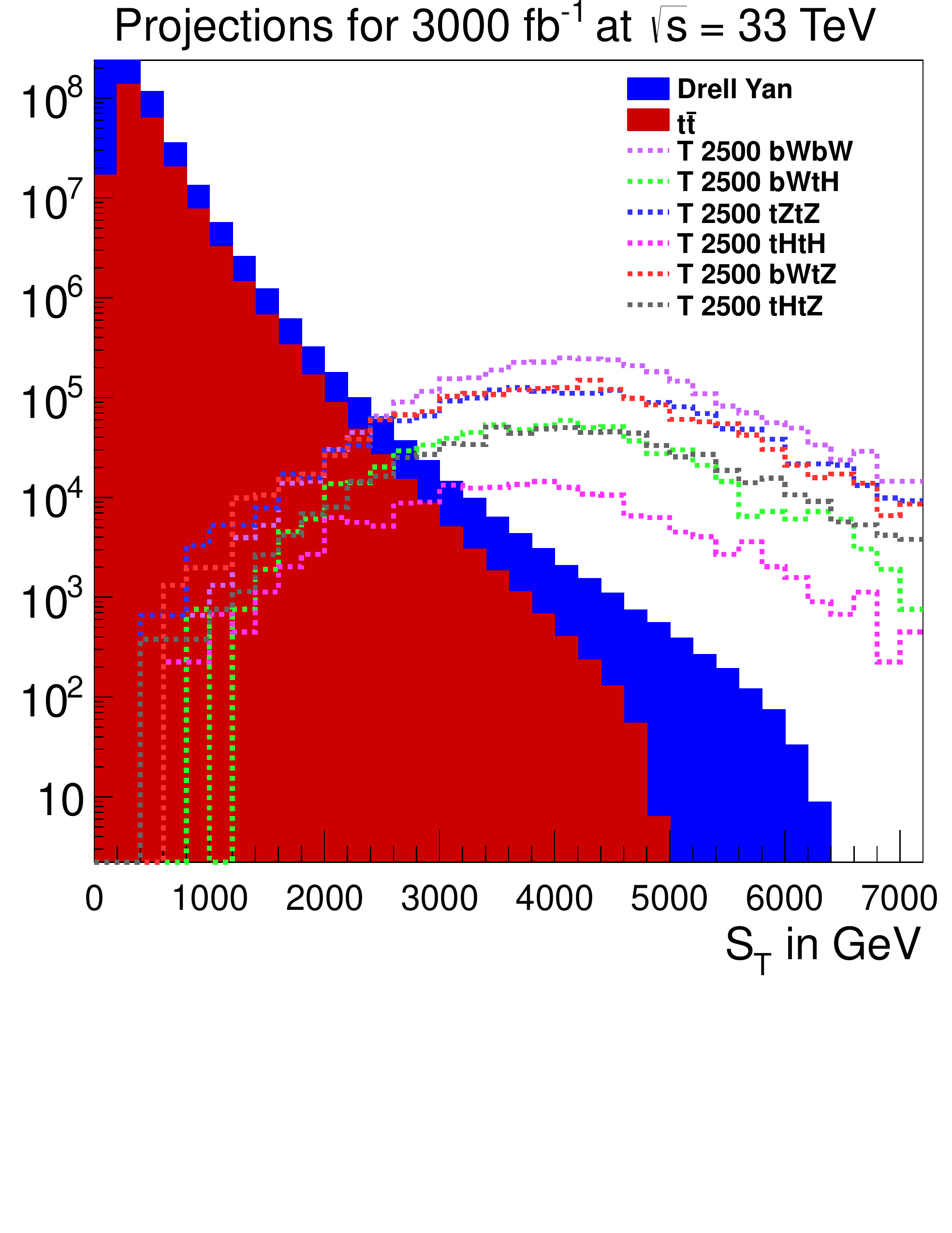}
\includegraphics[width=0.4\textwidth]{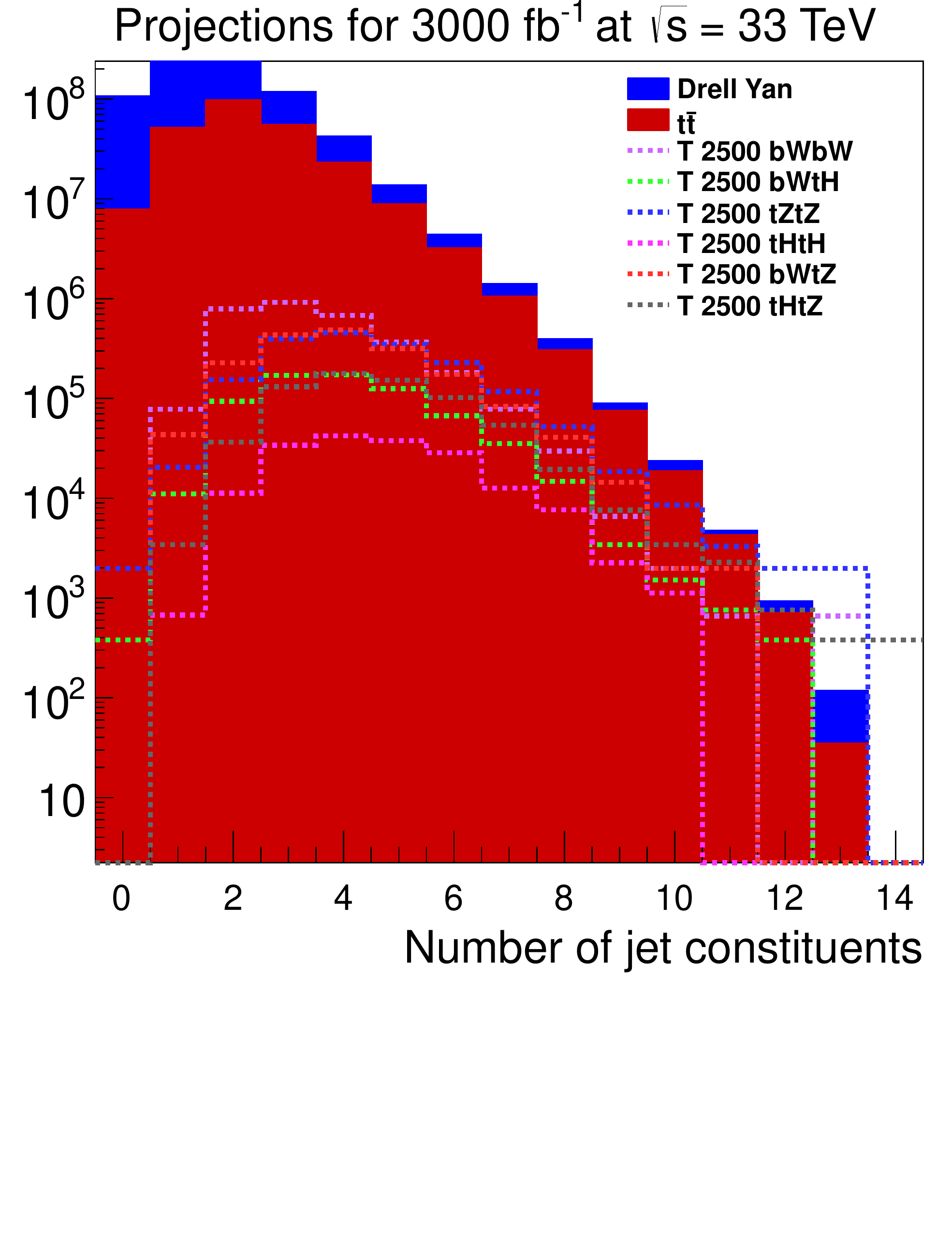}
\caption{Distributions of  H$_{T}$, minM$_{lb}$, S$_{T}$ and the number of jet constituents for the OS category. The signal is scaled by 5000.}
\end{figure}

\begin{figure}[!h]
\includegraphics[width=0.3\textwidth]{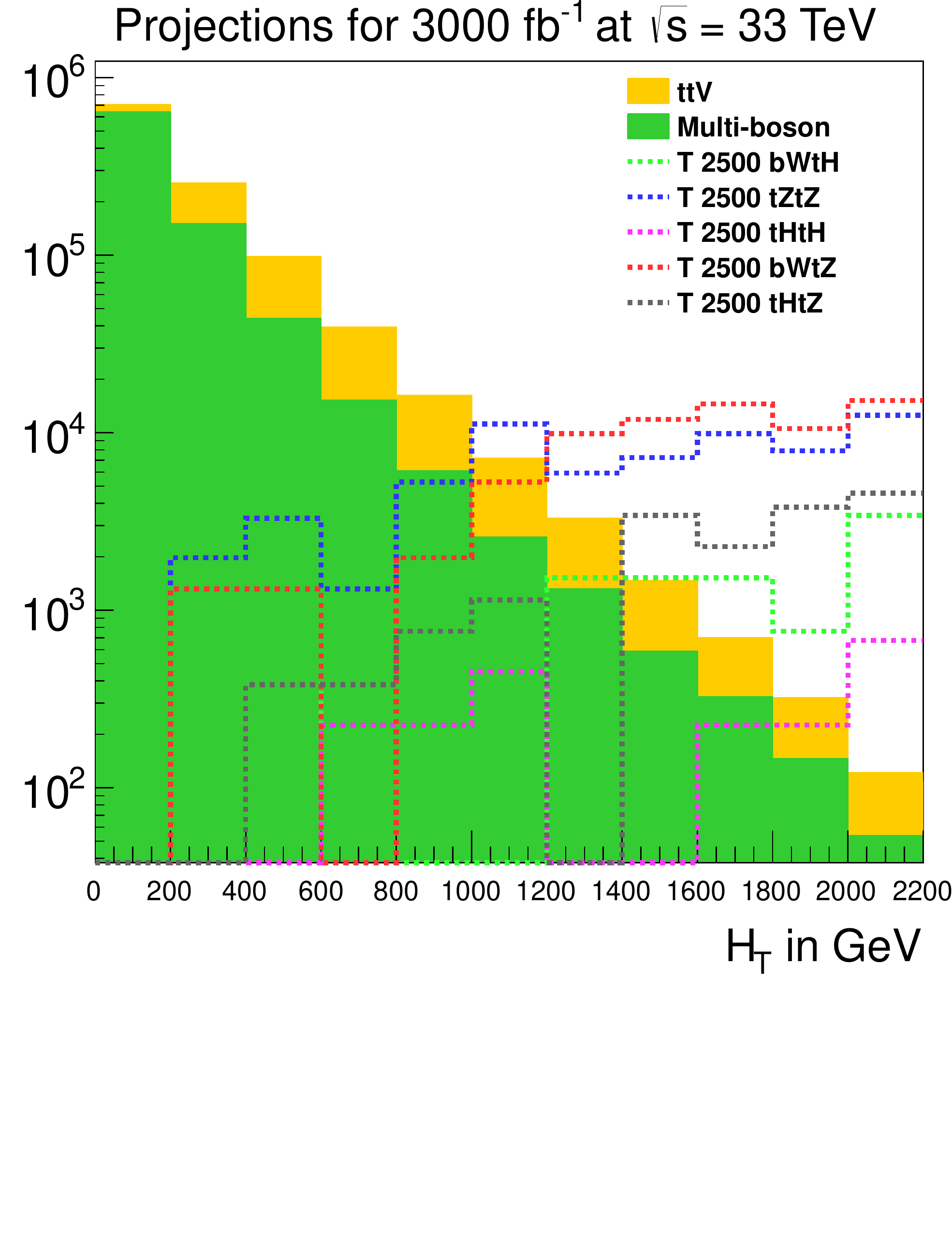}
\includegraphics[width=0.3\textwidth]{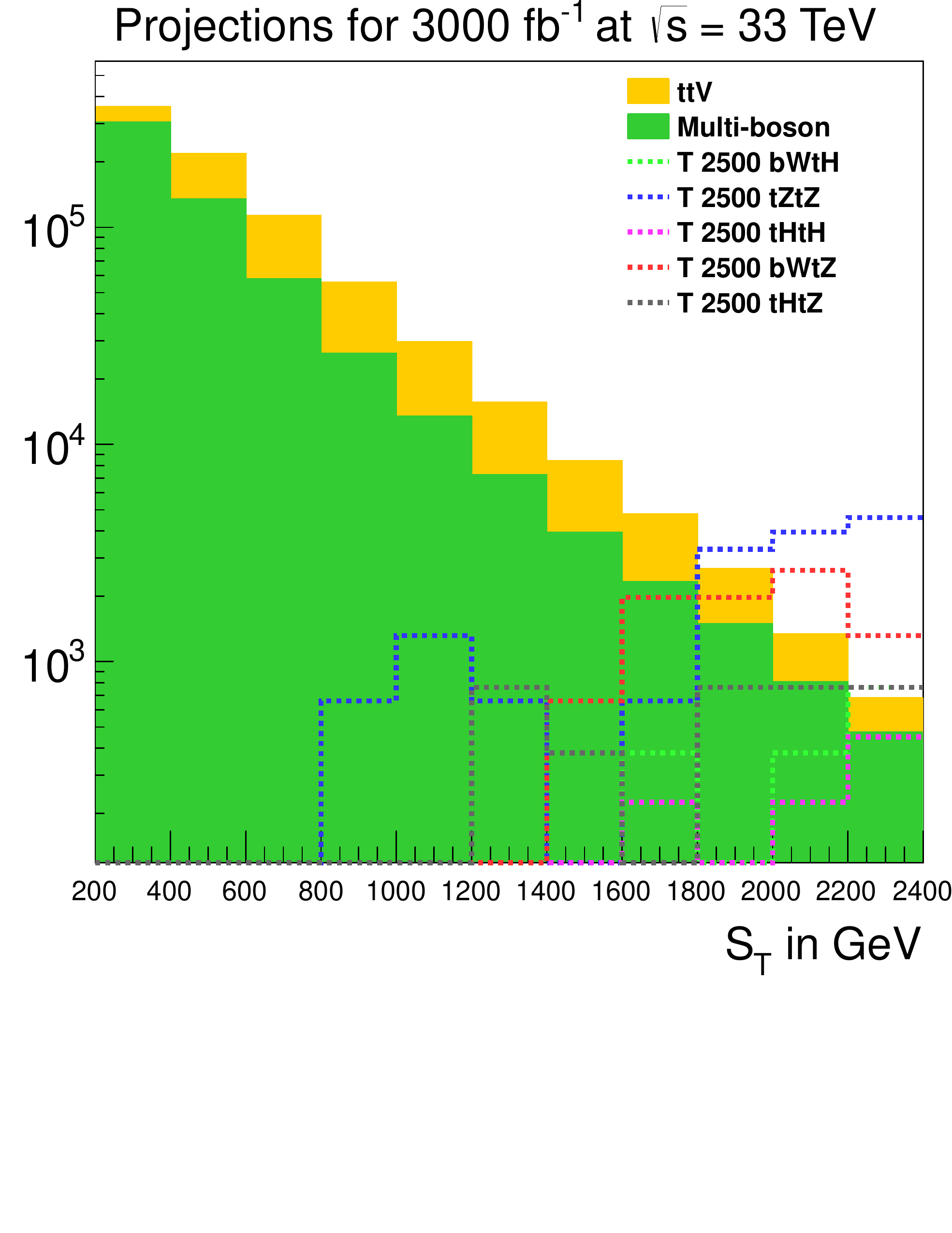}
\includegraphics[width=0.3\textwidth]{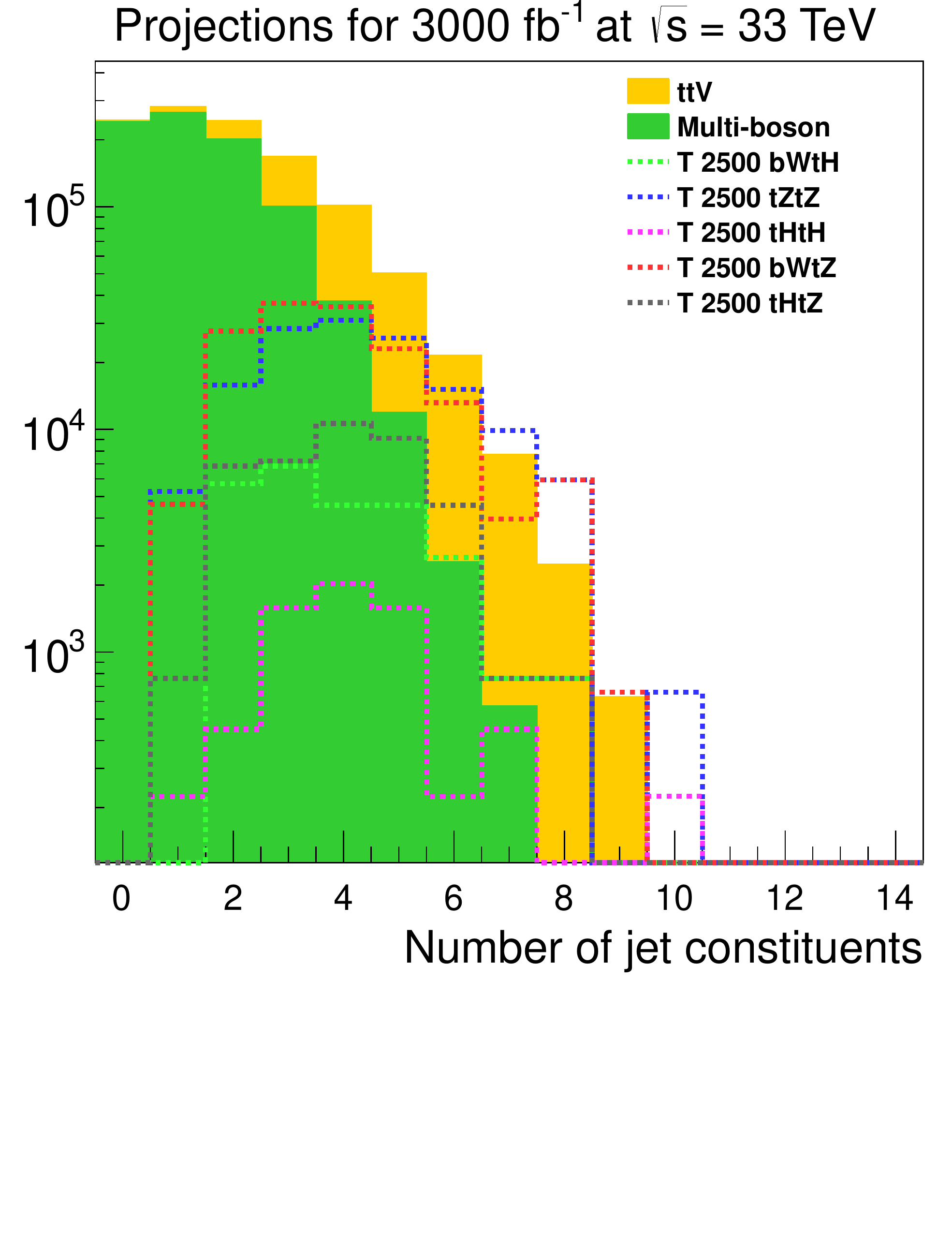}
\caption{Distributions of  H$_{T}$, S$_{T}$ and the number of jet constituents for the SS category. The signal is scaled by 5000.}
\end{figure}

\begin{figure}[!h]
\includegraphics[width=0.3\textwidth]{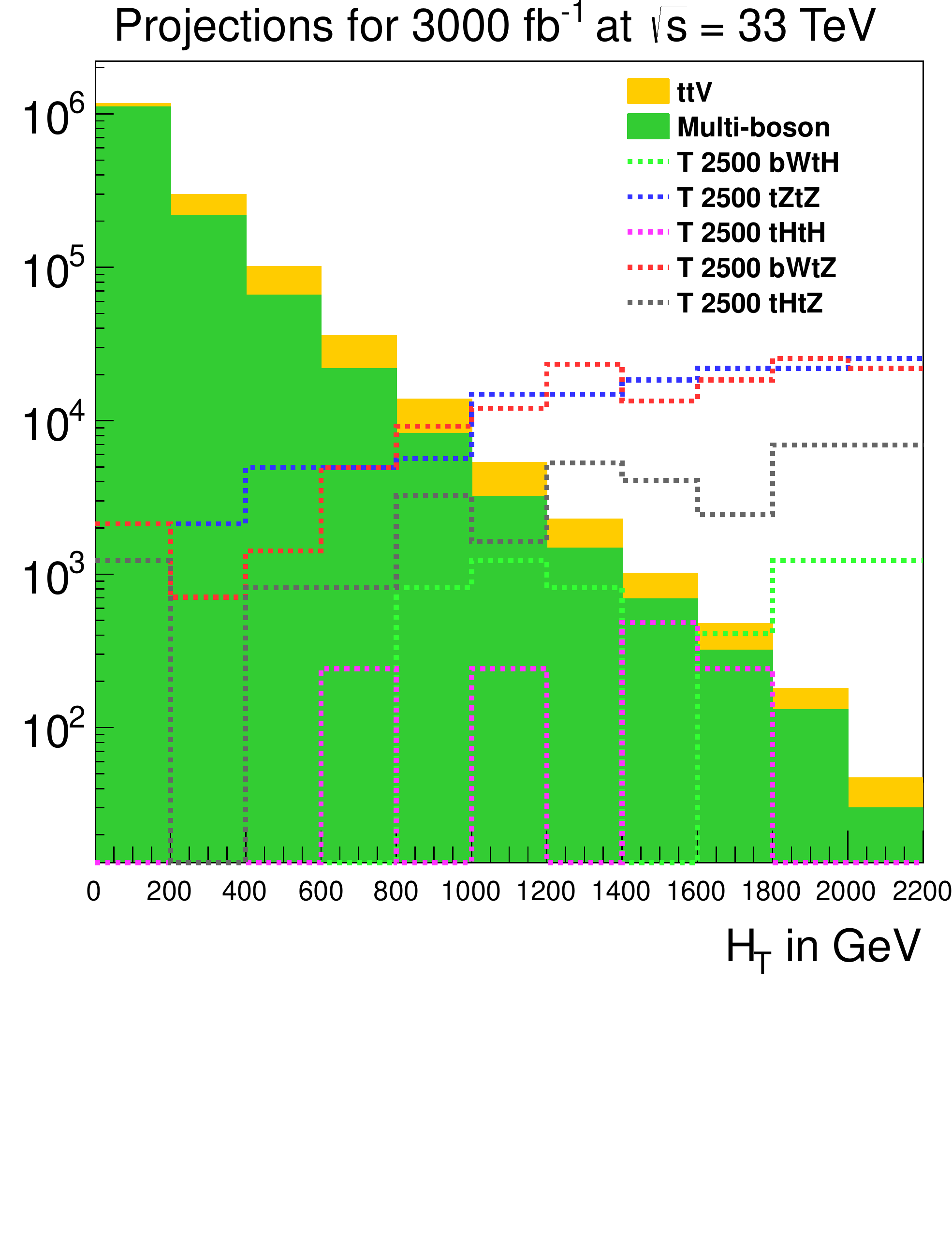}
\includegraphics[width=0.3\textwidth]{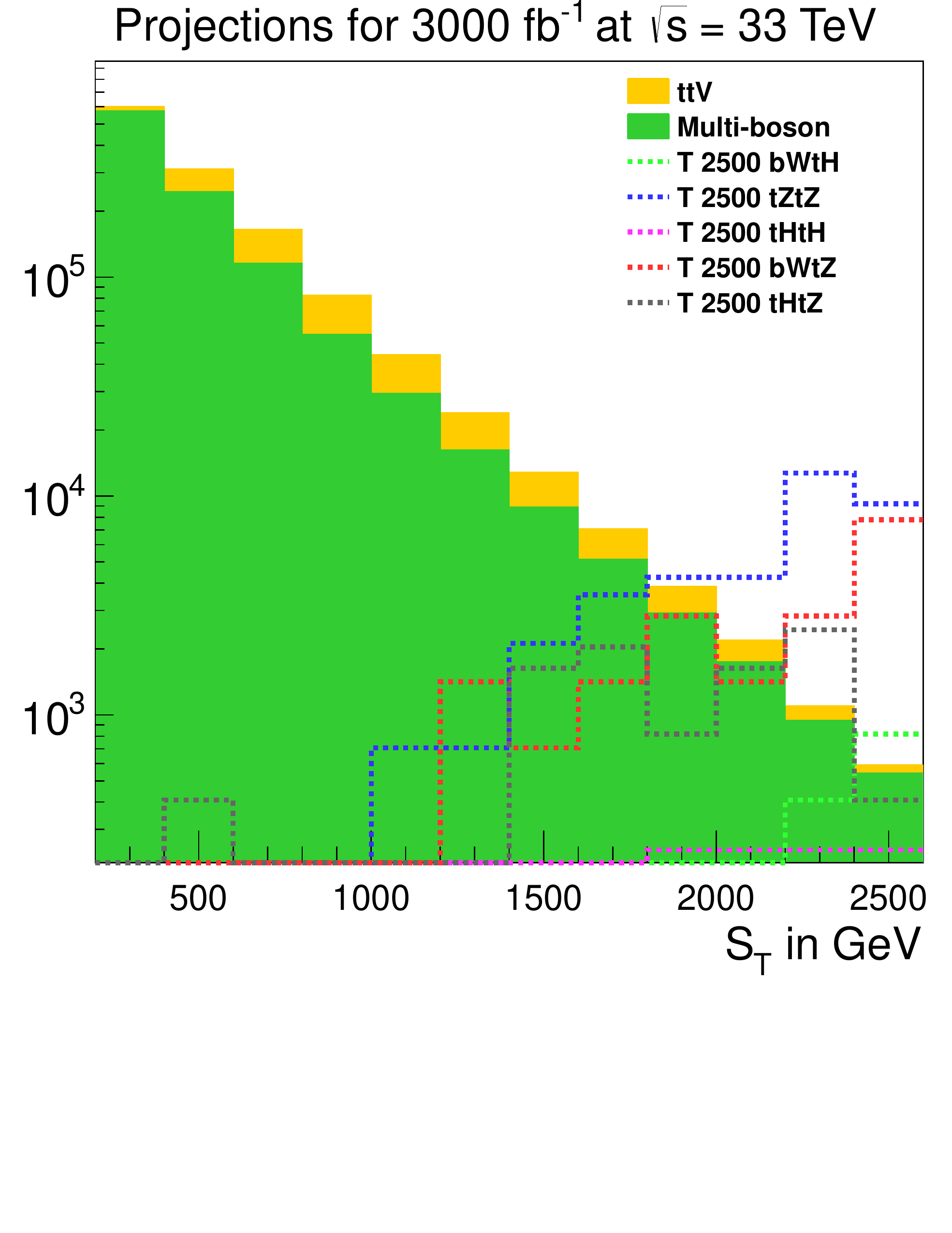}
\includegraphics[width=0.3\textwidth]{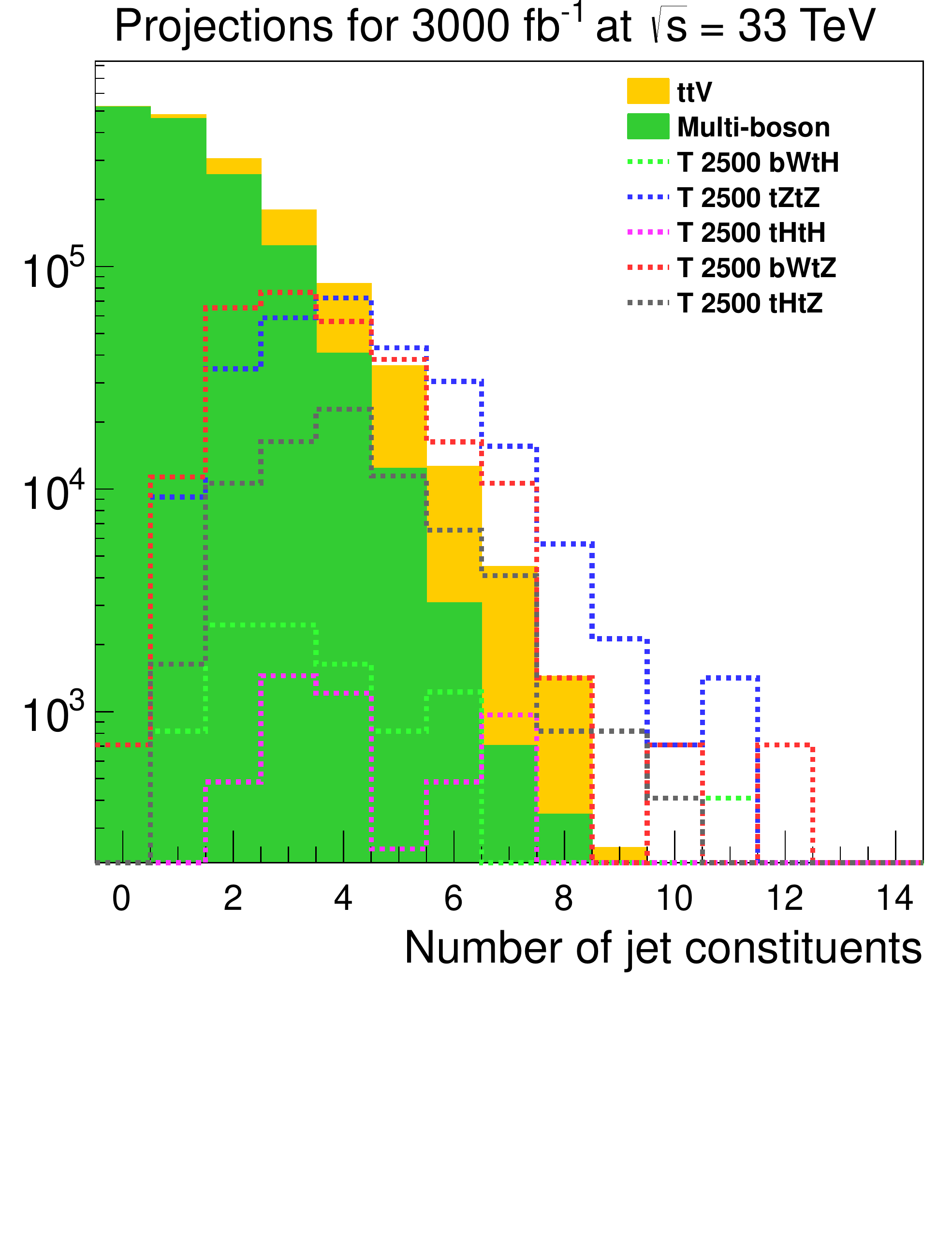}
\caption{Distributions for  H$_{T}$, S$_{T}$ and the number of jet constituents for events with $\geq$ 3 leptons. The signal is scaled by 5000.}
\end{figure}

\end{document}